\documentclass[fleqn,usenatbib]{mnras}

\usepackage{newtxtext,newtxmath}
\usepackage[T1]{fontenc}

\DeclareRobustCommand{\VAN}[3]{#2}
\let\VANthebibliography\thebibliography
\def\thebibliography{\DeclareRobustCommand{\VAN}[3]{##3}\VANthebibliography}

\usepackage{amsmath}	
\usepackage{graphicx}
\usepackage{booktabs}
\usepackage{natbib}
\usepackage{hyperref}
\usepackage{footnote}
\usepackage{rotating}
\usepackage{tablefootnote}
\usepackage{longtable}
\usepackage{multirow}
\usepackage{array}
\usepackage{makecell}
\usepackage{comment}
\usepackage{subcaption} 
\usepackage{tikz}

\usepackage{url}
\usepackage{color}
\usepackage{subcaption}
\usepackage{float}
\usepackage{xcolor}
\usepackage{svg}

\setcellgapes{3pt}
\newcommand{\teff}{T_{\rm eff}}
\newcommand{\logg}{\log g}
\newcommand{\vmic}{\xi_{\rm t}}
\newcommand{\vmac}{V_{\rm mac}}

\newcommand{\Elow}{E_{\rm low}}
\newcommand{\Eup}{E_{\rm up}}

\newcommand{\Eu}[5]{\mbox{$\rm #1\,^{\rm #2}{\rm #3}^{{\rm #4}}_{\rm #5}$}}
\newcommand{\Y}[5]{\mbox{$\rm #1\,^{\rm #2}{\rm #3}^{{\rm #4}}_{\rm #5}$}}
\newcommand{\Sr}[5]{\mbox{$\rm #1\,^{\rm #2}{\rm #3}^{{\rm #4}}_{\rm #5}$}}
\newcommand{\Ba}[5]{\mbox{$\rm #1\,^{\rm #2}{\rm #3}^{{\rm #4}}_{\rm #5}$}}
\newcommand{\Mn}[5]{\mbox{$\rm #1\,^{\rm #2}{\rm #3}^{{\rm #4}}_{\rm #5}$}}
\newcommand{\Ni}[5]{\mbox{${\rm #1}\,^#2{\rm #3}^{{\rm #4}}_{\rm #5}$}}

\newcommand{\Co}[5]{\mbox{$\rm #1\,^{\rm #2}{\rm #3}^{{\rm #4}}_{\rm #5}$}}

\title[3D NLTE abundances of metals]{Observational constraints on the origin of the elements. IX. 3D NLTE abundances of metals in the context of Galactic Chemical Evolution Models and 4MOST}

\author[N. Storm et al.]{Nicholas Storm$^{1}{}^{,2}$\thanks{0000-0002-5259-3974},
Maria Bergemann$^{1}$\thanks{0000-0002-9908-5571},
Philipp Eitner$^{1}{}^{,2}$,
Richard Hoppe$^{1}{}^{,2}$\thanks{0000-0002-8451-6260},
Alex J. Kemp$^{3}$\thanks{0000-0003-2059-5841},\newauthor
Ashley J. Ruiter$^{1}{}^{,4}{}^{,5}{}^{,6}$,
Hans-Thomas Janka$^{7}$\thanks{0000-0002-0831-3330}, 
Andre Sieverding$^{7}$,
Selma E. de Mink$^{7}$\thanks{0000-0001-9336-2825},
Ivo R. Seitenzahl$^{6}$,\newauthor
Evans K. Owusu$^{1}{}^{,4}{}^{,5}{}^{,6}$
\\
$^{1}$Max-Planck-Institut f\"{u}r Astronomie, K\"{o}nigstuhl 17, D-69117 Heidelberg, Germany;\\
$^{2}$Heidelberg University, Grabengasse 1, 69117 Heidelberg, Germany;\\
$^{3}$Institute of Astronomy (IvS), KU Leuven, Celestijnenlaan 200D, 3001 Leuven, Belgium;\\
$^{4}$School of Science, University of New South Wales Canberra, Australia Defence Force Academy, 2600, Australia Capital Territory, Australia;\\
$^{5}$ARC Centre of Excellence for All Sky Astrophysics in 3 Dimensions (ASTRO 3D), Canberra, ACT 2611, Australia;\\
$^{6}$Heidelberger Institut für Theoretische Studien, Schloss-Wolfsbrunnenweg 35, 69118 Heidelberg, Germany;\\
$^{7}$Max Planck Institute for Astrophysics, Karl-Schwarzschild-Str.~1, D-85748 Garching, Germany;\\
}

\date{Accepted XXX. Received YYY; in original form ZZZ}

\pubyear{2024}

\begin{document}
\label{firstpage}
\pagerange{\pageref{firstpage}--\pageref{lastpage}}
\maketitle

\begin{abstract}
\noindent Historically, various methods have been employed to understand the origin of the elements, including observations of elemental abundances which have been compared to Galactic Chemical Evolution (GCE) models. It is also well known that 1D Local Thermodynamic Equilibrium (LTE) measurements fail to accurately capture elemental abundances. Non-LTE (NLTE) effects may play a significant role, and neglecting them leads to erroneous implications in galaxy modelling. In this paper, we calculate 3D NLTE abundances of seven key iron-peak and neutron-capture elements (Mn, Co, Ni, Sr, Y, Ba, Eu) based on carefully assembled 1D LTE literature measurements, and investigate their impact within the context of the OMEGA+ GCE model. Our findings reveal that 3D NLTE abundances are significantly higher for iron-peak elements at [Fe/H]~$< -3$, with (for the first time ever) [Ni/Fe] and (confirming previous studies) [Co/Fe] on average reaching 0.6-0.8 dex, and [Mn/Fe] reaching $-0.1$ dex, which current 1D core-collapse supernova (CCSN) models cannot explain. We also observe a slightly higher production of neutron-capture elements at low metallicities, with 3D NLTE abundances of Eu being higher by +0.2 dex at [Fe/H]~$= -3$. 3D effects are most significant for iron-peak elements in the very metal-poor regime, with average differences between 3D NLTE and 1D NLTE {reaching} up to 0.15 dex. Thus, ignoring 3D NLTE effects introduces significant biases, so including {them} should be considered whenever possible.

\end{abstract}




\section{Introduction}

The investigation into the origin of chemical elements in the Milky Way has a rich and extensive history. While significant progress has been made on understanding the formation of lighter elements (Z $< 30$), the origins and production sites of the heavier neutron-capture elements remain less clear. One commonly used method to explore the production sites is through the so-called Galactic Chemical Evolution (GCE) models. These models use various calculated yields as inputs, along with assumptions on the star formation history, mass function, and gas flows, to generate theoretical predictions of stellar abundances, which can then be compared with observed data \citep[e.g.][]{Gibson2003, Cote2017, Kaur2019, Kobayashi2020a, Matteucci2021}. Nevertheless, many of the observed abundances still rely on measurements using simplified hydrostatic, one-dimensional (1D) model atmospheres and the assumption of local thermal equilibrium (LTE), under which particles are, for example, assumed to follow the Boltzmann distribution.

The astrophysical origin of carbon (C) is still an uncertain topic \citep{Bensby2006, Romano2017}. However, it is well established that there are several sites that are associated with the production of C in our Galaxy: stellar winds of AGB stars \citep[e.g.][]{Habing1996, Nissen2014} {and} supernovae type II (SNe II) of massive stars \citep[e.g.][]{Burbidge1957, Woosley1995, Farmer2021}. Iron-peak (Fe-peak) elements are typically associated with the SNe Ia \citep[see e.g][]{Timmes1995, Kobayashi2020a}. {However,} recent studies \citep[e.g.][]{Taubenberger2017, Ruiter2020} suggested a greater diversity of SN Ia types; we outline some of them briefly here. Firstly, the `textbook' canonical scenario involves the single-degenerate case leading to a Chandrasekhar-mass explosion, where a white dwarf (WD) accretes material from a non-degenerate star in a binary system, approaching the Chandrasekhar-mass limit (M$_\textrm{ch}$) thereby triggering the explosion in the dense, central region of the WD \citep{Whelan1973}. Secondly, another scenario involves two WDs in a close binary resulting in a (violent) merger in a double-degenerate system \citep{Iben1984, Pakmor2012}; the explosion in this case is likely to arise through a double-detonation \citep[e.g.][]{Pakmor2022}. Thirdly, a WD quiescently accreting He-rich material from a stellar companion could also result in a double-detonation (first in the He layer, rapidly followed by a detonation in the CO core) \citep{Livne1990, Fink2010, Shen2018, Goriely2018}. And lastly, (likely rare) collisions of two WDs in high-multiplicity systems due to the Lidov-Kozai mechanism \citep{Katz2012, Kushnir2013, Antognini2016, Toonen2018}. We note that most of the above-mentioned scenarios involve the explosion of a sub-Chandrasekhar mass WD. Recently some studies \citep[e.g.][]{Seitenzahl2013, Kirby2019, Eitner2020, Sanders2021, Palla2021, Eitner2023} suggested that a significant number of SNe~Ia are actually sub-M$_\textrm{ch}$ explosions. In contrast, the 1D LTE abundance measurements of Fe-group species are consistent with GCE models relying solely on canonical Chandrasekhar-mass SN Ia. These results highlight the importance of accurate stellar abundance measurements for unbiased interpretations. 

Neutron-capture elements are typically divided into three main groups: slow (s)-, intermediate (i)-, and rapid (r)-processes, depending on the timescale differences of $\beta$-decay and neutron capture, and on the flux (or density) of neutrons available in the system in given astrophysical conditions. S-process elements are thought to be mostly produced in intermediate-mass stars during the asymptotic giant branch (AGB) phase \citep[e.g.][]{Busso1999, Cristallo2011, Karakas2016}. A recent study by \citet{Guiglion2024} has shown the importance of {non-LTE (NLTE)} effects on the GCE interpretation of the stellar abundances. R-process sites are still heavily debated, but typically these include neutron-driven winds in {core-collapse supernovae (CCSNe)} \citep[e.g.][]{Takahashi1994, Woosley1994, Arcones2013, Bliss2018}, explosions of rapidly rotating magnetised massive stars (also known as magneto-rotational supernovae, MRSN) \citep[e.g.][]{Siegel2017, Halevi2018, Siegel2019, Reichert2023}, and mergers of two neutron stars in a binary system (NSM) \citep[e.g.][]{Rosswog1999, Halevi2018, Siegel2019, Watson2019}. Both \citet{Kobayashi2020a} and \citet{Lian2023} compared GCE tracks to {europium (Eu) abundances and} suggested that the MRSNe are a crucial r-process site. Lastly, i-process sources are considered to be low-metallicity AGB stars \citep{Karinkuzhi2021}, post-AGB stars \citep{Herwig2011}, accreting white dwarfs \citep{Cote2018c, Denissenkov2019} and super-AGB stars \citep{Jones2016}. 

Despite the best efforts of works such as \citet{Kobayashi2020a} on origin of elements, there is still a discrepancy between GCE models and observational data for many elements, including some iron-peak and neutron-capture ones. {Therefore, in this work we approach this question from the observational side. Using previously developed and validated NLTE atomic models and 3D radiation-hydrodynamics (RHD) models atmospheres from the \texttt{Stagger} grid \citep{Magic2013}, we compute 3D NLTE abundances of several important elements of the
{iron-group elements — manganese (Mn), cobalt (Co), and nickel (Ni) — and neutron-capture elements — strontium (Sr), yttrium (Y), barium (Ba), and Eu}. The observed chemical abundance trends corrected for 3D NLTE effect are explored in the context of GCE models}.

{The paper is organised as follows. We describe our input stellar sample and methods for 3D NLTE calculations in the Section \ref{sec:data_selection}. Section \ref{sec:results} discusses the physics of 3D NLTE effects on line formation and the resulting chemical abundances. We compare our [X/Fe] results to the GCE models and explore the astrophysical impact of 3D NLTE abundances in the context of stellar nucleosynthesis and evolution of stripped massive binaries in Section \ref{sec:gce_model}. Finally, we discuss and conclude our results in Sections \ref{sec:discussion} and \ref{sec:conclusion}.}

\section{Methods}
\label{sec:data_selection}

\begin{figure}
\includegraphics[width=1\columnwidth]{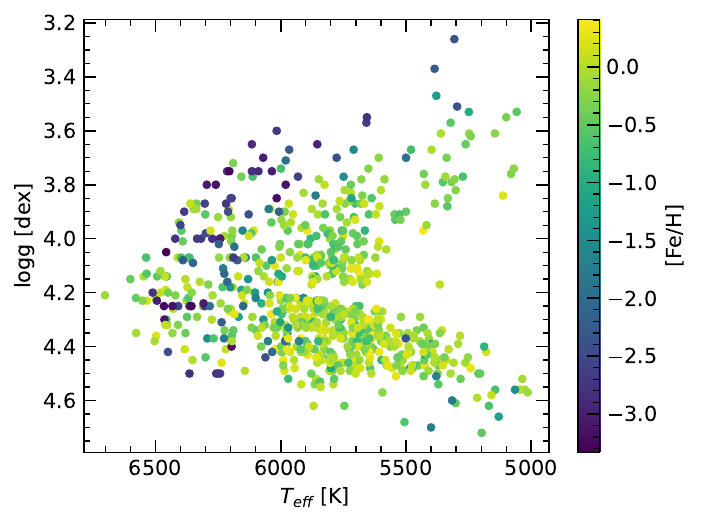}
\caption{HR diagram of the stars in our final sample with [Fe/H] in colour. Note that not all elemental abundances were available or were computed for all of the plotted stars.}
 \label{fig:hr_diagram}
\end{figure}

In this paper, we focus on main-sequence, turn-off and subgiant branch stars, because they are the prime targets for detailed studies of the evolution in the Galactic disc in the 4MIDABLE-HR (4MOST consortium survey 4: MIlky way Disc And BuLgE High-Resolution) survey \citep{Bensby2019}. These stars are most relevant {because their} ages can be determined {more precisely} \citep{Miglio2013, Serenelli2017}. 

\subsection{Input stellar sample}

{Our input stellar sample contains abundance measurements for $746$ stars (Fig. \ref{fig:hr_diagram}). The sample was selected to cover stars in the following parameter space: $5000 \textrm{ K} \lesssim \teff \lesssim 6700\textrm{ K}$, $3.2\textrm{ dex} \lesssim \logg \lesssim 4.7\textrm{ dex}$, -3.4 dex $\lesssim$ [Fe/H] \footnote{We use the standard notation [X/Y] = log($N_{\rm X}$/$N_{\rm Y}$) - log($N_{\rm X}$/$N_{\rm Y})_{\odot}$ and A(X) = log$_{10}$($N_{\rm X}/N_{\rm H}$) + 12 for elements X and Y, where $N_{\rm X}$ is the number of atoms per unit volume.}, and microturbulence values representative of dwarfs and subgiants \citep{Smiljanic2014}, $0.6 \textrm{ km s}^{-1} \leq \vmic \leq 3.0 \textrm{ km s}^{-1}$. In short, the input data sources are as follows:} 

- {The study by \citet{Bonifacio2009} is a part of the "First Stars" paper series, and it provides abundance measurements for low-metallicity stars (-3.8 $\lesssim$ [Fe/H] $\lesssim$ -2.5), mostly for F-type dwarfs. The sample contains abundances of C (based on the G-band of the CH molecule), Mn, Co, Ni, Sr and Ba}.

- \citet{Hansen2013} derived Sr abundances for 21 main-sequence and red giants in the metallicity range -3.1 $\lesssim$ [Fe/H] $\lesssim$ -0.5. {We adopt [Sr/Fe] 1D LTE measurements for 7 main-sequence stars from this sample.}

- \citet{Bensby2014, Battistini2015, Battistini2016} analysed the same sample of 714 F and G dwarf and subgiant stars in the Solar neighbourhood{, with 94\% of the sample having metallicity -1 $\lesssim$ [Fe/H] $\lesssim$ 0.4, with their lowest star having [Fe/H]$ = -2.62$}. They determined both stellar parameters and elemental abundances of 24 elements, providing a well-sampled distribution of {Mn, Co, Ni, Sr, Y, Ba and Eu elements} from metal poor to super-solar metallicity stars.

- \citet{Zhao2016} derived abundances for 17 chemical elements in 51 kinematically selected Galactic thin and thick disc stars, and halo F- and G-type dwarfs in {the metallicity} range, -2.7 $\lesssim$ [Fe/H] $\lesssim$ 0.3. We {adopted C (based on CH G-band), Sr, Ba and Eu abundances for 47 stars}.

- \citet{Li2022} {analysed high-resolution SUBARU spectra of 385 metal-poor stars selected from the LAMOST survey \citep[][their data releases DR1 to DR5]{Cui2012, Luo2012, Zhao2012}}. They obtained both stellar parameters and abundance measurements for {$21$} elements. We decided to introduce a cut on the SNR $>$ 75 to only use the best abundance measurements. {As a result, we include 52 main sequence and subgiant stars in the metallicity range} of -3.8 $\lesssim$ [Fe/H] $\lesssim$ -1.8. 

- Recently \citet{Mardini2024} derived abundances for 27 near main-sequence turnoff stars in the metal-poor regime -3.6 $\lesssim$ [Fe/H] $\lesssim$ -2.5. {We used their derived C (based on the CH G-band), Mn, Co, Ni, Sr and Ba abundances.}
\subsection{Metallicities}

{For metallicities ([Fe/H]) we adopted values based on singly ionized (Fe~II) lines. As demonstrated in the literature, Fe II lines in spectra of main-sequence stars and subgiants are almost unaffected by NLTE \citep{Mashonkina2011, Bergemann2012a, Lind2012} and these lines are weakly sensitive to effects of 3D convection  \citep{Amarsi2016a, Lind2017, Amarsi2022}. Only two of our chosen literature sources \citep{Bensby2014, Hansen2013} provided exclusively Fe~I-based abundances. However, both studies included 1D NLTE corrections based on the calculations by \citet{Lind2012}. Moreover, the sample from \citet{Bensby2014} only contributes in a minor way at the lowest metallicity. In the regime of $-2.7 \lesssim $ [Fe/H] $ \lesssim -2$, we have only 3 values of [Ba/Fe] and 2 values of [Ni/Fe].}

\begin{figure}
\includegraphics[width=1\columnwidth]{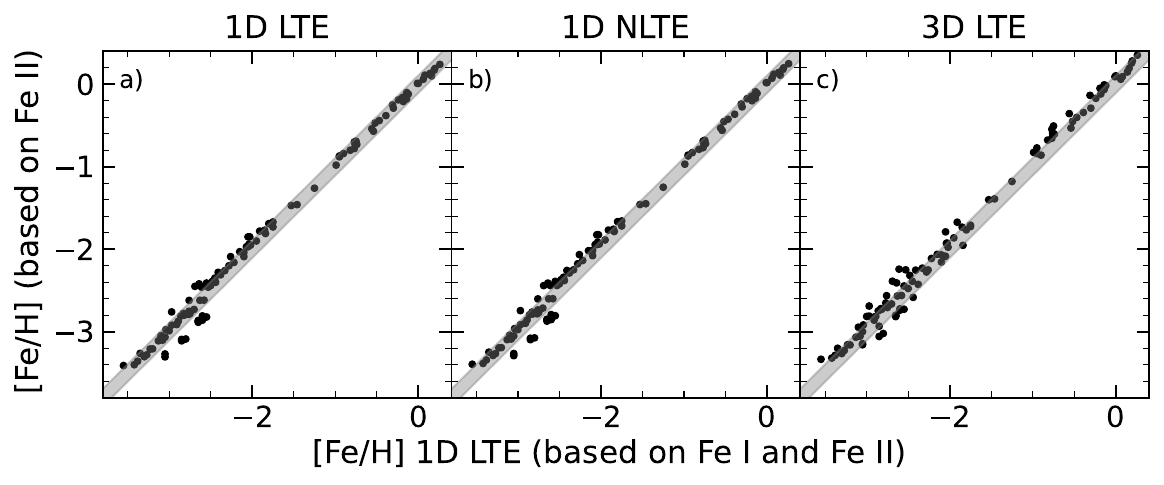}
\caption{{Metallicities for the selected sample of stars determined using Fe~I and Fe~II lines (x-axis) versus [Fe/H] determined using Fe~II lines only (y-axis). The panels (b) and (c) illustrate the effect of correcting the Fe~II-based values for 1D NLTE and 3D LTE, respectively.}}
 \label{fig:fe2h_vs_feh}
\end{figure}

{To reassure that our metallicity scale is not affected by NLTE and 3D, we have also explored the influence of these effects in Fig. \ref{fig:fe2h_vs_feh}. Clearly, the agreement between the [Fe/H] values based on Fe~I and Fe~II lines and that based on Fe~II lines is excellent, which is partly because of the chosen analysis methodology in these studies. However, as we show in the middle panel, 1D NLTE corrections to Fe~II lines are negligibly small and do not exceed $\leq0.02$ dex for all stars in our sample. These corrections were calculated using the results from \citet{Bergemann2012a}\footnote{\url{nlte.mpia.de}}. We also computed 3D LTE line formation for a representative sample of diagnostic Fe~II lines (right panel) and we find that 3D effects, as expected, have a slightly larger influence on Fe abundance derived from the Fe~II lines. The 3D LTE - 1D LTE difference is of $\approx 0.05$ dex, which is very small compared to other uncertainties in the chemical abundance analysis.}

{Finally, we have opted to adopt the literature values based on Fe II lines but we corrected them for 3D effects. We have also verified that our chemical abundance distributions and conclusions do not change depending on the choice of metallicity scale.}
\subsection{Spectrum synthesis calculations}
\label{subsec:spectrum_synthesis}

\begin{figure}
\includegraphics[width=1\columnwidth]{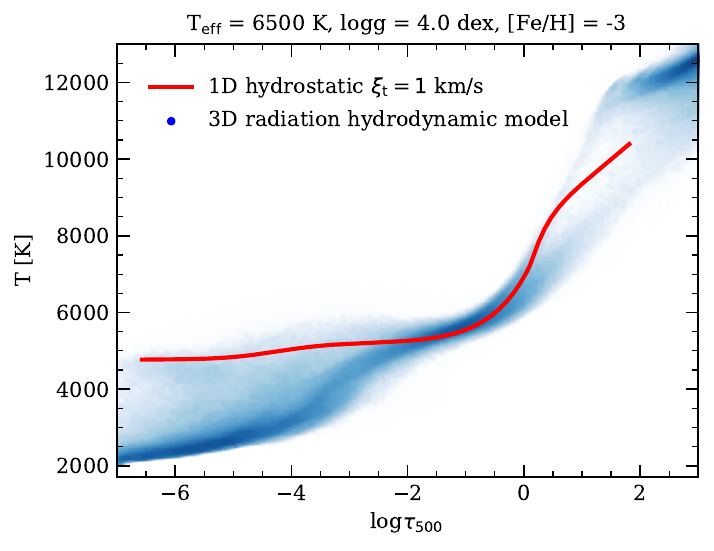}
\caption{{Temperature structure of the 1D hydrostatic MARCS (red) and 3D RHD \texttt{Stagger} (blue) model atmospheres as a function of $\log \tau_{500}$ for $\teff = 6500$~K, $\logg = 4.0$~dex and [Fe/H] $= -3$.}}
 \label{fig:1d_vs_3d_atmo}
\end{figure}

\begin{figure*}
     \centering
    \begin{subfigure}[t]{0.33\textwidth}
        \raisebox{-\height}{\includegraphics[width=\textwidth]{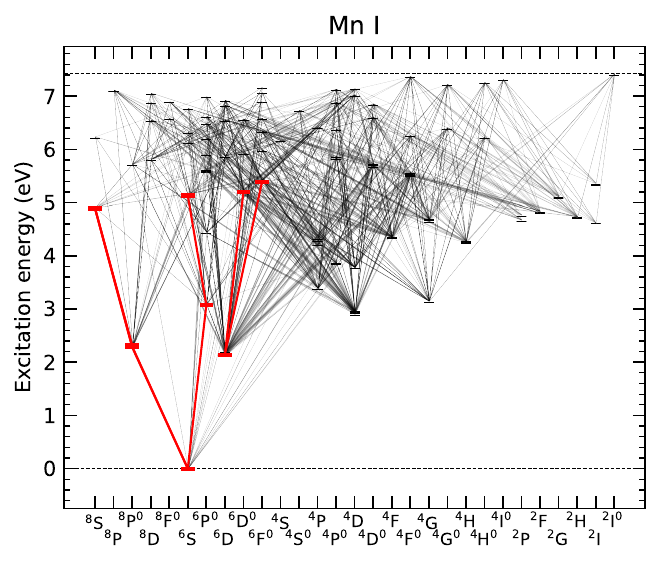}}
    \end{subfigure}
    \hfill
    \begin{subfigure}[t]{0.33\textwidth}
        \raisebox{-\height}{\includegraphics[width=\textwidth]{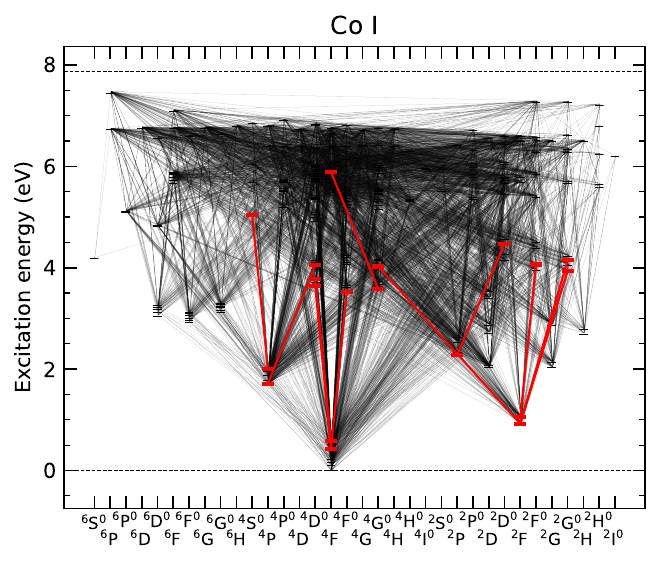}}
    \end{subfigure}
    \begin{subfigure}[t]{0.33\textwidth}
        \raisebox{-\height}{\includegraphics[width=\textwidth]{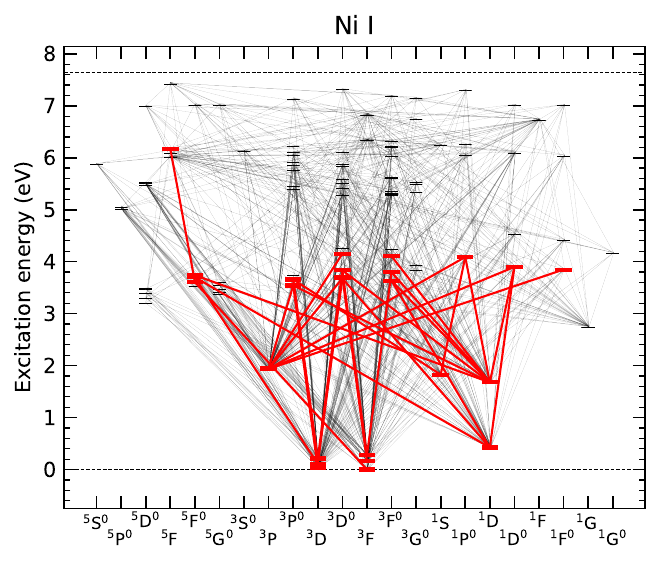}}
    \end{subfigure}
    \begin{subfigure}[t]{0.33\textwidth}
        \raisebox{-\height}{\includegraphics[width=\textwidth]{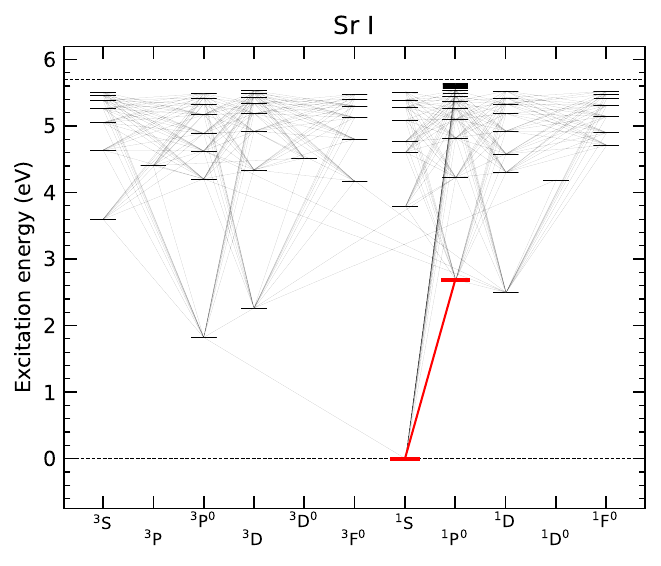}}
    \end{subfigure}
    \hfill
    \begin{subfigure}[t]{0.33\textwidth}
        \raisebox{-\height}{\includegraphics[width=\textwidth]{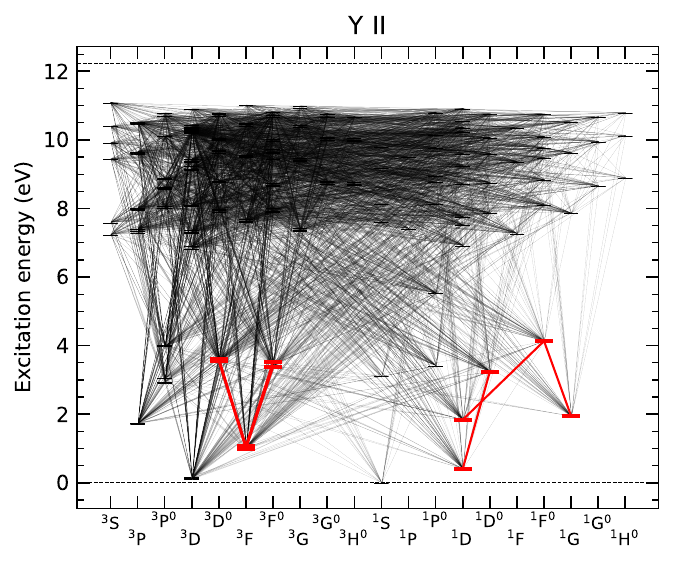}}
    \end{subfigure}
    \begin{subfigure}[t]{0.33\textwidth}
        \raisebox{-\height}{\includegraphics[width=\textwidth]{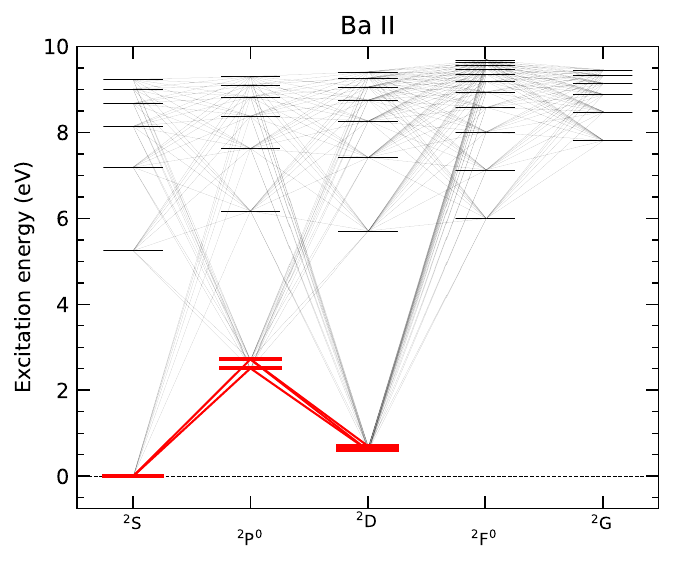}}
    \end{subfigure}
    \begin{subfigure}[t]{0.33\textwidth}
        \raisebox{-\height}{\includegraphics[width=\textwidth]{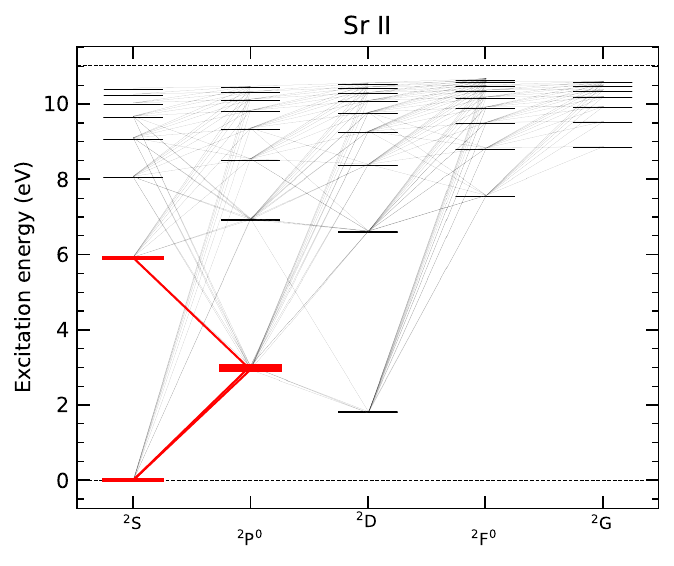}}
    \end{subfigure}
    \hfill
    \begin{subfigure}[t]{0.33\textwidth}
        \raisebox{-\height}{\includegraphics[width=\textwidth]{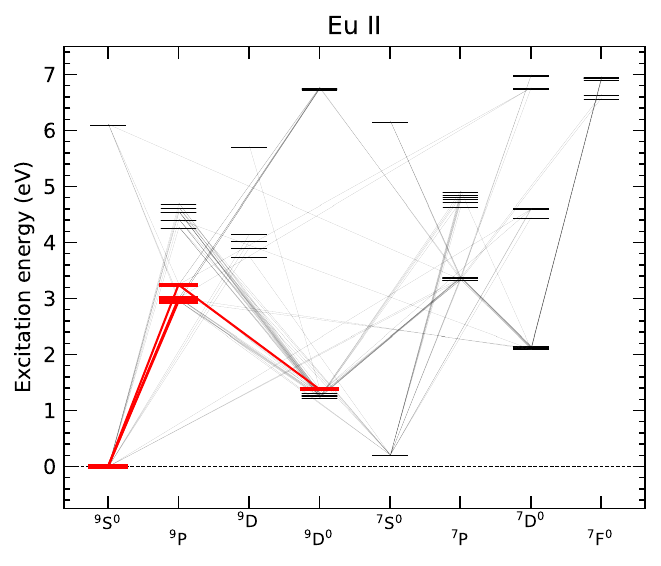}}
    \end{subfigure}
    \begin{subfigure}[t]{0.33\textwidth}
        \raisebox{-\height}{\includegraphics[width=\textwidth]{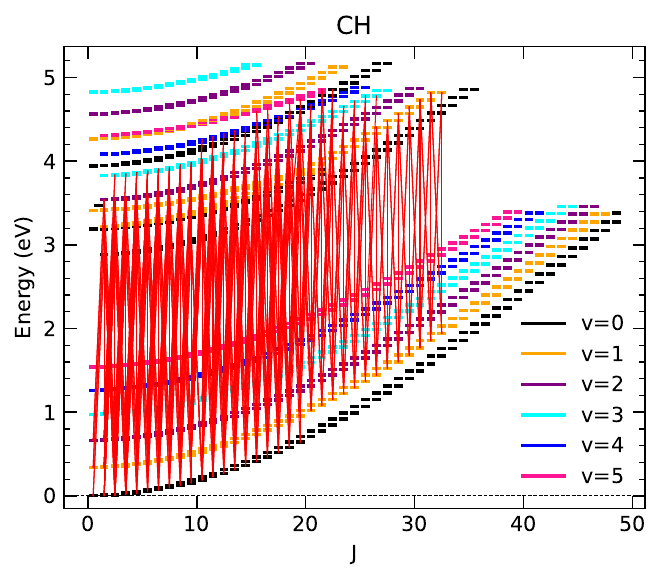}}
    \end{subfigure}
    \caption{Grotrian diagrams of the model atoms with all bound-bound transitions in black, and diagnostic lines used in this study highlighted in red. The last subplot is the model molecule of CH, where we did not plot unused bound-bound transitions due to their sheer amount. For CH $\nu$ represents vibrational quantum number.}
    \label{fig:grotrian}
\end{figure*}

{The 1D LTE, 1D NLTE, and 3D NLTE synthetic profiles and their equivalent widths (EW) were computed using the \textsc{multi3d at dispatch} code \citep{Eitner2024}. 3D RHD model atmospheres were adopted from the \texttt{Stagger} grid \citep{Magic2013}. For 1D calculations, we utilised the MARCS 1D hydrostatic model atmospheres \citep{Gustafsson2008}. Fig. \ref{fig:1d_vs_3d_atmo} shows the temperature structures of the 1D MARCS and the 3D RHD models as a function of optical depth in the continuum at 500 nm, $\log \tau_{500}$. Both models have the same parameters: $\teff = 6500$ K, $\logg = 4.0$ dex and [Fe/H] $= -3$. In outermost atmospheric layers, the 1D MARCS is significantly hotter than its 3D RHD counterpart, by over $1\,000$ K \citep[see also][]{Bergemann2017, Bergemann2019}. The difference in the physical structure of 1D hydrostatic vs 3D RHD models is at the origin of large 3D NLTE effects in abundances, as described in Sect. \ref{subsec:3d_nlte_vs_1d_lte_atmo} below.}

{The NLTE models were adopted from the following studies: CH \citep[][]{Popa2023}, Mn \citep{Bergemann2019}, Co \citep{Bergemann2010b, Yakovleva2020}, Ni \citep{Bergemann2021, Voronov2022}, Sr \citep[][]{Bergemann2012b, Gerber2023}, Y \citep{Storm2023, Storm2024}, Ba \citep{Gallagher2020} and Eu \citep{Storm2024}. For CH, we updated the dissociation energy of the molecule to 3.47 eV. Fig. \ref{fig:grotrian} shows the Grotrian diagrams for the chemical elements modelled in NLTE in this work.}

{In 3D NLTE statistical equilibrium was solved in cubes with the resolution of $(x,y,z) = (30,30,230)$ for Ba and $(x,y,z) = (20,20,230)$ for other chemical elements. The geometric resolution was chosen as a compromise between the precision of 3D NLTE \citep[see also][]{Bergemann2019, Bergemann2021} and the computational expense of 3D NLTE calculations. We have also verified that abundance difference in the spectra compared to spatial resolution of $(x,y,z) = (90,90,230)$ does not exceed $0.05$ dex for the spectral lines with EW $< 100$ m\AA. This is the case for the majority lines used for abundance analysis in our project, especially in metal-poor stars.}

{Line-by-line abundance corrections were computed by first computing EWs for each stellar model with the following steps: $\teff$, $\logg$, [Fe/H] and [X/Fe] respectively 500 K, 0.5 dex, 1.0 dex (0.5 at [Fe/H] = +0.5) and 0.2-0.3 dex (0.1 dex for 1D LTE). For 1D models the microturbulence $\vmic$ was varied from $0.50$ to $3.00$ km s$^{-1}$ in steps of 0.25 km s$^{-1}$. Next, the 1D LTE EW  was matched to the corresponding 1D or 3D NLTE EW to determine the NLTE correction. Thus, the line-by-line abundance corrections were determined by interpolating between corrections corresponding to different $\teff$, $\logg$, [Fe/H] and $\vmic$, and the closest [X/Fe] value. When individual line data were available, abundance corrections were computed separately for each line; otherwise, an average NLTE correction was applied for all lines used in the relevant literature source. All lines used for different elements from each source is provided in Appendix \ref{app:lines}. For CH, only 1D NLTE calculations are explored in this work, owing to additional complexities of handling molecules in 3D \citep{Gallagher2020}.}

\begin{figure*}
\includegraphics[width=1\textwidth]{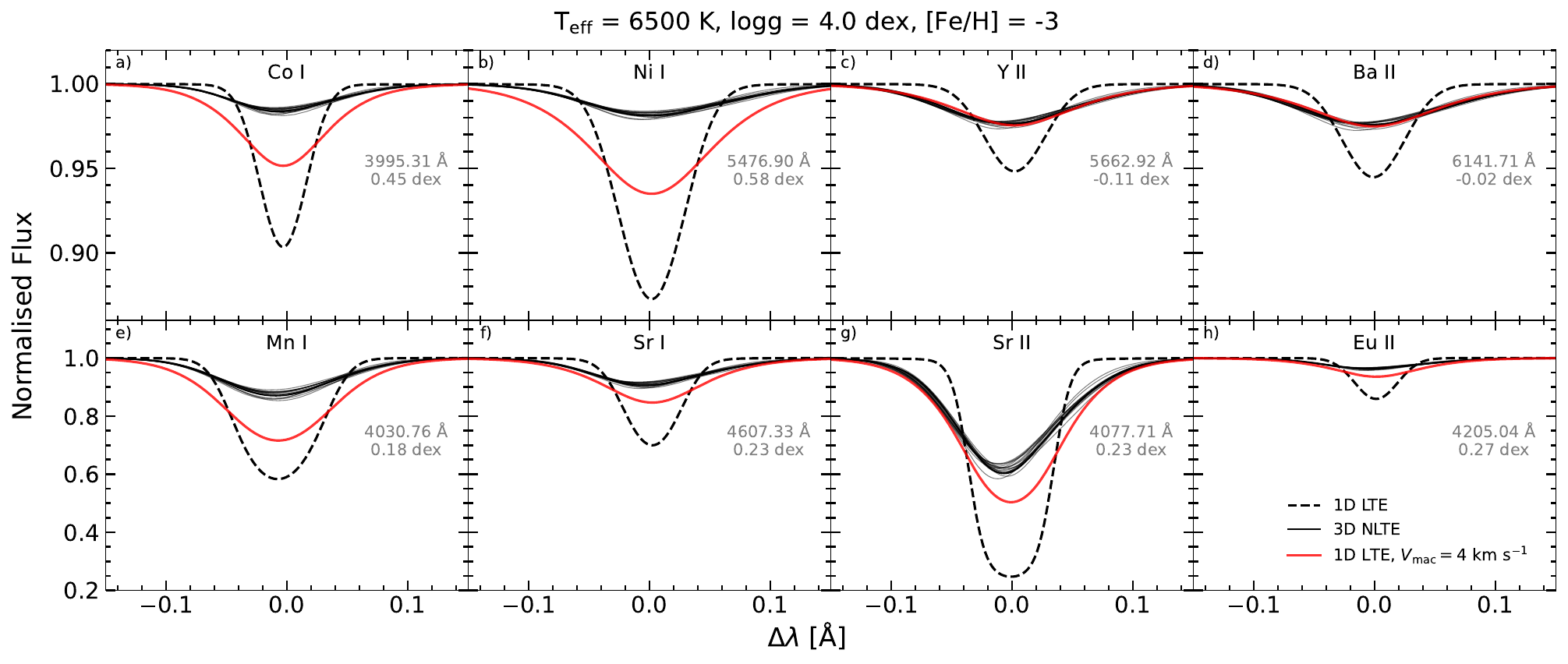}
\caption{Line profiles of some diagnostic lines for several elements for the model atmosphere $\teff = 6500$ K, $\logg = 4.0$ dex and [Fe/H] = -3. Black dashed line is for the 1D LTE MARCS atmosphere with $\vmic = 1$ km s$^{-1}$, while the red solid line is with an additional broadening of $\vmac = 4$ km s$^{-1}$. The solid black lines correspond to profiles from individual 3D NLTE \texttt{Stagger} snapshots. Numbers represent respectively top to bottom: the wavelength of the line and by how much 3D NLTE abundance would need to be increased to match 1D LTE EW. All spectra were computed with [X/Fe] = 0, except Sr I, Y II and Eu II abundances were increased to get a stronger line for a more clear illustration, by respectively 2.5, 1.3 and 1.75 dex.}
 \label{fig:spectra_3d_1d}
\end{figure*}

\section{Results}\label{sec:results}

{The main scientific novelty of our work is the use of 3D NLTE corrected elemental abundances for studies of Galactic chemical evolution. We begin with the qualitative analysis of 3D NLTE effects on abundances of the chemical elements (Sec. \ref{subsec:4most_lines_3d_nlte}), and then proceed with the quantitative analysis of 3D NLTE [X/Fe] trends in the context of GCE (Sec. \ref{sec:gce_model}).}
\subsection{Physical effects of 3D and NLTE on abundances}
\label{subsec:3d_nlte_vs_1d_lte_atmo}
{In Fig. \ref{fig:spectra_3d_1d}, we demonstrate the impact of 3D NLTE line formation on selected diagnostic lines of Mn I, Co I, Ni I, Sr I, Sr II, Y II, Ba II and Eu II for a typical model of a metal-poor turn-off star.}

{As expected from previous studies \citep{Bergemann2010b, Bergemann2019, Eitner2023}, our calculations confirm that 3D NLTE line profiles of Fe-peak elements (Co I, Ni I, Mn I) are significantly weaker compared to 1D LTE. This effect is caused by stronger NLTE over-ionisation in metal-poor 3D models, which arises due to stronger UV radiation fluxes owing to a reduced line blanketing \citep[see e.g.][]{Bergemann2012a, Nordlander2017, Bergemann2019}. Also for Sr I lines, the NLTE over-ionisation is prominent, as shown in \citet{Bergemann2013} in 1D. For the diagnostic lines of singly-ionised species, Y II, Ba II, Sr II, and Eu II, the effects of 3D NLTE are smaller. The resulting synthetic profiles in 3D NLTE are not too different from 1D LTE profiles computed with a $\vmac$ of 4 km~s$^{-1}$. In addition, an interesting effect of 3D RHD line formation is that line profiles are asymmetric and typically blue-shifted in the cores, which is due to the complex distribution of velocities associated with up-flows in the granules and down-flows in the inter-granular lanes \citep[see section 4.2.1 in ][]{Bergemann2019}. A more in-depth comparison of the differences between 1D and 3D atmospheric model can also be found in Appendix \ref{app:1d_vs_3d}.}

\begin{figure}
\includegraphics[width=1\columnwidth]{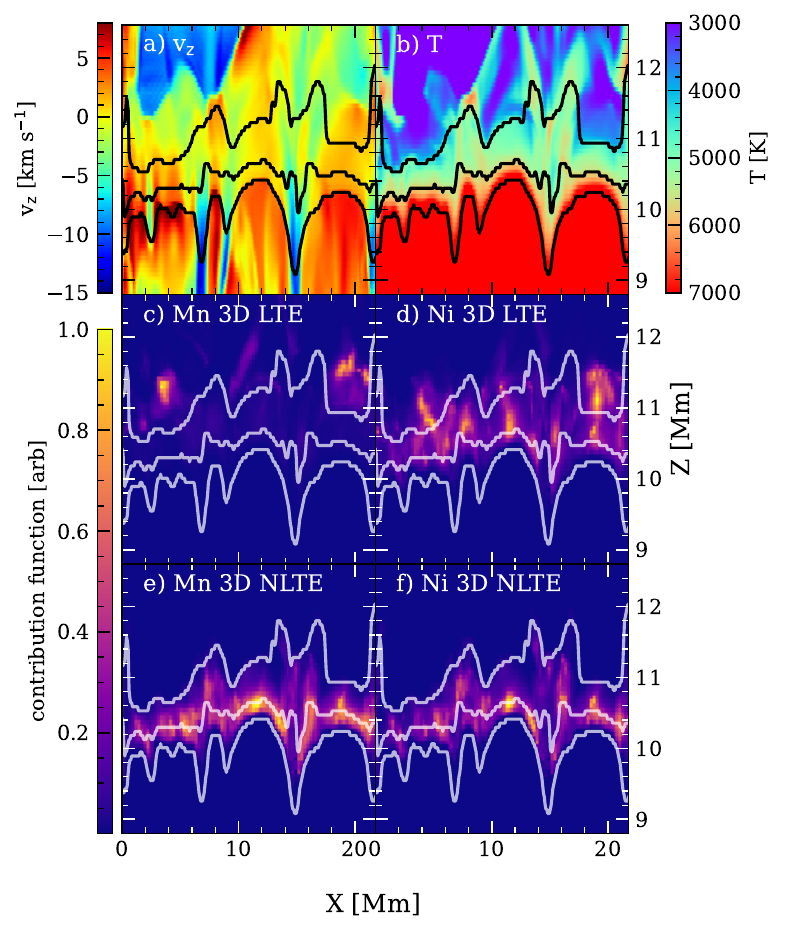}
\caption{{2D slices in the x-z plane of the 3D \texttt{Stagger} model atmosphere with parameters $\teff = 6500$ K, $\logg = 4.0$ dex, and [Fe/H] = -3. Top left: vertical velocity v$_{\rm z}$ with black contours denoting $\log \tau_{500}$ values of $-3, -1, 1$, from top to bottom (contours are preserved throughout other panels). Top right: temperature structure. Middle panels: 3D LTE of the contribution function (defined in the text) in arbitrary units for Mn line at 4030.76 \AA~with [Mn/Fe] = 0.0 (left) and Ni line 5476.90 \AA~with [Ni/Fe] = 0.5 (right). Bottom panels: same as middle panels but for the 3D NLTE case.}}
 \label{fig:cont_func_3d_both}
\end{figure}

{The most intuitive way to understand the properties of line formation in 3D RHD is to explore the contribution function (CF) of lines \citep[see for example, ][]{Amarsi2015, Bergemann2017}. In Fig. \ref{fig:cont_func_3d_both}, we show the normalised CF values for the line cores, defined the same as in eq. 3 in \citet{Bergemann2017}:

\begin{equation}
    CF_{\tau, \nu} = \dfrac{(\ln{10})\tau_{500}}{\kappa_{500}}\kappa_{\textrm{l},\nu} \int_0^1(I_{\rm c} - S_{\rm l}) \exp^{-\tau/\mu} d\mu,   
\end{equation}

where $\tau_{500}$ and $\kappa_{500}$ are optical depth and opacity at 500 nm, $\kappa_{\textrm{l},\nu}$ is the line opacity at a given frequency, $S_{\rm l}$ is the line source function, $I_{\rm c}$ is the intensity in the continuum and $\mu = \cos{\theta}$ is the angle between the ray and the direction to the observer. The CF is plotted for Mn I line 4030.76 \AA~with excitation potential (E$_{\rm low}$) of 0.0 eV and Ni I line 5476.90 \AA~with E$_{\rm low}$ of 1.83 eV in the x-z slice of a 3D RHD metal-poor turnoff model atmosphere. We also show the vertical velocity (panel a) and temperature distributions (panel b) of the same slice, in order to highlight the correlations with these physical properties. The colour scheme is chosen such that the brighter area highlights regions with the maximum CF, that is where most of radiation at the line core of the diagnostic line forms. As seen in Fig. \ref{fig:cont_func_3d_both} (panels c, d), in 3D LTE both Ni I and Mn I features preferentially form in the upper part of the atmosphere. This is because in LTE, the line opacity and the source functions are coupled to local temperature, and hence the line formation is confined to cooler regions above the granules. As a consequence, these lines are significantly stronger in 3D LTE. In contrast, in NLTE the opacities and source functions are set by the radiation field and the CF peaks around the optical surface $\log \tau_{500} \approx -1$. As a result, the lines are much weaker and the 3D NLTE corrections are large and positive. The line formation of other diagnostic lines is qualitatively similar, however, the amplitude of NLTE effects depends strongly on atomic parameters of each transition \citep{Bergemann2014}.}

\subsection{Effects of 3D NLTE on 4MOST lines}
\label{subsec:4most_lines_3d_nlte}

\begin{figure}
\includegraphics[width=1\columnwidth]{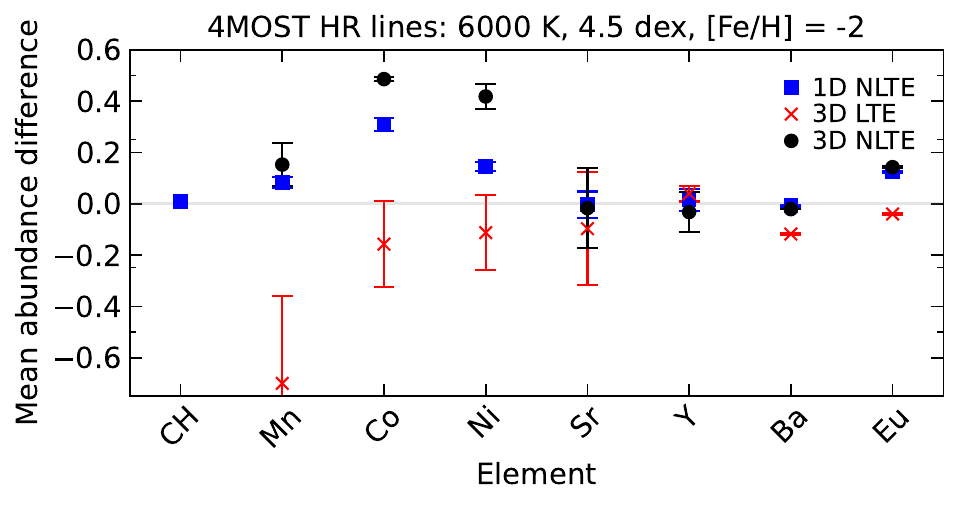}
\includegraphics[width=1\columnwidth]{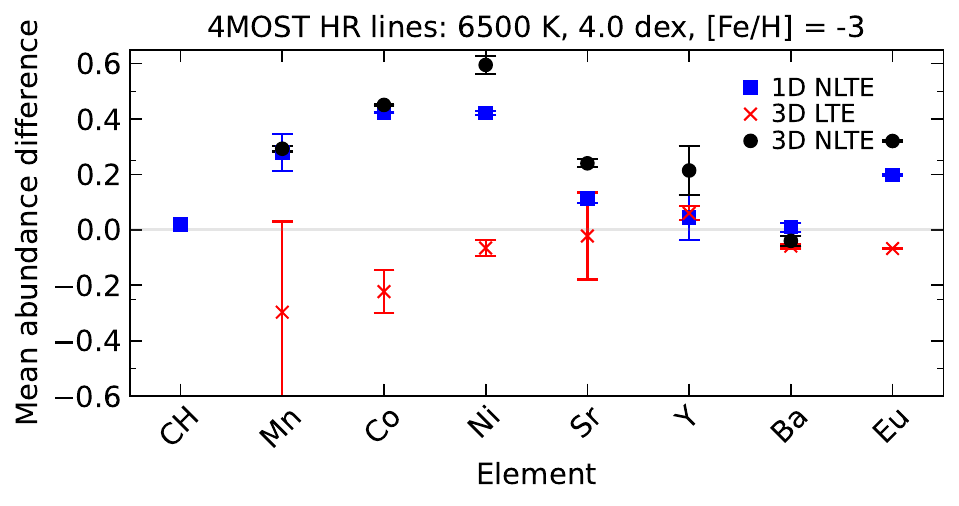}
\includegraphics[width=1\columnwidth]{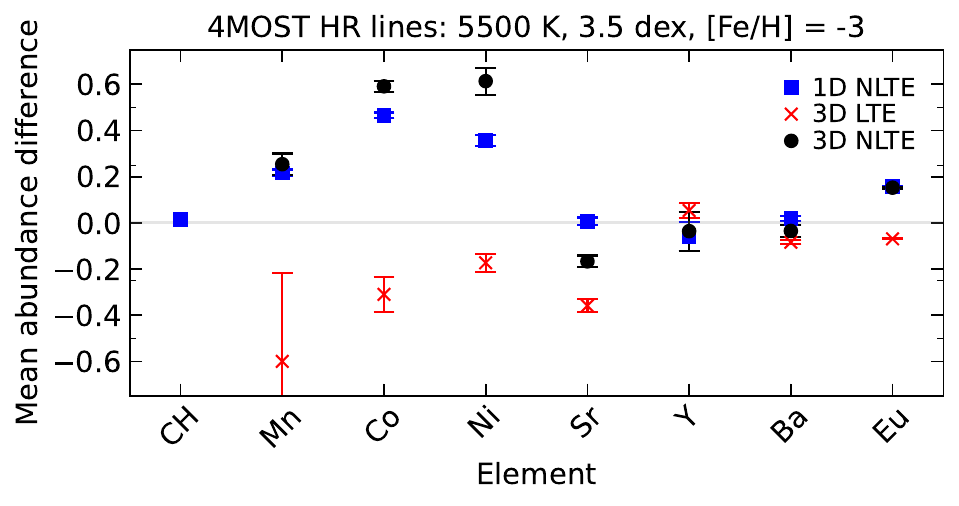}
\caption{The average difference between 1D NLTE, 3D LTE, and 3D NLTE abundances (shown as blue squares, red crosses, and black circles, respectively) compared to 1D LTE values for diagnostic lines in the 4MOST HR windows for three model atmospheres. Error bars indicate the standard deviation across different lines.}
 \label{fig:all_corrections_one_ms}
\end{figure}

In this section we explore the effects on abundances for the representative stars to be observed within the scope of the 4MIDABLE-HR survey. In 4MOST HR windows the wavelength coverage {are} roughly 3926-4355, 5160-5730 and 6100-6790 \AA. Out of analysed diagnostic lines in this project, we have 4 Mn I lines, 8 Co I lines, 17 Ni I lines, 4 Sr II lines, 3 Y II lines, 2 Ba II lines and 3 Eu II lines (see also Tables \ref{app:lines} and \ref{tab:lines_data_ni}).

The resulting average abundance difference between 1D LTE and other cases for 4MOST lines in the HR window are shown in {Fig. \ref{fig:all_corrections_one_ms}} for respectively representative main-sequence, turn-off and subgiant stars. The standard deviation of the difference for different lines are plotted as error bars. 1D NLTE (in blue) and 3D NLTE (in black) have similar NLTE trends in most cases. 3D NLTE in general amplifies the NLTE effects and can result in biases of up to 0.3 dex compared to 1D NLTE. 3D NLTE effects are the strongest for iron-peak elements with the differences reaching 0.3-0.6 dex. Sr and Eu are the most NLTE affected among neutron-capture elements with differences compared to 1D LTE of up to 0.2 dex. 3D LTE (in red) clearly has different behaviour, and should generally not be trusted over NLTE results. This plot also indicates that the combination of 1D NLTE and 3D LTE is not equivalent to the self-consistent 3D NLTE analysis, which is sometimes done in the literature. For a more detailed analysis, we also plotted 1D NLTE and 3D NLTE corrections for typical turn-off, main-sequence and subgiant stars for some diagnostic lines in Appendix \ref{app:nlte_corrections}.

In summary, we find that for high-resolution studies of the Galactic disc and bulge it is essential to use 3D NLTE models for high-quality abundance diagnostics. The most sensitive elements are {Mn, Co, Ni} and Eu, for which 3D NLTE abundance corrections may exceed 0.3 to 0.6 dex at metallicity $\lesssim -2$. But the effects are also substantial at higher metallicities, which will also be the focus of observations of 4MIDABLE-HR survey.

\begin{figure*}
\includegraphics[width=1\textwidth]{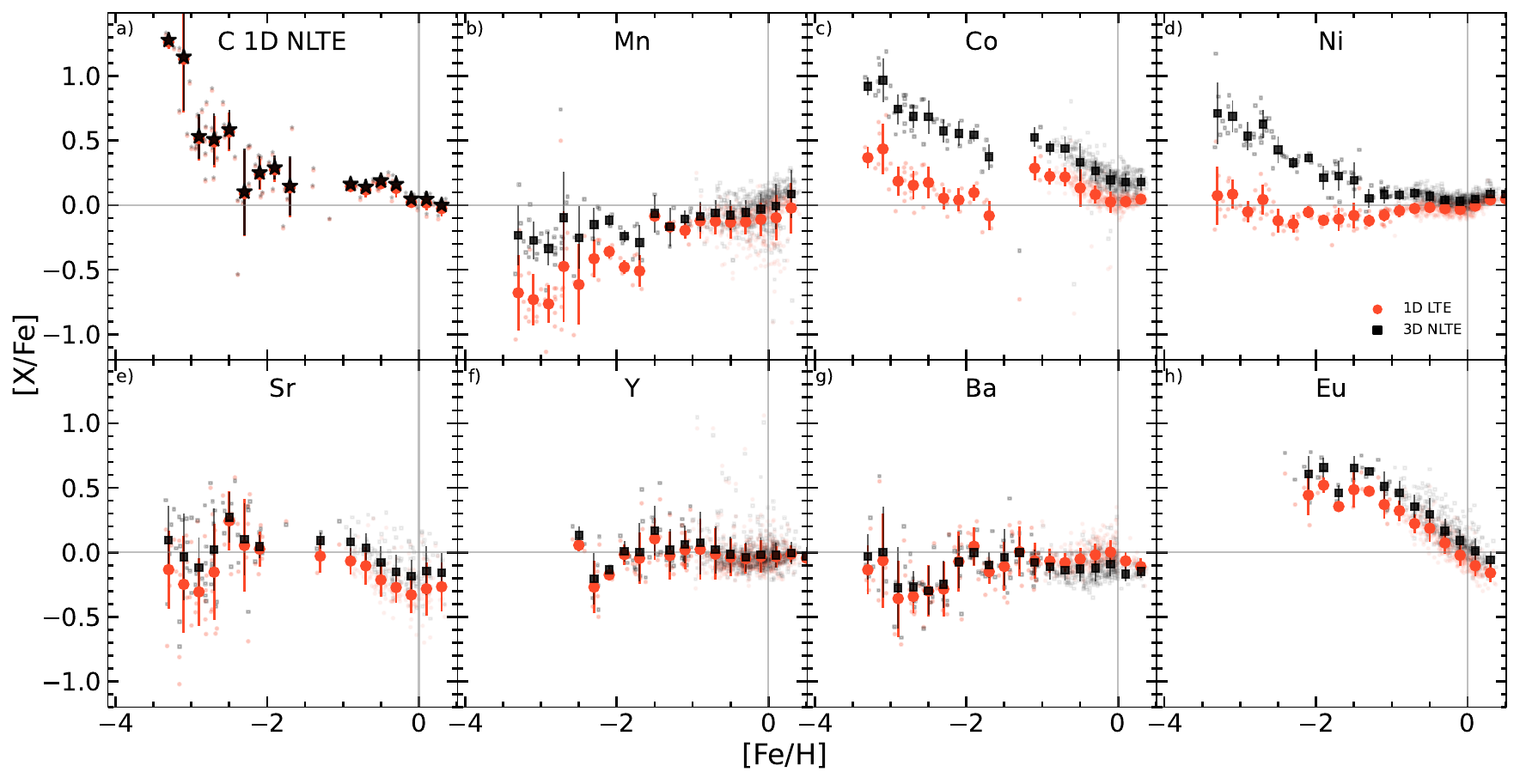}
\caption{Average binned abundances of the original literature 1D LTE abundances (red circles) together with {1D NLTE for C (black stars)} and 3D NLTE for others (black squares) abundances. {Points in the background represent individual stellar measurements.} Error bars represent the standard deviation for the respective bins.}
 \label{fig:ltenlte_gce_corrections_v2}
\end{figure*}

\subsection{Iron-peak elements}

In Fig. \ref{fig:ltenlte_gce_corrections_v2} we plotted the binned original 1D LTE literature abundance as red circles together with 3D NLTE abundance values as black squares (1D NLTE as black stars for C {based on CH G-band}). In general, all three iron-peak elements have much higher 3D NLTE abundances at lower metallicities. [Mn/Fe] has a downwards trend in 1D LTE {towards lower [Fe/H]}, that approaches a flatter one in 3D NLTE, similar to 1D NLTE findings in \citet{Eitner2020}. In 3D NLTE Co has a much steeper decreasing trend with metallicity, especially at [Fe/H] $\gtrapprox -1.8$. Finally, Ni has a generally flat trend in 1D LTE. However, for 3D NLTE abundances, that are even higher than 1D NLTE ones found in \citet{Eitner2023}, the values {of [Ni/Fe]} are $\approx 0.5$ dex at [Fe/H] {$\lesssim -3$}. Thus neglecting 3D NLTE effects results in biases of 0.4-0.5 dex at [Fe/H] < -2. 

\subsection{Neutron-capture elements}

{3D NLTE corrections for all neutron-capture elements result in a slightly increased abundance, of up to $0.1$ to $0.2$ dex at the lowest [Fe/H], except for Ba at solar metallicity}. Sr and Ba show scatter of up to 1.5 dex at [Fe/H] $< -3$, with a slightly reduced one at solar metallicity. Y has a slight increase in its abundances, up to 0.07 dex. Eu, on the other hand, shows a much higher 3D NLTE abundance by up to 0.2 dex. Higher neutron-capture element abundance at low [Fe/H] would imply that the r-process produces more of them, happens more frequently or there is another production site for it.

\subsection{C abundances}

Due to computational challenges, which we refer to a subsequent paper, we computed only the 1D NLTE abundances for CH G-band. NLTE results in higher values of up to +0.1 dex at [Fe/H] = 0, though staying close to LTE at lower metallicity. The spread of NLTE abundances stays the same as the LTE ones at all metallicities.

\section{Galactic Chemical Evolution}
\label{sec:gce_model}

We used OMEGA+ GCE model\footnote{\url{https://github.com/becot85/JINAPyCEE/}} \citep{Cote2017, Cote2018b} to briefly interpret our results. As the so-called "baseline" model, we use the same prescription and parameters as in \citet{Lian2023}. We refer the reader to that work for details and only briefly summarise our chosen GCE inputs here. While our chosen model and parameters are mostly only applicable for the disc, we use it at all metallicities as a general guideline.

\subsection{Input yields}

We followed the same method as in \citet{Eitner2023} and used Type Ia supernovae from 4 channels: single degenerate Chandrasekhar mass SNIa with H-transfer from the companion via stable Roche-lobe overflow, fainter SNeIax, sub-M$_{\text{ch}}$ with double-detonation of a C-O WD and sub-M$_{\text{ch}}$ SNIa due to a merger of two WDs. The delay time distributions (DTD) are based on StarTrack binary population synthesis code \citep[see Fig. 6 in][]{Eitner2023}. 

For the AGB stars we firstly adopted yields from \citet{Cristallo2015} as the baseline model. To see an effect of a different set of AGB yields, we also adopted the updated \citet{Karakas2010} yields that include neutron-capture elements based on \citet{Karakas2014, Karakas2016, Karakas2018, Karakas2022} \citep[see further comparisons between the two sets of yields in Section 4.2 in][]{Lian2023}. For the latter ones we used yields that assume partial mixing zone (PMZ) with a mass of $10^{-3}$ M$_\odot$ for stars with masses less than 4 M$_\odot$; PMZ of $10^{-4}$ M$_\odot$ for stars between 4 and 5 M$_\odot$; and no PMZ otherwise.

For the CCSNe we adopted the recommended yields from \citet[][set 'R']{Limongi2018}. For that set any stars more massive than 25 $M_\odot$ are assumed to fully collapse to a black hole and their yields include only the stellar wind. Since rotation is an important parameter impacting yields, we used the average yields of all rotations based on the velocity distribution derived by \citet{Prantzos2018}. 

For MRSN, we used the same approach as in \citet{Lian2023} (that is based on the similar approach as in \citet{Kobayashi2020a}), by replacing 0.01 per cent of CCSN yields by yields of MRSN from \citet{Nishimura2015}, which are based on a post-processed simulation on a magnetohydrodynamic of a 25 M$_\odot$ star \citep{Takiwaki2009}. 

For neutron-star mergers, we used the default configuration of NSM in OMEGA+: an ejecta mass of 0.025 M$_\odot$ and a power-law DTD with a slope of $\alpha=-1$ \citep[again, referring to these choices in][]{Lian2023} with the yield set from \citet{Arnould2007}. Alternative yield source from \citet{Rosswog2014} produces almost an identical amount of Eu with slightly increased amount of Ba, which would not affect our results. However, more recent NSM yields from \citet{Just2015} have a higher production of Eu. Thus we also calculated GCE tracks based on those yields to briefly analyse their impact. That paper has 6 different NS-NS configurations with each having two distinct remnant models. We decided to go with their "SFHO\_145145" configuration, which is a merger of two NS with masses of 1.45 $M_{\odot}$ (although note that a different mass configuration would not have a significant different impact on the GCE trend). This set was chosen since most double NS binaries have a mass of around 1.4 $M_{\odot}$ \citet{Lattimer2012}. We adopted the "M3A8m03a5" remnant model with a higher yield of the two to see the maximum extent of the effects. 

\subsection{Massive binaries}
\label{subsubsec:binary_imp}

\begin{figure}
\includegraphics[width=1\columnwidth]{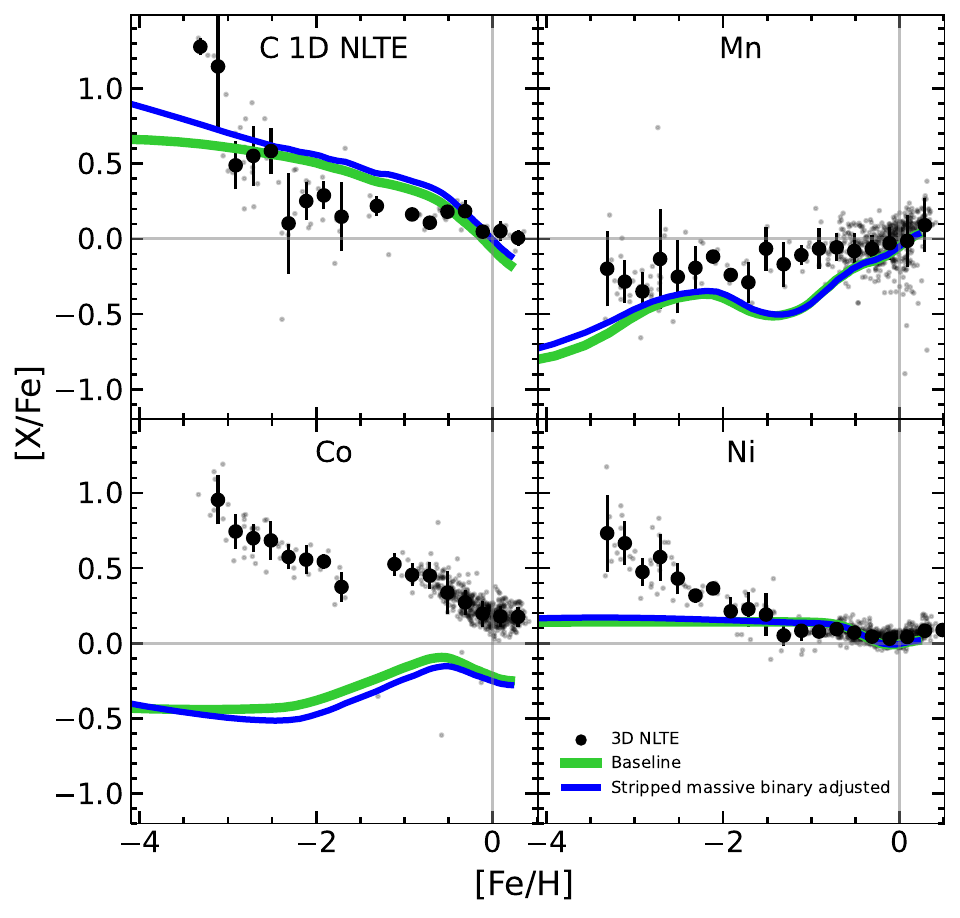}
\caption{GCE models overplotted over our 3D NLTE corrected abundances. The baseline GCE model is plotted in green with another one adjusted for the relative ratio of {stripped massive} binaries to single from \citet{Farmer2023} in blue (see text Sect. \ref{subsubsec:binary_imp}).}
 \label{fig:gce_binary_adjusted}
\end{figure}

\begin{figure}
\includegraphics[width=1\columnwidth]{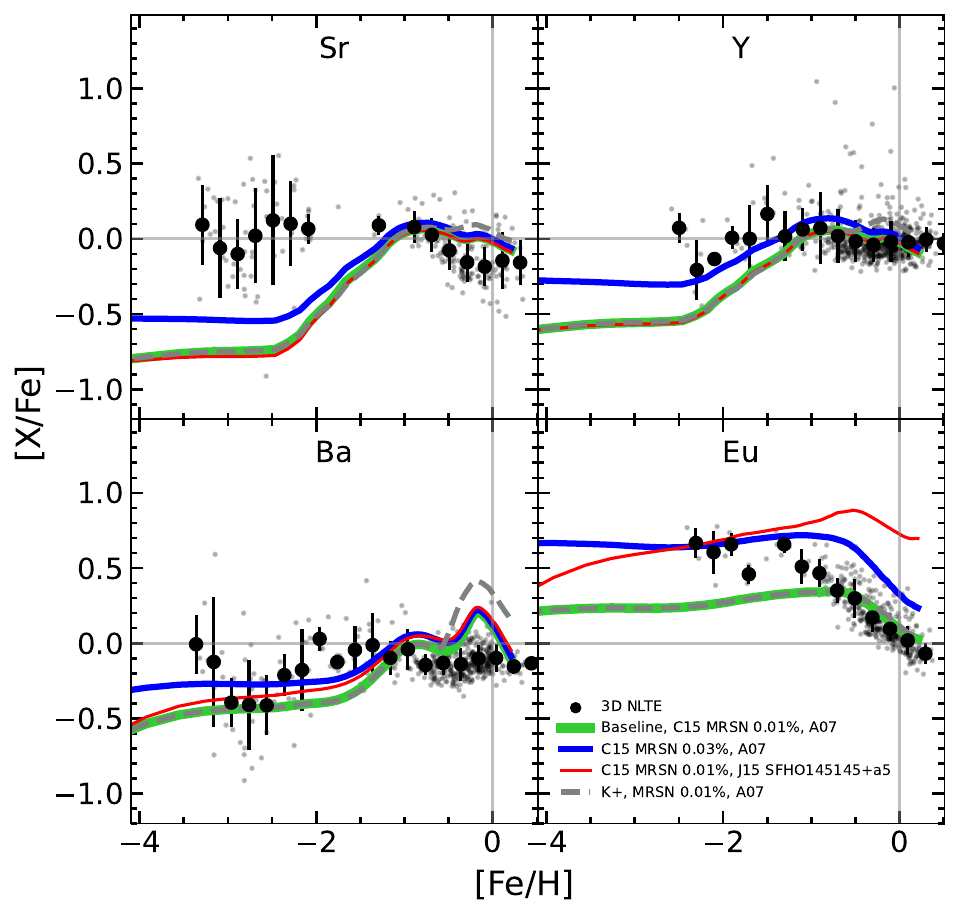}
\caption{GCE models overplotted over our 3D NLTE corrected abundances. The baseline model is plotted in green with \citet{Arnould2007} (A07) NSM yields and 0.01 per cent MRSN. The model in blue has tripled amount of MRSN, the model in red uses \citet{Just2015} (J15) NSM yields instead (their model "SFHO\_145145+M3A8m03a5"), while the grey dashed line (K+) uses AGB yields from Karakas group's AGB yields.}
 \label{fig:gce_neutron_capture_mrsn}
\end{figure}

The vast majority of the GCE literature uses massive {star yields} that are exclusively based on {calculations of structure and evolution of single massive stars}. However, it is now well established that most massive stars are actually in binary systems where stars can interact and change each others further life and final fate \citep[e.g. ][]{Sana2012, Moe2017}. Despite this, the existing literature studying the effect of binary stellar evolution on stellar yields is sparse.

For an initial qualitative exploration, we use results from the recently published work on binary-stripped {massive} stars from \citet{Farmer2023}. In that work, the authors use their consistently computed single and binary yields to study the impact of binarity on the stellar yields of the initially more massive star (primary). In this case "massive stripped star" refers to a scenario where a massive star loses a significant portion of its atmosphere due to the interaction with its companion star. \citet{Farmer2023} {focused} mostly on the relative differences between their binary and single stellar yields. Their yields have not been explicitly calibrated. A direct comparison of their yields to \citet{Limongi2018} and \citet{Sukhbold2016, Woosley2018} can be found in their Fig. 5, highlighting the vastly different production of iron-peak elements between all three works. \citet{Farmer2023} only provides predictions for solar metallicity stellar models. Detailed GCE modeling with complete yields for {stripped} massive binary stars is not possible with the limited yields predictions that are available to date. Nevertheless, we wish to take a first step and explore the possible impact of binary stellar evolution in broad terms. To achieve this we take the following approach.  We use the ratios predicted in \citet{Farmer2023} for each stable element between binary and single yields, and apply this ratio to 50 percent of our CCSN yields from \citet{Limongi2018}, thereby approximating a mixed stellar population of single and binary-stripped {massive} stars.  

The results are shown in Fig. \ref{fig:gce_binary_adjusted}, where we can see the baseline (in green) GCE tracks plotted together with the {stripped massive} binary-adjusted ones (in blue) over our 3D NLTE abundances in black. C was the most significantly affected element, resulting in increased production, \citep[see also][]{Farmer2021}. This increased abundance of carbon may explain some of the most C-rich stars. Mn and Ni elements have a small overall effect, since those elements are affected in a similar way as Fe. There is less Co production overall, decreasing its abundance by 0.1 dex at most metallicity regimes. All of these 3 GCE tracks underproduce iron-peak elements. Stripped {massive} binary yields do not improve the situation much, as the general slope of these elements is not captured either.

\subsection{Chemical evolution: models versus data}

{Stripped} massive binaries yields from \citet{Farmer2023} lack nuclear network beyond Zn, so we could not test their impact on the heavier elements. Instead we opted to briefly explore how different AGB and r-process sources and yields impact our GCE models. 

In Fig. \ref{fig:gce_neutron_capture_mrsn} we plotted our baseline model in green. For the s-process elements it underestimates their abundances at [Fe/H] < -2, and for europium already at [Fe/H] < -1. To compensate for the seemingly low production of Eu, we tripled the amount of MRSN to get the curve in blue. This value matches the observational data of Eu and Ba at [Fe/H] = -2, although it overproduces them for solar metallicity stars. Perhaps a higher percentage of MRSN would match Sr and Y elements, but that would lead to a poorer match of both Ba and Eu. This does show the need for more r-process in the early Galaxy and increasing MRSN rate is {one of the plausible explanations}, especially given that we do not have unambiguous observation of one yet. 

As another comparison, we also tried to substitute NSM yields from the baseline model \citet{Arnould2007} to \citet{Just2015} seen in the red thin line. Eu has a much higher yield and even though it reaches a roughly expected [Eu/Fe] of 0.6 at [Fe/H] = -3, there is a significant overproduction at solar metallicity. On the other hand Ba only has a slightly higher yield, while Sr and Y are not impacted at all, thus the GCE model is still much lower than the data for those elements. While it is possible to change the DTD of NSM \citep[e.g. with a higher slope to get more of them at lower metallicity, see e.g.][]{Maoz2024} to better fit the Eu distribution, the GCE models still do not produce enough of the other 3 elements. This situation doesn't change no matter the yield configuration from \citet{Just2015}. Only Eu is significantly impacted, with yields being different significantly only for Y or Ba (and not all 4 elements at once).

Finally, as an alternative AGB model, we used Karakas group's yields, plotted in a grey dashed line. Those yields only result in the increased production of s-process elements at the solar metallicity, especially for Ba. Since most of the discrepancies are at the lower metallicity regime, a different set of AGB yields has minimal effect on our analysis. 

\section{Discussion}
\label{sec:discussion}

In this project, we have primarily focused on determining the accurate 3D NLTE abundances in order to provide as unbiased as possible constraints on the Galactic chemical evolution of different chemical elements. We used carefully  determined 1D LTE abundances from the literature and corrected these for the critical effects of non-local thermodynamic equilibrium and convection using NLTE radiation transfer and 3D {RHD} simulations of stellar atmospheres. We also emphasise that full 3D NLTE calculations are, however, extremely computationally expensive, requiring 4 to 5 orders of magnitude in CPU time compared to standard 1D LTE abundance calculations. Therefore, only a selected stellar sample, comprised of main-sequence and subgiant stars, was used and only the most reliable diagnostic features were employed. While more 3D {RHD} models would be desirable in the future to further improve the precision of diagnostics, we show via comparisons of independent diagnostic codes (NLTE, spectrum synthesis, see App. \ref{app:ni_comparison} and \ref{app:multi3d_vs_multi1d}) that our results can be trusted. We also demonstrate that at lowest metallicities the effects of 3D NLTE on abundances reach 0.25 to 0.5 dex, which is much beyond a typical uncertainty of the NLTE corrections (e.g. for Ba see \citealt{Gallagher2020} or for Sun see Table A1 in \citealt{Asplund2021}). Thus any systematic trends reported in this work are statistically significant. 

We performed a comparison of our observed 3D NLTE abundances of iron-peak elements with predictions of GCE models, updated using state-of-the-art data from nucleosynthesis calculations. This analysis suggests that whereas overall the trends with [Fe/H] in the Galactic disc agree well for carbon, as well as for s- and r-process elements, for Fe-group species the GCE models are not consistent with the data. Specifically, the GCE models predict a mild down-turn from the [Fe/H] $\sim -1$  to solar [Fe/H] for the 1st peak s-process elements (Sr, Y), and a stronger increase of [Eu/Fe] with decreasing [Fe/H], which is fully in line with the observed abundances. However, in the domain of low metallicities, [Fe/H] $\lesssim -1$, {our tested} GCE model tracks fall systematically below the observed values. The 3D NLTE trends of [Sr/Fe], [Y/Fe], and [Ba/Fe] are closer to zero, which may either suggest that the contribution of exotic events, such as MRSNe, to the production of these species should be higher, or that the GCE model OMEGA$+$ is not suitable to describe the chemical enrichment in this [Fe/H] regime representative of the transition from the disc to the halo. {For this regime, a stochastic GCE model, such as the one used in \citet{Cescutti2014, Guiglion2024}, might be more appropriate to reproduce the observed large scatter in abundances of neutron-capture elements. Additionally, including a hierarchical halo formation from merging sub-halos \citep{Ishimaru2015} or a steeper slope of DTD \citep{Maoz2024} could increase the contribution of NSM at lower metallicity.}

For Fe-group elements, an interesting discrepancy between 3D NLTE data and GCE models is present at all metallicities below [Fe/H] $\lesssim -1$. All GCE models predict systematically and significantly \textit{lower} values of [Mn/Fe], [Co/Fe], and [Ni/Fe], by up to 1 dex for [Co/Fe] at [Fe/H] $\approx -3$ and up to $\sim 0.5$ dex for [Mn/Fe] and [Ni/Fe]. This problem is well-known for Mn and Co and it was a subject of careful literature studies \citep[e.g.][]{Bergemann2010b, Eitner2020, Sanders2021, Palla2021}. The GCE models of \citet{Kobayashi2006} and \citet{Kobayashi2020a} are lower than [Co/Fe] data (even in LTE) at [Fe/H] $\lesssim -2$ by at least $0.3$ to $0.4$ dex. This puzzling discrepancy is amplified with our new 3D NLTE abundances of Mn, Co, and Ni. The updated GCE models of \citet{Kobayashi2020b} are capable of reproducing the LTE trends of [Mn/Fe], however, their GCE models are also slightly discrepant with the LTE data for Ni at [Fe/H] $\lesssim -1$. We note, however, that they relied on 1D LTE measurements of Mn abundances, whereas for Cr they adopted Cr II-based data (which are, as shown in \citealt{Bergemann2010a}, fully consistent with NLTE values of Cr I). The 3D NLTE corrected data for [Ni/Fe] show a characteristic increase with decreasing [Fe/H], in fact resembling closely the trend for [Co/Fe]. This is intriguing, especially, in light of other independent evidence. For example, the JWST spectral analysis of the Ni/Fe ratios of the SN remnant Crab Nebula suggest highly super-solar [Ni/Fe] values, in excess of 3--5 times of the solar value \citep{Temim2024}. As we demonstrated in App. \ref{app:ni_comparison}, we find consistent NLTE results from different statistical equilibrium codes (both for Mn see \citet{Bergemann2019} and for Ni in this paper) and the 3D effects are small (see App. \ref{app:nlte_corrections} for details), which suggests that the problem is unlikely due to our 3D NLTE data.

It is more plausible that the offsets between models and data for Ni, Mn, and Co are associated with issues of GCE models or underlying yields. This is not unrealistic, since, except for SNeIa, all stellar structure and explosion models used in nucleosynthesis calculations implemented in standard GCE models are 1D hydrostatic and highly-parametrised \citep[e.g.][]{Woosley1995, Nishimura2015, Limongi2018, Fragos2019, Roepke2023}. Iron-peak isotopes have contributions from different channels and sites, including incomplete ($^{55}$Mn) and complete (Fe, Co) Si burning, $\alpha$-rich freeze-out (Co, Ni), and neutron capture (heavier n-rich Ni and Fe isotopes) \citep{Woosley2002}. The production of these isotopes is thus highly dependent on the entropy, mass cut shifts, and piston locations in 1D CCSN models \citep{Woosley1995}. This possibility is also addressed in GCE studies. For example, \citet{Kobayashi2006} suggest that a better fit of GCE models to the data of Cr, Mn, Co and Ni may potentially be obtained via fine-tuning of electron excess $Y_e$, explosion energy, and the mixing-fallback process. We {emphasise}, however, that these are free parameters and any ad-hoc scaling does not reveal the physics of element production, but may only help to narrow down the parameter space of more likely sites (assuming the overall physics and geometry is realistic, which is not the case in 1D).

In this regard, it is unfortunate that there is a severe lack of quantitative predictions of 3D CCSN explosion models, especially at low metallicities. It is well-known that multi-D effects in modelling CCSN explosions are relevant for the Fe-peak nucleosynthesis (see further discussions in \citealt{Sieverding2023}, \citealt{Wang2024a} and also \citealt{Stockinger2020,Wongwathanarat2017}; for 2D results, see \citealt{Wanajo2011,Bruenn2016,Wanajo2018,Pakmor2024}). It this context, is is particularly interesting that some of the Electron-Capture SN (ECSN) models (e.g. e8.8, z9.6 and u8.1 {in} \citealt{Wanajo2018}) and lower-mass lower-metallicity CCSN have highly super-solar ratios [Co/Fe] and [Ni/Fe] $\gg 0$. Here ECSN refers to a subclass of CCSNe arising from collapsing O-Ne-Mg cores in a mass ranges 8-10 $M_{\odot}$, where neon ignition is not triggered resulting in a degenerative core, with subsequent electron captures and in turn a CCSN \citep{Nomoto1987, Janka2008}. However, it is worth noting that to achieve a significant contribution from these massive stars at [Fe/H] $\leq -3$ may require a different IMF, e.g. the bottom-heavy IMF as known from models of dusty gas clouds \citep{Sharda2022}. At this point, comprehensive yields based on metallicity-dependent multi-D explosion models of CCSNe would be essential {in} order to test whether realistic 3D explosion models of CCSNe can explain the puzzling observed enhancement of the key Fe-peak elements, Mn, Co, and Ni at low [Fe/H].

In the attempt to better understand the origin of the model-data differences, we explored the effect of {stripped} massive binaries in GCE. We did this by scaling 50 per cent of the standard massive stellar yields (as described in Sect. \ref{subsubsec:binary_imp}) using the yield predictions from the solar-metallicity stripped {massive} star models of \cite{Farmer2023}. However, the impact of these systems on the GCE as implemented in our study turned out to be small. Only [C/Fe] ratios were affected at the level of up to $\sim 0.1-0.2$ dex, whereas for the other species the effect is within $0.05-0.1$ dex. \cite{Farmer2021} noted that the difference between binary and single yields are of the same order of magnitude as the differences between 1D CCSNe yields for single stars \citep{Pignatari2016, Limongi2018, Griffith2021}. For Fe-group elements, massive binary {yields} from \cite{Farmer2023} do not change the GCE model significantly enough to resolve the systematic under-production of Mn and Ni at [Fe/H] $\lesssim -1$. However, this could be due to the lack of binary stellar evolution channels, metallicity effects (which were not included here), or the simplified supernova physics that the underlying binary stellar yields rely upon. Further discussion on this can be found in the Appendix \ref{app:binary_yields_discussion}.

\section{Conclusions}\label{sec:conclusion}

In this work, we are concerned with the question of how does our understanding of Galactic Chemical Evolution model change, when abundances measured using state-of-the-art NLTE models in 3D RHD simulations of stellar atmospheres are used. 3D NLTE calculations are not new and have been presented and validated in numerous recent studies \citep[e.g.][]{Steffen2015, Lind2017, Bergemann2019, Bergemann2021, Pietrow2023, Storm2024, Wang2024b}. However, in this work we apply the 3D NLTE methods self-consistently to a large number of Galactic main-sequence and subgiant stars, and explore the effects of 3D NLTE on GCE tracks of the key elements, including Mn, Ni, Co and selected s- and r-process elements (Ba, Sr, Y, and Eu). Specifically, we focus on the chemical elements for which robust and carefully tested NLTE model atoms with quantum-mechanical data for H charge transfer reactions are available (Mn \citep{Bergemann2019}, Co \citep{Yakovleva2020}, Ni \citep{Voronov2022}, Sr \citep{Belyaev2013, Yakovleva2016}, Y \citep{Storm2024}, Ba \citep{Gallagher2020}, Eu \citep{Storm2024}). The 3D NLTE corrections are computed separately for each line of each elements and applied to 1D LTE measurements carefully assembled from literature \citep{Bonifacio2009, Hansen2013, Bensby2014, Battistini2015, Battistini2016, Zhao2016, Li2022, Mardini2024}. We use the new 3D NLTE data to compare with predictions of the OMEGA$+$ GCE model. This model includes different enrichment sources, various types of SN Ia, AGB, and compact binary mergers as described in our recent works \citep{Eitner2023, Lian2023}. Here, we further update the model with new yields for stripped massive binaries using the theoretical predictions from \citet{Farmer2023}. Given the focus of this work on 3D NLTE abundance calculations, no attempt is made here to cover other more exotic nucleosynthesis sites, such as the pair-instability SNe or electron capture SNe. From the differential analysis of the 3D NLTE corrected abundance measurements and the predictions of GCE models, we find the following with respect to individual chemical elements and their evolution in the Galaxy disc:

- \textbf{Carbon}: Here, we explore the effects of NLTE on the abundance of C derived from the G-band around 430 nm, the key diagnostic at low metallicity \citep[e.g.][]{Bessell2015, Ji2016, Sneden2016, Hansen2018, Holmbeck2020}. We find that the differences between NLTE and LTE abundance of carbon are small, at the level of $\sim 0.1$ dex, yet systematic and positive. The measured [C/Fe] trend in NLTE shows a stable increase with decreasing [Fe/H], and the spread of [C/Fe] values increases significantly (by a factor of 3) below [Fe/H] $\lesssim -2$ dex. The GCE models including standard yields for core-collapse SNe \citep{Limongi2018} and for SNe I{a} \citep{Eitner2023} are consistent with NLTE data through the entire range of [Fe/H] from $-4$ dex to solar. We also note that massive stripped stars that result in CCSN slightly increase the [C/Fe] ratios by up to 0.1-0.2 dex. This is the consequence of the reduced He-burning zone in the massive star due to stronger atmosphere stripping from interaction in a binary system, resulting in decreased carbon burning \citep{Farmer2021}. Such sites associated with {stripped} massive binaries may help to explain some of the metal-poor Galactic stars with clear enhancements of [C/Fe].

- \textbf{Manganese}: Using 3D NLTE calculations, we find a rather flat trend, with only slight upturn, of Galactic [Mn/Fe] ratios with [Fe/H], confirming previous results \citep{Bergemann2008, Eitner2020}. The 3D NLTE abundance corrections are substantial and reach $\sim +0.3$ dex or more at [Fe/H] $= -$3 dex, thus the measured GCE slope is close to zero. This is in strong contrast to the [Mn/Fe] depletion, as predicted by highly simplified 1D LTE models. We strongly advise against using 1D LTE abundances of Mn, which is sometimes done in the literature, as physically there is no reason to expect conditions in which 1D LTE is satisfied and the element, in virtue of its atomic structure is very sensitive to 3D NLTE effects \citep{Bergemann2019}. The constant [Mn/Fe] ratios could be explained by e.g. higher production of Mn in core-collapse SNe (see the discussion in \citealt[][]{Eitner2020} and \citealt{Palla2021}). {Stripped} massive binaries do show a slight increase in [Mn/Fe] production, but not sufficient to account for the flat trend. 

- \textbf{Cobalt}: We find an increasing trend of 3D NLTE abundance ratios of [Co/Fe] with decreasing [Fe/H], reaching [Co/Fe] $\geq 0.5$ dex at [Fe/H] $ \lesssim -2$ dex. These are similar to the strong corrections found previously \citep{Bergemann2010b, Yakovleva2020}. All GCE models, also the literature values \citep{Kobayashi2020a}, fail to explain the enhancement of [Co/Fe] at [Fe/H] $\lesssim -1$ dex even in LTE. Here we show that this discrepancy increases further with 3D NLTE effects on Co abundances taken into account. The super-solar [Co/Fe] ratios at low [Fe/H] may be associated with ECSN \citep{Wanajo2018}, possibly hinting at a bottom-heavy initial mass function function during the early phases of Galaxy formation.

- \textbf{Nickel}: Here we report the first 3D NLTE estimates of [Ni/Fe] values over the entire [Fe/H] range from {$-3.5$} dex to solar. We find a strong increase of [Ni/Fe] ratios with decreasing [Fe/H], reaching [Ni/Fe] $\approx 0.5$ dex at [Fe/H] {$\lesssim -3$}. The 3D NLTE abundance corrections for Ni range from zero at solar [Fe/H], but grow with decresaing metallicity and reach up to $0.5-0.7$ dex at [Fe/H] $=-4$ dex (although the effect highly depends on the diagnostic Ni I line), compared to 1D LTE values. This is physically expected based on radiation-driven NLTE line formation of this minority species in 3D inhomogeneous atmospheres at low metallicity. The effect is also similar that of other iron-peak elements, Mn I and Co I, and supports our previous 1D NLTE results for Ni in \citet{Eitner2023}. Similar to Co, yields from low-metallicity CCSN and ECSN models from \citet{Wanajo2018} have [Ni/Fe] $\gg 0$ dex.

- \textbf{Strontium}: We find typically small positive up to +0.3 dex 3D NLTE corrections for diagnostic Sr I and Sr II lines, although negative for 4077 and 4215 \AA~lines at [Fe/H] $\approx -2$ dex. There is a big scatter at [Fe/H] $\lesssim -3$ dex, which doesn't generally decrease for 3D NLTE abundances. In general, we predict a higher abundance of [Sr/Fe] in 3D NLTE compared to 1D LTE measurements. The expected GCE trend is more flat at all metallicities, requiring higher or more frequent r-process to account for Sr abundance at the lowest metallicity or a steeper NS-NS DTD.

- \textbf{Yttrium}: The literature sources we used employed lines with medium excitation potential, such as 4883 \AA, {and those} have 3D NLTE corrections of typically only up to 0.1 dex. The weaker higher excitation potential lines, such as 5402 and 5662 \AA, instead have slightly negative corrections, down to -0.1 dex. There is a {slightly} higher increase in 3D NLTE [Y/Fe] abundance at the lower metallicity regime, which flattens the overall trend at all [Fe/H]. Our tested GCE models reach only up to -0.2 dex at [Fe/H] $\lesssim -3$ dex, which is below the observed trend. 

- \textbf{Barium}: {Our 3D NLTE abundances are lower than 1D LTE values at subsolar metallicities, with effects reaching down to $-0.2$ dex at} [Fe/H] $\approx -1$ dex, which then flip sign and increase to +0.2 dex at [Fe/H] $\lesssim -3$ dex. However, $\vmic$ has a significant effect on the 1D EW and thus the 3D corrections. At very low metallicities around [Fe/H] $\approx -3$ dex, the scatter in [Ba/Fe] becomes quite large, extending up to 1.5 dex. At solar metallicity, 3D NLTE corrections slightly reduce and flatten the [Ba/Fe] ratios. This discrepancy suggests an overproduction of Ba in GCE models at solar metallicity. Conversely, at low metallicity, the models underproduce Ba.

- \textbf{Europium}: Our 3D NLTE calculations indicate abundance corrections of $+0.1$ to $0.2$ dex at [Fe/H] $\lesssim -1$ dex for the key diagnostic resonance lines at 3819, 4129, 4205 \AA~of Eu II. The {corrections for} subordinate line at 6645 \AA\ in the optical are usually negative, down to -0.05 dex. By comparing with GCE models that include yields from NSM and MRSN, we find that 3D NLTE [Eu/Fe] values are highly supersolar and reach 0.5 dex at [Fe/H] $\lesssim -2$ dex. Compared to our previous work using 1D LTE abundances in \citet{Lian2023}, we find that a tripled fraction of MRSN, compared to their 0.01 per cent, can help explain the observed 3D NLTE [Eu/Fe] abundances at low metallicities. But with higher fraction of MRSNe, the GCE model cannot reproduce the slope of [Eu/Fe] values at [Fe/H] $\gtrapprox -1.2$ dex. This discrepancy is amplified, if GCE models based on compact binary merger yields from \citet{Just2015} are used. These findings underscore the importance of considering multiple neutron-capture elements simultaneously to better understand r-process sites.

Our study highlights the importance of accurate NLTE analysis for elemental abundances. We emphasise that full 3D NLTE analysis is not equivalent to simply combining 3D LTE and 1D NLTE corrections. Our GCE models struggle to reproduce the significantly higher 3D NLTE abundances of iron-peak elements compared to the commonly used 1D LTE values in the literature. Similarly, the persistent discrepancies between neutron-capture element abundances in the metal-poor regime and our GCE models, especially when compared to 3D NLTE data, underscore our incomplete understanding of the r-process. While stripped massive binaries increase carbon abundances, they do not significantly affect the yields of other elements. More advanced models, such as low-metallicity multi-dimensional CCSN or ECSN and including models that incorporate modified stellar structure due to binary interactions, may provide insight into these other discrepancies. On the observational front, new facilities like 4MOST are expected to offer a more comprehensive view of abundance variations in our Galaxy and provide more accurate data for refining GCE models.

\section*{Acknowledgements}

We sincerely thank the referee for the positive feedback and comments that improved the readability of the paper.

We thank Jonas Klevas, Adam Burrows, Alexander Heger, and Heitor Ernandes for their valuable discussions.

NS and MB acknowledge funding from the European Research Council (ERC) under the European Union’s Horizon 2020 research and innovation programme (Grant agreement No. 949173).

MB is supported through the Lise Meitner grant from the Max Planck Society. We acknowledge support by the Collaborative Research centre SFB 881 (projects A5, A10), Heidelberg University, of the Deutsche Forschungsgemeinschaft (DFG, German Research Foundation). 

HTJ and AS acknowledge support by the European Union's
Framework Programme for Research and Innovation Horizon
Europe under Marie Sklodowska-Curie grant agreement
No. 101065891, and by the German Research Foundation (DFG) through the Collaborative Research Centre ``Neutrinos
and Dark Matter in Astro- and Particle Physics (NDM),'' grant
No. SFB-1258-283604770, and under Germany's Excellence
Strategy through the Cluster of Excellence ORIGINS EXC-2094-390783311. 
Contributions from AS are prepared by LLNL under Contract DE-AC52-07NA27344.

AJR, MB, EKO, PE, and NS acknowledge funding support from the DAAD Australia-Germany Joint Research Cooperation Scheme, grant Project Number 57654415. 
AJR acknowledges support from the Australian Research Council under award number FT170100243. 

Computations were performed on the HPC systems Cobra, Raven and Viper at the Max Planck Computing and Data Facility.

\section*{Data Availability}

The 3D NLTE abundances and stellar parameters for the stars analysed in this study will be made available in machine-readable format via CDS upon publication of the paper. NLTE models atoms are available at \url{https://keeper.mpdl.mpg.de/d/6c2033ef5c5d4c9ca8d1/}.



\bibliographystyle{mnras}
\bibliography{references} 

\begin{thebibliography}{}
\makeatletter
\relax
\def\mn@urlcharsother{\let\do\@makeother \do\$\do\&\do\#\do\^\do\_\do\%\do\~}
\def\mn@doi{\begingroup\mn@urlcharsother \@ifnextchar [ {\mn@doi@} {\mn@doi@[]}}
\def\mn@doi@[#1]#2{\def\@tempa{#1}\ifx\@tempa\@empty \href {http://dx.doi.org/#2} {doi:#2}\else \href {http://dx.doi.org/#2} {#1}\fi \endgroup}
\def\mn@eprint#1#2{\mn@eprint@#1:#2::\@nil}
\def\mn@eprint@arXiv#1{\href {http://arxiv.org/abs/#1} {{\tt arXiv:#1}}}
\def\mn@eprint@dblp#1{\href {http://dblp.uni-trier.de/rec/bibtex/#1.xml} {dblp:#1}}
\def\mn@eprint@#1:#2:#3:#4\@nil{\def\@tempa {#1}\def\@tempb {#2}\def\@tempc {#3}\ifx \@tempc \@empty \let \@tempc \@tempb \let \@tempb \@tempa \fi \ifx \@tempb \@empty \def\@tempb {arXiv}\fi \@ifundefined {mn@eprint@\@tempb}{\@tempb:\@tempc}{\expandafter \expandafter \csname mn@eprint@\@tempb\endcsname \expandafter{\@tempc}}}

\bibitem[\protect\citeauthoryear{{Amarsi}}{{Amarsi}}{2015}]{Amarsi2015}
{Amarsi} A.~M.,  2015, \mn@doi [\mnras] {10.1093/mnras/stv1392}, \href {https://ui.adsabs.harvard.edu/abs/2015MNRAS.452.1612A} {452, 1612}

\bibitem[\protect\citeauthoryear{{Amarsi}, {Lind}, {Asplund}, {Barklem}  \& {Collet}}{{Amarsi} et~al.}{2016}]{Amarsi2016a}
{Amarsi} A.~M.,  {Lind} K.,  {Asplund} M.,  {Barklem} P.~S.,   {Collet} R.,  2016, \mn@doi [\mnras] {10.1093/mnras/stw2077}, \href {https://ui.adsabs.harvard.edu/abs/2016MNRAS.463.1518A} {463, 1518}

\bibitem[\protect\citeauthoryear{{Amarsi}, {Liljegren}  \& {Nissen}}{{Amarsi} et~al.}{2022}]{Amarsi2022}
{Amarsi} A.~M.,  {Liljegren} S.,   {Nissen} P.~E.,  2022, \mn@doi [\aap] {10.1051/0004-6361/202244542}, \href {https://ui.adsabs.harvard.edu/abs/2022A&A...668A..68A} {668, A68}

\bibitem[\protect\citeauthoryear{{Antognini} \& {Thompson}}{{Antognini} \& {Thompson}}{2016}]{Antognini2016}
{Antognini} J. M.~O.,  {Thompson} T.~A.,  2016, \mn@doi [\mnras] {10.1093/mnras/stv2938}, \href {https://ui.adsabs.harvard.edu/abs/2016MNRAS.456.4219A} {456, 4219}

\bibitem[\protect\citeauthoryear{{Arcones} \& {Thielemann}}{{Arcones} \& {Thielemann}}{2013}]{Arcones2013}
{Arcones} A.,  {Thielemann} F.~K.,  2013, \mn@doi [Journal of Physics G Nuclear Physics] {10.1088/0954-3899/40/1/013201}, \href {https://ui.adsabs.harvard.edu/abs/2013JPhG...40a3201A} {40, 013201}

\bibitem[\protect\citeauthoryear{{Arnould}, {Goriely}  \& {Takahashi}}{{Arnould} et~al.}{2007}]{Arnould2007}
{Arnould} M.,  {Goriely} S.,   {Takahashi} K.,  2007, \mn@doi [\physrep] {10.1016/j.physrep.2007.06.002}, \href {https://ui.adsabs.harvard.edu/abs/2007PhR...450...97A} {450, 97}

\bibitem[\protect\citeauthoryear{{Asplund}}{{Asplund}}{2005}]{Asplund2005}
{Asplund} M.,  2005, \mn@doi [\araa] {10.1146/annurev.astro.42.053102.134001}, \href {https://ui.adsabs.harvard.edu/abs/2005ARA&A..43..481A} {43, 481}

\bibitem[\protect\citeauthoryear{{Asplund}, {Amarsi}  \& {Grevesse}}{{Asplund} et~al.}{2021}]{Asplund2021}
{Asplund} M.,  {Amarsi} A.~M.,   {Grevesse} N.,  2021, \mn@doi [\aap] {10.1051/0004-6361/202140445}, \href {https://ui.adsabs.harvard.edu/abs/2021A&A...653A.141A} {653, A141}

\bibitem[\protect\citeauthoryear{{Battistini} \& {Bensby}}{{Battistini} \& {Bensby}}{2015}]{Battistini2015}
{Battistini} C.,  {Bensby} T.,  2015, \mn@doi [\aap] {10.1051/0004-6361/201425327}, \href {https://ui.adsabs.harvard.edu/abs/2015A&A...577A...9B} {577, A9}

\bibitem[\protect\citeauthoryear{{Battistini} \& {Bensby}}{{Battistini} \& {Bensby}}{2016}]{Battistini2016}
{Battistini} C.,  {Bensby} T.,  2016, \mn@doi [\aap] {10.1051/0004-6361/201527385}, \href {https://ui.adsabs.harvard.edu/abs/2016A&A...586A..49B} {586, A49}

\bibitem[\protect\citeauthoryear{Belyaev}{Belyaev}{2013}]{Belyaev2013}
Belyaev A.~K.,  2013, \mn@doi [Phys. Rev. A] {10.1103/PhysRevA.88.052704}, 88, 052704

\bibitem[\protect\citeauthoryear{{Bensby} \& {Feltzing}}{{Bensby} \& {Feltzing}}{2006}]{Bensby2006}
{Bensby} T.,  {Feltzing} S.,  2006, \mn@doi [\mnras] {10.1111/j.1365-2966.2006.10037.x}, \href {https://ui.adsabs.harvard.edu/abs/2006MNRAS.367.1181B} {367, 1181}

\bibitem[\protect\citeauthoryear{{Bensby}, {Feltzing}  \& {Oey}}{{Bensby} et~al.}{2014}]{Bensby2014}
{Bensby} T.,  {Feltzing} S.,   {Oey} M.~S.,  2014, \mn@doi [\aap] {10.1051/0004-6361/201322631}, \href {https://ui.adsabs.harvard.edu/abs/2014A&A...562A..71B} {562, A71}

\bibitem[\protect\citeauthoryear{{Bensby} et~al.,}{{Bensby} et~al.}{2019}]{Bensby2019}
{Bensby} T.,  et~al., 2019, \mn@doi [The Messenger] {10.18727/0722-6691/5123}, \href {https://ui.adsabs.harvard.edu/abs/2019Msngr.175...35B} {175, 35}

\bibitem[\protect\citeauthoryear{{Bergemann} \& {Cescutti}}{{Bergemann} \& {Cescutti}}{2010}]{Bergemann2010a}
{Bergemann} M.,  {Cescutti} G.,  2010, \mn@doi [\aap] {10.1051/0004-6361/201014250}, \href {https://ui.adsabs.harvard.edu/abs/2010A&A...522A...9B} {522, A9}

\bibitem[\protect\citeauthoryear{{Bergemann} \& {Gehren}}{{Bergemann} \& {Gehren}}{2008}]{Bergemann2008}
{Bergemann} M.,  {Gehren} T.,  2008, \mn@doi [\aap] {10.1051/0004-6361:200810098}, \href {https://ui.adsabs.harvard.edu/abs/2008A&A...492..823B} {492, 823}

\bibitem[\protect\citeauthoryear{{Bergemann} \& {Nordlander}}{{Bergemann} \& {Nordlander}}{2014}]{Bergemann2014}
{Bergemann} M.,  {Nordlander} T.,  2014, in {Niemczura} E.,  {Smalley} B.,   {Pych} W.,  eds, , Determination of Atmospheric Parameters of B.
pp 169--185, \mn@doi{10.1007/978-3-319-06956-2_16}

\bibitem[\protect\citeauthoryear{{Bergemann}, {Pickering}  \& {Gehren}}{{Bergemann} et~al.}{2010}]{Bergemann2010b}
{Bergemann} M.,  {Pickering} J.~C.,   {Gehren} T.,  2010, \mn@doi [\mnras] {10.1111/j.1365-2966.2009.15736.x}, \href {https://ui.adsabs.harvard.edu/abs/2010MNRAS.401.1334B} {401, 1334}

\bibitem[\protect\citeauthoryear{{Bergemann}, {Lind}, {Collet}, {Magic}  \& {Asplund}}{{Bergemann} et~al.}{2012a}]{Bergemann2012a}
{Bergemann} M.,  {Lind} K.,  {Collet} R.,  {Magic} Z.,   {Asplund} M.,  2012a, \mn@doi [\mnras] {10.1111/j.1365-2966.2012.21687.x}, \href {https://ui.adsabs.harvard.edu/abs/2012MNRAS.427...27B} {427, 27}

\bibitem[\protect\citeauthoryear{{Bergemann}, {Hansen}, {Bautista}  \& {Ruchti}}{{Bergemann} et~al.}{2012b}]{Bergemann2012b}
{Bergemann} M.,  {Hansen} C.~J.,  {Bautista} M.,   {Ruchti} G.,  2012b, \mn@doi [\aap] {10.1051/0004-6361/201219406}, \href {https://ui.adsabs.harvard.edu/abs/2012A&A...546A..90B} {546, A90}

\bibitem[\protect\citeauthoryear{{Bergemann}, {Kudritzki}, {W{\"u}rl}, {Plez}, {Davies}  \& {Gazak}}{{Bergemann} et~al.}{2013}]{Bergemann2013}
{Bergemann} M.,  {Kudritzki} R.-P.,  {W{\"u}rl} M.,  {Plez} B.,  {Davies} B.,   {Gazak} Z.,  2013, \mn@doi [\apj] {10.1088/0004-637X/764/2/115}, \href {https://ui.adsabs.harvard.edu/abs/2013ApJ...764..115B} {764, 115}

\bibitem[\protect\citeauthoryear{{Bergemann}, {Collet}, {Amarsi}, {Kovalev}, {Ruchti}  \& {Magic}}{{Bergemann} et~al.}{2017}]{Bergemann2017}
{Bergemann} M.,  {Collet} R.,  {Amarsi} A.~M.,  {Kovalev} M.,  {Ruchti} G.,   {Magic} Z.,  2017, \mn@doi [\apj] {10.3847/1538-4357/aa88cb}, \href {https://ui.adsabs.harvard.edu/abs/2017ApJ...847...15B} {847, 15}

\bibitem[\protect\citeauthoryear{{Bergemann} et~al.,}{{Bergemann} et~al.}{2019}]{Bergemann2019}
{Bergemann} M.,  et~al., 2019, \mn@doi [\aap] {10.1051/0004-6361/201935811}, \href {https://ui.adsabs.harvard.edu/abs/2019A&A...631A..80B} {631, A80}

\bibitem[\protect\citeauthoryear{{Bergemann} et~al.,}{{Bergemann} et~al.}{2021}]{Bergemann2021}
{Bergemann} M.,  et~al., 2021, \mn@doi [\mnras] {10.1093/mnras/stab2160}, \href {https://ui.adsabs.harvard.edu/abs/2021MNRAS.508.2236B} {508, 2236}

\bibitem[\protect\citeauthoryear{{Bessell} et~al.,}{{Bessell} et~al.}{2015}]{Bessell2015}
{Bessell} M.~S.,  et~al., 2015, \mn@doi [\apjl] {10.1088/2041-8205/806/1/L16}, \href {https://ui.adsabs.harvard.edu/abs/2015ApJ...806L..16B} {806, L16}

\bibitem[\protect\citeauthoryear{{Bliss}, {Arcones}  \& {Qian}}{{Bliss} et~al.}{2018}]{Bliss2018}
{Bliss} J.,  {Arcones} A.,   {Qian} Y.~Z.,  2018, \mn@doi [\apj] {10.3847/1538-4357/aade8d}, \href {https://ui.adsabs.harvard.edu/abs/2018ApJ...866..105B} {866, 105}

\bibitem[\protect\citeauthoryear{{Boeltzig} et~al.,}{{Boeltzig} et~al.}{2016}]{Boeltzig2016}
{Boeltzig} A.,  et~al., 2016, \mn@doi [European Physical Journal A] {10.1140/epja/i2016-16075-4}, \href {https://ui.adsabs.harvard.edu/abs/2016EPJA...52...75B} {52, 75}

\bibitem[\protect\citeauthoryear{{Bonifacio} et~al.,}{{Bonifacio} et~al.}{2009}]{Bonifacio2009}
{Bonifacio} P.,  et~al., 2009, \mn@doi [\aap] {10.1051/0004-6361/200810610}, \href {https://ui.adsabs.harvard.edu/abs/2009A&A...501..519B} {501, 519}

\bibitem[\protect\citeauthoryear{{Bruenn} et~al.,}{{Bruenn} et~al.}{2016}]{Bruenn2016}
{Bruenn} S.~W.,  et~al., 2016, \mn@doi [\apj] {10.3847/0004-637X/818/2/123}, \href {https://ui.adsabs.harvard.edu/abs/2016ApJ...818..123B} {818, 123}

\bibitem[\protect\citeauthoryear{{Bruls}, {Rutten}  \& {Shchukina}}{{Bruls} et~al.}{1992}]{Bruls1992}
{Bruls} J.~H.~M.~J.,  {Rutten} R.~J.,   {Shchukina} N.~G.,  1992, \aap, \href {https://ui.adsabs.harvard.edu/abs/1992A&A...265..237B} {265, 237}

\bibitem[\protect\citeauthoryear{{Burbidge}, {Burbidge}, {Fowler}  \& {Hoyle}}{{Burbidge} et~al.}{1957}]{Burbidge1957}
{Burbidge} E.~M.,  {Burbidge} G.~R.,  {Fowler} W.~A.,   {Hoyle} F.,  1957, \mn@doi [Reviews of Modern Physics] {10.1103/RevModPhys.29.547}, \href {https://ui.adsabs.harvard.edu/abs/1957RvMP...29..547B} {29, 547}

\bibitem[\protect\citeauthoryear{{Busso}, {Gallino}  \& {Wasserburg}}{{Busso} et~al.}{1999}]{Busso1999}
{Busso} M.,  {Gallino} R.,   {Wasserburg} G.~J.,  1999, \mn@doi [\araa] {10.1146/annurev.astro.37.1.239}, \href {https://ui.adsabs.harvard.edu/abs/1999ARA&A..37..239B} {37, 239}

\bibitem[\protect\citeauthoryear{{Carlsson}}{{Carlsson}}{1986}]{Carlsson1986}
{Carlsson} M.,  1986, Uppsala Astronomical Observatory Reports, \href {https://ui.adsabs.harvard.edu/abs/1986UppOR..33.....C} {33}

\bibitem[\protect\citeauthoryear{{Cescutti} \& {Chiappini}}{{Cescutti} \& {Chiappini}}{2014}]{Cescutti2014}
{Cescutti} G.,  {Chiappini} C.,  2014, \mn@doi [\aap] {10.1051/0004-6361/201423432}, \href {https://ui.adsabs.harvard.edu/abs/2014A&A...565A..51C} {565, A51}

\bibitem[\protect\citeauthoryear{{Clayton}}{{Clayton}}{1968}]{Clayton1968}
{Clayton} D.~D.,  1968, {Principles of stellar evolution and nucleosynthesis}

\bibitem[\protect\citeauthoryear{{C{\^o}t{\'e}}, {O'Shea}, {Ritter}, {Herwig}  \& {Venn}}{{C{\^o}t{\'e}} et~al.}{2017}]{Cote2017}
{C{\^o}t{\'e}} B.,  {O'Shea} B.~W.,  {Ritter} C.,  {Herwig} F.,   {Venn} K.~A.,  2017, \mn@doi [\apj] {10.3847/1538-4357/835/2/128}, \href {https://ui.adsabs.harvard.edu/abs/2017ApJ...835..128C} {835, 128}

\bibitem[\protect\citeauthoryear{{C{\^o}t{\'e}}, {Denissenkov}, {Herwig}, {Ruiter}, {Ritter}, {Pignatari}  \& {Belczynski}}{{C{\^o}t{\'e}} et~al.}{2018a}]{Cote2018c}
{C{\^o}t{\'e}} B.,  {Denissenkov} P.,  {Herwig} F.,  {Ruiter} A.~J.,  {Ritter} C.,  {Pignatari} M.,   {Belczynski} K.,  2018a, \mn@doi [\apj] {10.3847/1538-4357/aaaae8}, \href {https://ui.adsabs.harvard.edu/abs/2018ApJ...854..105C} {854, 105}

\bibitem[\protect\citeauthoryear{{C{\^o}t{\'e}}, {Silvia}, {O'Shea}, {Smith}  \& {Wise}}{{C{\^o}t{\'e}} et~al.}{2018b}]{Cote2018b}
{C{\^o}t{\'e}} B.,  {Silvia} D.~W.,  {O'Shea} B.~W.,  {Smith} B.,   {Wise} J.~H.,  2018b, \mn@doi [\apj] {10.3847/1538-4357/aabe8f}, \href {https://ui.adsabs.harvard.edu/abs/2018ApJ...859...67C} {859, 67}

\bibitem[\protect\citeauthoryear{{Cristallo} et~al.,}{{Cristallo} et~al.}{2011}]{Cristallo2011}
{Cristallo} S.,  et~al., 2011, \mn@doi [\apjs] {10.1088/0067-0049/197/2/17}, \href {https://ui.adsabs.harvard.edu/abs/2011ApJS..197...17C} {197, 17}

\bibitem[\protect\citeauthoryear{{Cristallo}, {Straniero}, {Piersanti}  \& {Gobrecht}}{{Cristallo} et~al.}{2015}]{Cristallo2015}
{Cristallo} S.,  {Straniero} O.,  {Piersanti} L.,   {Gobrecht} D.,  2015, \mn@doi [\apjs] {10.1088/0067-0049/219/2/40}, \href {https://ui.adsabs.harvard.edu/abs/2015ApJS..219...40C} {219, 40}

\bibitem[\protect\citeauthoryear{{Cui} et~al.,}{{Cui} et~al.}{2012}]{Cui2012}
{Cui} X.-Q.,  et~al., 2012, \mn@doi [Research in Astronomy and Astrophysics] {10.1088/1674-4527/12/9/003}, \href {https://ui.adsabs.harvard.edu/abs/2012RAA....12.1197C} {12, 1197}

\bibitem[\protect\citeauthoryear{{Denissenkov}, {Herwig}, {Woodward}, {Andrassy}, {Pignatari}  \& {Jones}}{{Denissenkov} et~al.}{2019}]{Denissenkov2019}
{Denissenkov} P.~A.,  {Herwig} F.,  {Woodward} P.,  {Andrassy} R.,  {Pignatari} M.,   {Jones} S.,  2019, \mn@doi [\mnras] {10.1093/mnras/stz1921}, \href {https://ui.adsabs.harvard.edu/abs/2019MNRAS.488.4258D} {488, 4258}

\bibitem[\protect\citeauthoryear{{Drawin}}{{Drawin}}{1968}]{Drawin1968}
{Drawin} H.-W.,  1968, \mn@doi [Zeitschrift fur Physik] {10.1007/BF01379963}, \href {https://ui.adsabs.harvard.edu/abs/1968ZPhy..211..404D} {211, 404}

\bibitem[\protect\citeauthoryear{{Eitner}, {Bergemann}, {Hansen}, {Cescutti}, {Seitenzahl}, {Larsen}  \& {Plez}}{{Eitner} et~al.}{2020}]{Eitner2020}
{Eitner} P.,  {Bergemann} M.,  {Hansen} C.~J.,  {Cescutti} G.,  {Seitenzahl} I.~R.,  {Larsen} S.,   {Plez} B.,  2020, \mn@doi [\aap] {10.1051/0004-6361/201936603}, \href {https://ui.adsabs.harvard.edu/abs/2020A&A...635A..38E} {635, A38}

\bibitem[\protect\citeauthoryear{{Eitner}, {Bergemann}, {Ruiter}, {Avril}, {Seitenzahl}, {Gent}  \& {C{\^o}t{\'e}}}{{Eitner} et~al.}{2023}]{Eitner2023}
{Eitner} P.,  {Bergemann} M.,  {Ruiter} A.~J.,  {Avril} O.,  {Seitenzahl} I.~R.,  {Gent} M.~R.,   {C{\^o}t{\'e}} B.,  2023, \mn@doi [\aap] {10.1051/0004-6361/202244286}, \href {https://ui.adsabs.harvard.edu/abs/2023A&A...677A.151E} {677, A151}

\bibitem[\protect\citeauthoryear{{Eitner}, {Bergemann}, {Hoppe}, {Nordlund}, {Plez}  \& {Klevas}}{{Eitner} et~al.}{2024}]{Eitner2024}
{Eitner} P.,  {Bergemann} M.,  {Hoppe} R.,  {Nordlund} {\r{A}}.,  {Plez} B.,   {Klevas} J.,  2024, \mn@doi [\aap] {10.1051/0004-6361/202348448}, \href {https://ui.adsabs.harvard.edu/abs/2024A&A...688A..52E} {688, A52}

\bibitem[\protect\citeauthoryear{{Farmer}, {Laplace}, {de Mink}  \& {Justham}}{{Farmer} et~al.}{2021}]{Farmer2021}
{Farmer} R.,  {Laplace} E.,  {de Mink} S.~E.,   {Justham} S.,  2021, \mn@doi [\apj] {10.3847/1538-4357/ac2f44}, \href {https://ui.adsabs.harvard.edu/abs/2021ApJ...923..214F} {923, 214}

\bibitem[\protect\citeauthoryear{{Farmer}, {Laplace}, {Ma}, {de Mink}  \& {Justham}}{{Farmer} et~al.}{2023}]{Farmer2023}
{Farmer} R.,  {Laplace} E.,  {Ma} J.-z.,  {de Mink} S.~E.,   {Justham} S.,  2023, \mn@doi [\apj] {10.3847/1538-4357/acc315}, \href {https://ui.adsabs.harvard.edu/abs/2023ApJ...948..111F} {948, 111}

\bibitem[\protect\citeauthoryear{{Fink}, {R{\"o}pke}, {Hillebrandt}, {Seitenzahl}, {Sim}  \& {Kromer}}{{Fink} et~al.}{2010}]{Fink2010}
{Fink} M.,  {R{\"o}pke} F.~K.,  {Hillebrandt} W.,  {Seitenzahl} I.~R.,  {Sim} S.~A.,   {Kromer} M.,  2010, \mn@doi [\aap] {10.1051/0004-6361/200913892}, \href {https://ui.adsabs.harvard.edu/abs/2010A&A...514A..53F} {514, A53}

\bibitem[\protect\citeauthoryear{{Fragos}, {Andrews}, {Ramirez-Ruiz}, {Meynet}, {Kalogera}, {Taam}  \& {Zezas}}{{Fragos} et~al.}{2019}]{Fragos2019}
{Fragos} T.,  {Andrews} J.~J.,  {Ramirez-Ruiz} E.,  {Meynet} G.,  {Kalogera} V.,  {Taam} R.~E.,   {Zezas} A.,  2019, \mn@doi [\apjl] {10.3847/2041-8213/ab40d1}, \href {https://ui.adsabs.harvard.edu/abs/2019ApJ...883L..45F} {883, L45}

\bibitem[\protect\citeauthoryear{{Gallagher}, {Bergemann}, {Collet}, {Plez}, {Leenaarts}, {Carlsson}, {Yakovleva}  \& {Belyaev}}{{Gallagher} et~al.}{2020}]{Gallagher2020}
{Gallagher} A.~J.,  {Bergemann} M.,  {Collet} R.,  {Plez} B.,  {Leenaarts} J.,  {Carlsson} M.,  {Yakovleva} S.~A.,   {Belyaev} A.~K.,  2020, \mn@doi [\aap] {10.1051/0004-6361/201936104}, \href {https://ui.adsabs.harvard.edu/abs/2020A&A...634A..55G} {634, A55}

\bibitem[\protect\citeauthoryear{{Gerber}, {Magg}, {Plez}, {Bergemann}, {Heiter}, {Olander}  \& {Hoppe}}{{Gerber} et~al.}{2023}]{Gerber2023}
{Gerber} J.~M.,  {Magg} E.,  {Plez} B.,  {Bergemann} M.,  {Heiter} U.,  {Olander} T.,   {Hoppe} R.,  2023, \mn@doi [\aap] {10.1051/0004-6361/202243673}, \href {https://ui.adsabs.harvard.edu/abs/2023A&A...669A..43G} {669, A43}

\bibitem[\protect\citeauthoryear{{Gibson}, {Fenner}, {Renda}, {Kawata}  \& {Lee}}{{Gibson} et~al.}{2003}]{Gibson2003}
{Gibson} B.~K.,  {Fenner} Y.,  {Renda} A.,  {Kawata} D.,   {Lee} H.-c.,  2003, \mn@doi [\pasa] {10.1071/AS03052}, \href {https://ui.adsabs.harvard.edu/abs/2003PASA...20..401G} {20, 401}

\bibitem[\protect\citeauthoryear{{Goriely}, {Bauswein}, {Janka}, {Just}  \& {Pllumbi}}{{Goriely} et~al.}{2018}]{Goriely2018}
{Goriely} S.,  {Bauswein} A.,  {Janka} H.-T.,  {Just} O.,   {Pllumbi} E.,  2018, in European Physical Journal Web of Conferences. p. 01025, \mn@doi{10.1051/epjconf/201716501025}

\bibitem[\protect\citeauthoryear{{Griffith}, {Sukhbold}, {Weinberg}, {Johnson}, {Johnson}  \& {Vincenzo}}{{Griffith} et~al.}{2021}]{Griffith2021}
{Griffith} E.~J.,  {Sukhbold} T.,  {Weinberg} D.~H.,  {Johnson} J.~A.,  {Johnson} J.~W.,   {Vincenzo} F.,  2021, \mn@doi [\apj] {10.3847/1538-4357/ac1bac}, \href {https://ui.adsabs.harvard.edu/abs/2021ApJ...921...73G} {921, 73}

\bibitem[\protect\citeauthoryear{{Guiglion}, {Bergemann}, {Storm}, {Lian}, {Cescutti}  \& {Serenelli}}{{Guiglion} et~al.}{2024}]{Guiglion2024}
{Guiglion} G.,  {Bergemann} M.,  {Storm} N.,  {Lian} J.,  {Cescutti} G.,   {Serenelli} A.,  2024, \mn@doi [\aap] {10.1051/0004-6361/202348522}, \href {https://ui.adsabs.harvard.edu/abs/2024A&A...683A..73G} {683, A73}

\bibitem[\protect\citeauthoryear{{Guo} et~al.,}{{Guo} et~al.}{2025}]{Guo2025}
{Guo} Y.,  et~al., 2025, \mn@doi [\aap] {10.1051/0004-6361/202451536}, \href {https://ui.adsabs.harvard.edu/abs/2025A&A...693A.211G} {693, A211}

\bibitem[\protect\citeauthoryear{{Gustafsson}, {Edvardsson}, {Eriksson}, {J{\o}rgensen}, {Nordlund}  \& {Plez}}{{Gustafsson} et~al.}{2008}]{Gustafsson2008}
{Gustafsson} B.,  {Edvardsson} B.,  {Eriksson} K.,  {J{\o}rgensen} U.~G.,  {Nordlund} {\r{A}}.,   {Plez} B.,  2008, \mn@doi [\aap] {10.1051/0004-6361:200809724}, \href {https://ui.adsabs.harvard.edu/abs/2008A&A...486..951G} {486, 951}

\bibitem[\protect\citeauthoryear{{Habing}}{{Habing}}{1996}]{Habing1996}
{Habing} H.~J.,  1996, \mn@doi [\aapr] {10.1007/PL00013287}, \href {https://ui.adsabs.harvard.edu/abs/1996A&ARv...7...97H} {7, 97}

\bibitem[\protect\citeauthoryear{{Halevi} \& {M{\"o}sta}}{{Halevi} \& {M{\"o}sta}}{2018}]{Halevi2018}
{Halevi} G.,  {M{\"o}sta} P.,  2018, \mn@doi [\mnras] {10.1093/mnras/sty797}, \href {https://ui.adsabs.harvard.edu/abs/2018MNRAS.477.2366H} {477, 2366}

\bibitem[\protect\citeauthoryear{{Hansen}, {Bergemann}, {Cescutti}, {Fran{\c{c}}ois}, {Arcones}, {Karakas}, {Lind}  \& {Chiappini}}{{Hansen} et~al.}{2013}]{Hansen2013}
{Hansen} C.~J.,  {Bergemann} M.,  {Cescutti} G.,  {Fran{\c{c}}ois} P.,  {Arcones} A.,  {Karakas} A.~I.,  {Lind} K.,   {Chiappini} C.,  2013, \mn@doi [\aap] {10.1051/0004-6361/201220584}, \href {https://ui.adsabs.harvard.edu/abs/2013A&A...551A..57H} {551, A57}

\bibitem[\protect\citeauthoryear{{Hansen} et~al.,}{{Hansen} et~al.}{2018}]{Hansen2018}
{Hansen} T.~T.,  et~al., 2018, \mn@doi [\apj] {10.3847/1538-4357/aabacc}, \href {https://ui.adsabs.harvard.edu/abs/2018ApJ...858...92H} {858, 92}

\bibitem[\protect\citeauthoryear{{Herwig}, {Pignatari}, {Woodward}, {Porter}, {Rockefeller}, {Fryer}, {Bennett}  \& {Hirschi}}{{Herwig} et~al.}{2011}]{Herwig2011}
{Herwig} F.,  {Pignatari} M.,  {Woodward} P.~R.,  {Porter} D.~H.,  {Rockefeller} G.,  {Fryer} C.~L.,  {Bennett} M.,   {Hirschi} R.,  2011, \mn@doi [\apj] {10.1088/0004-637X/727/2/89}, \href {https://ui.adsabs.harvard.edu/abs/2011ApJ...727...89H} {727, 89}

\bibitem[\protect\citeauthoryear{{Holmbeck} et~al.,}{{Holmbeck} et~al.}{2020}]{Holmbeck2020}
{Holmbeck} E.~M.,  et~al., 2020, \mn@doi [\apjs] {10.3847/1538-4365/ab9c19}, \href {https://ui.adsabs.harvard.edu/abs/2020ApJS..249...30H} {249, 30}

\bibitem[\protect\citeauthoryear{{Iben} \& {Tutukov}}{{Iben} \& {Tutukov}}{1984}]{Iben1984}
{Iben} I. J.,  {Tutukov} A.~V.,  1984, \mn@doi [\apjs] {10.1086/190932}, \href {https://ui.adsabs.harvard.edu/abs/1984ApJS...54..335I} {54, 335}

\bibitem[\protect\citeauthoryear{{Ishimaru}, {Wanajo}  \& {Prantzos}}{{Ishimaru} et~al.}{2015}]{Ishimaru2015}
{Ishimaru} Y.,  {Wanajo} S.,   {Prantzos} N.,  2015, \mn@doi [\apjl] {10.1088/2041-8205/804/2/L35}, \href {https://ui.adsabs.harvard.edu/abs/2015ApJ...804L..35I} {804, L35}

\bibitem[\protect\citeauthoryear{{Janka}, {M{\"u}ller}, {Kitaura}  \& {Buras}}{{Janka} et~al.}{2008}]{Janka2008}
{Janka} H.~T.,  {M{\"u}ller} B.,  {Kitaura} F.~S.,   {Buras} R.,  2008, \mn@doi [\aap] {10.1051/0004-6361:20079334}, \href {https://ui.adsabs.harvard.edu/abs/2008A&A...485..199J} {485, 199}

\bibitem[\protect\citeauthoryear{{Ji}, {Frebel}, {Simon}  \& {Chiti}}{{Ji} et~al.}{2016}]{Ji2016}
{Ji} A.~P.,  {Frebel} A.,  {Simon} J.~D.,   {Chiti} A.,  2016, \mn@doi [\apj] {10.3847/0004-637X/830/2/93}, \href {https://ui.adsabs.harvard.edu/abs/2016ApJ...830...93J} {830, 93}

\bibitem[\protect\citeauthoryear{{Jones}, {Ritter}, {Herwig}, {Fryer}, {Pignatari}, {Bertolli}  \& {Paxton}}{{Jones} et~al.}{2016}]{Jones2016}
{Jones} S.,  {Ritter} C.,  {Herwig} F.,  {Fryer} C.,  {Pignatari} M.,  {Bertolli} M.~G.,   {Paxton} B.,  2016, \mn@doi [\mnras] {10.1093/mnras/stv2488}, \href {https://ui.adsabs.harvard.edu/abs/2016MNRAS.455.3848J} {455, 3848}

\bibitem[\protect\citeauthoryear{{Just}, {Bauswein}, {Ardevol Pulpillo}, {Goriely}  \& {Janka}}{{Just} et~al.}{2015}]{Just2015}
{Just} O.,  {Bauswein} A.,  {Ardevol Pulpillo} R.,  {Goriely} S.,   {Janka} H.~T.,  2015, \mn@doi [\mnras] {10.1093/mnras/stv009}, \href {https://ui.adsabs.harvard.edu/abs/2015MNRAS.448..541J} {448, 541}

\bibitem[\protect\citeauthoryear{{Karakas}}{{Karakas}}{2010}]{Karakas2010}
{Karakas} A.~I.,  2010, \mn@doi [\mnras] {10.1111/j.1365-2966.2009.16198.x}, \href {https://ui.adsabs.harvard.edu/abs/2010MNRAS.403.1413K} {403, 1413}

\bibitem[\protect\citeauthoryear{{Karakas} \& {Lattanzio}}{{Karakas} \& {Lattanzio}}{2014}]{Karakas2014}
{Karakas} A.~I.,  {Lattanzio} J.~C.,  2014, \mn@doi [\pasa] {10.1017/pasa.2014.21}, \href {https://ui.adsabs.harvard.edu/abs/2014PASA...31...30K} {31, e030}

\bibitem[\protect\citeauthoryear{{Karakas} \& {Lugaro}}{{Karakas} \& {Lugaro}}{2016}]{Karakas2016}
{Karakas} A.~I.,  {Lugaro} M.,  2016, \mn@doi [\apj] {10.3847/0004-637X/825/1/26}, \href {https://ui.adsabs.harvard.edu/abs/2016ApJ...825...26K} {825, 26}

\bibitem[\protect\citeauthoryear{{Karakas}, {Lugaro}, {Carlos}, {Cseh}, {Kamath}  \& {Garc{\'\i}a-Hern{\'a}ndez}}{{Karakas} et~al.}{2018}]{Karakas2018}
{Karakas} A.~I.,  {Lugaro} M.,  {Carlos} M.,  {Cseh} B.,  {Kamath} D.,   {Garc{\'\i}a-Hern{\'a}ndez} D.~A.,  2018, \mn@doi [\mnras] {10.1093/mnras/sty625}, \href {https://ui.adsabs.harvard.edu/abs/2018MNRAS.477..421K} {477, 421}

\bibitem[\protect\citeauthoryear{{Karakas}, {Cinquegrana}  \& {Joyce}}{{Karakas} et~al.}{2022}]{Karakas2022}
{Karakas} A.~I.,  {Cinquegrana} G.,   {Joyce} M.,  2022, \mn@doi [\mnras] {10.1093/mnras/stab3205}, \href {https://ui.adsabs.harvard.edu/abs/2022MNRAS.509.4430K} {509, 4430}

\bibitem[\protect\citeauthoryear{{Karinkuzhi}, {Van Eck}, {Goriely}, {Siess}, {Jorissen}, {Merle}, {Escorza}  \& {Masseron}}{{Karinkuzhi} et~al.}{2021}]{Karinkuzhi2021}
{Karinkuzhi} D.,  {Van Eck} S.,  {Goriely} S.,  {Siess} L.,  {Jorissen} A.,  {Merle} T.,  {Escorza} A.,   {Masseron} T.,  2021, \mn@doi [\aap] {10.1051/0004-6361/202038891}, \href {https://ui.adsabs.harvard.edu/abs/2021A&A...645A..61K} {645, A61}

\bibitem[\protect\citeauthoryear{{Katz} \& {Dong}}{{Katz} \& {Dong}}{2012}]{Katz2012}
{Katz} B.,  {Dong} S.,  2012, \mn@doi [arXiv e-prints] {10.48550/arXiv.1211.4584}, \href {https://ui.adsabs.harvard.edu/abs/2012arXiv1211.4584K} {p. arXiv:1211.4584}

\bibitem[\protect\citeauthoryear{{Kaur} \& {Sahijpal}}{{Kaur} \& {Sahijpal}}{2019}]{Kaur2019}
{Kaur} T.,  {Sahijpal} S.,  2019, \mn@doi [\mnras] {10.1093/mnras/stz2720}, \href {https://ui.adsabs.harvard.edu/abs/2019MNRAS.490.1620K} {490, 1620}

\bibitem[\protect\citeauthoryear{{Kirby} et~al.,}{{Kirby} et~al.}{2019}]{Kirby2019}
{Kirby} E.~N.,  et~al., 2019, \mn@doi [\apj] {10.3847/1538-4357/ab2c02}, \href {https://ui.adsabs.harvard.edu/abs/2019ApJ...881...45K} {881, 45}

\bibitem[\protect\citeauthoryear{{Kobayashi}, {Umeda}, {Nomoto}, {Tominaga}  \& {Ohkubo}}{{Kobayashi} et~al.}{2006}]{Kobayashi2006}
{Kobayashi} C.,  {Umeda} H.,  {Nomoto} K.,  {Tominaga} N.,   {Ohkubo} T.,  2006, \mn@doi [\apj] {10.1086/508914}, \href {https://ui.adsabs.harvard.edu/abs/2006ApJ...653.1145K} {653, 1145}

\bibitem[\protect\citeauthoryear{{Kobayashi}, {Leung}  \& {Nomoto}}{{Kobayashi} et~al.}{2020a}]{Kobayashi2020b}
{Kobayashi} C.,  {Leung} S.-C.,   {Nomoto} K.,  2020a, \mn@doi [\apj] {10.3847/1538-4357/ab8e44}, \href {https://ui.adsabs.harvard.edu/abs/2020ApJ...895..138K} {895, 138}

\bibitem[\protect\citeauthoryear{{Kobayashi}, {Karakas}  \& {Lugaro}}{{Kobayashi} et~al.}{2020b}]{Kobayashi2020a}
{Kobayashi} C.,  {Karakas} A.~I.,   {Lugaro} M.,  2020b, \mn@doi [\apj] {10.3847/1538-4357/abae65}, \href {https://ui.adsabs.harvard.edu/abs/2020ApJ...900..179K} {900, 179}

\bibitem[\protect\citeauthoryear{{Korotin}, {Andrievsky}, {Hansen}, {Caffau}, {Bonifacio}, {Spite}, {Spite}  \& {Fran{\c{c}}ois}}{{Korotin} et~al.}{2015}]{Korotin2015}
{Korotin} S.~A.,  {Andrievsky} S.~M.,  {Hansen} C.~J.,  {Caffau} E.,  {Bonifacio} P.,  {Spite} M.,  {Spite} F.,   {Fran{\c{c}}ois} P.,  2015, \mn@doi [\aap] {10.1051/0004-6361/201526558}, \href {https://ui.adsabs.harvard.edu/abs/2015A&A...581A..70K} {581, A70}

\bibitem[\protect\citeauthoryear{Krane}{Krane}{1988}]{Krane1988}
Krane K.~S.,  1988, {Introductory nuclear physics}.
Wiley, New York, NY, \url {https://cds.cern.ch/record/359790}

\bibitem[\protect\citeauthoryear{{Kub{\'a}t}}{{Kub{\'a}t}}{2014}]{Kubat2014}
{Kub{\'a}t} J.,  2014, in {Niemczura} E.,  {Smalley} B.,   {Pych} W.,  eds, , Determination of Atmospheric Parameters of B.
pp 149--157, \mn@doi{10.1007/978-3-319-06956-2_14}

\bibitem[\protect\citeauthoryear{{Kushnir}, {Katz}, {Dong}, {Livne}  \& {Fern{\'a}ndez}}{{Kushnir} et~al.}{2013}]{Kushnir2013}
{Kushnir} D.,  {Katz} B.,  {Dong} S.,  {Livne} E.,   {Fern{\'a}ndez} R.,  2013, \mn@doi [\apjl] {10.1088/2041-8205/778/2/L37}, \href {https://ui.adsabs.harvard.edu/abs/2013ApJ...778L..37K} {778, L37}

\bibitem[\protect\citeauthoryear{{Lattimer}}{{Lattimer}}{2012}]{Lattimer2012}
{Lattimer} J.~M.,  2012, \mn@doi [Annual Review of Nuclear and Particle Science] {10.1146/annurev-nucl-102711-095018}, \href {https://ui.adsabs.harvard.edu/abs/2012ARNPS..62..485L} {62, 485}

\bibitem[\protect\citeauthoryear{{Li} et~al.,}{{Li} et~al.}{2022}]{Li2022}
{Li} H.,  et~al., 2022, \mn@doi [\apj] {10.3847/1538-4357/ac6514}, \href {https://ui.adsabs.harvard.edu/abs/2022ApJ...931..147L} {931, 147}

\bibitem[\protect\citeauthoryear{{Lian}, {Storm}, {Guiglion}, {Serenelli}, {Cote}, {Karakas}, {Boardman}  \& {Bergemann}}{{Lian} et~al.}{2023}]{Lian2023}
{Lian} J.,  {Storm} N.,  {Guiglion} G.,  {Serenelli} A.,  {Cote} B.,  {Karakas} A.~I.,  {Boardman} N.,   {Bergemann} M.,  2023, \mn@doi [\mnras] {10.1093/mnras/stad2390}, \href {https://ui.adsabs.harvard.edu/abs/2023MNRAS.525.1329L} {525, 1329}

\bibitem[\protect\citeauthoryear{{Limongi} \& {Chieffi}}{{Limongi} \& {Chieffi}}{2018}]{Limongi2018}
{Limongi} M.,  {Chieffi} A.,  2018, \mn@doi [\apjs] {10.3847/1538-4365/aacb24}, \href {https://ui.adsabs.harvard.edu/abs/2018ApJS..237...13L} {237, 13}

\bibitem[\protect\citeauthoryear{{Lind}, {Bergemann}  \& {Asplund}}{{Lind} et~al.}{2012}]{Lind2012}
{Lind} K.,  {Bergemann} M.,   {Asplund} M.,  2012, \mn@doi [\mnras] {10.1111/j.1365-2966.2012.21686.x}, \href {https://ui.adsabs.harvard.edu/abs/2012MNRAS.427...50L} {427, 50}

\bibitem[\protect\citeauthoryear{{Lind} et~al.,}{{Lind} et~al.}{2017}]{Lind2017}
{Lind} K.,  et~al., 2017, \mn@doi [\mnras] {10.1093/mnras/stx673}, \href {https://ui.adsabs.harvard.edu/abs/2017MNRAS.468.4311L} {468, 4311}

\bibitem[\protect\citeauthoryear{{Livne} \& {Glasner}}{{Livne} \& {Glasner}}{1990}]{Livne1990}
{Livne} E.,  {Glasner} A.~S.,  1990, \mn@doi [\apj] {10.1086/169189}, \href {https://ui.adsabs.harvard.edu/abs/1990ApJ...361..244L} {361, 244}

\bibitem[\protect\citeauthoryear{{Luo} et~al.,}{{Luo} et~al.}{2012}]{Luo2012}
{Luo} A.~L.,  et~al., 2012, \mn@doi [Research in Astronomy and Astrophysics] {10.1088/1674-4527/12/9/004}, \href {https://ui.adsabs.harvard.edu/abs/2012RAA....12.1243L} {12, 1243}

\bibitem[\protect\citeauthoryear{{Magic}, {Collet}, {Asplund}, {Trampedach}, {Hayek}, {Chiavassa}, {Stein}  \& {Nordlund}}{{Magic} et~al.}{2013}]{Magic2013}
{Magic} Z.,  {Collet} R.,  {Asplund} M.,  {Trampedach} R.,  {Hayek} W.,  {Chiavassa} A.,  {Stein} R.~F.,   {Nordlund} {\r{A}}.,  2013, \mn@doi [\aap] {10.1051/0004-6361/201321274}, \href {https://ui.adsabs.harvard.edu/abs/2013A&A...557A..26M} {557, A26}

\bibitem[\protect\citeauthoryear{{Maoz} \& {Nakar}}{{Maoz} \& {Nakar}}{2024}]{Maoz2024}
{Maoz} D.,  {Nakar} E.,  2024, \mn@doi [arXiv e-prints] {10.48550/arXiv.2406.08630}, \href {https://ui.adsabs.harvard.edu/abs/2024arXiv240608630M} {p. arXiv:2406.08630}

\bibitem[\protect\citeauthoryear{{Mardini}, {Frebel}, {Betre}, {Jacobson}, {Norris}  \& {Christlieb}}{{Mardini} et~al.}{2024}]{Mardini2024}
{Mardini} M.~K.,  {Frebel} A.,  {Betre} L.,  {Jacobson} H.,  {Norris} J.~E.,   {Christlieb} N.,  2024, \mn@doi [\mnras] {10.1093/mnras/stad3925}, \href {https://ui.adsabs.harvard.edu/abs/2024MNRAS.528.2912M} {528, 2912}

\bibitem[\protect\citeauthoryear{{Mashonkina}, {Gehren}, {Shi}, {Korn}  \& {Grupp}}{{Mashonkina} et~al.}{2011}]{Mashonkina2011}
{Mashonkina} L.,  {Gehren} T.,  {Shi} J.~R.,  {Korn} A.~J.,   {Grupp} F.,  2011, \mn@doi [\aap] {10.1051/0004-6361/201015336}, \href {https://ui.adsabs.harvard.edu/abs/2011A&A...528A..87M} {528, A87}

\bibitem[\protect\citeauthoryear{{Matteucci}}{{Matteucci}}{2021}]{Matteucci2021}
{Matteucci} F.,  2021, \mn@doi [\aapr] {10.1007/s00159-021-00133-8}, \href {https://ui.adsabs.harvard.edu/abs/2021A&ARv..29....5M} {29, 5}

\bibitem[\protect\citeauthoryear{{Miglio} et~al.,}{{Miglio} et~al.}{2013}]{Miglio2013}
{Miglio} A.,  et~al., 2013, \mn@doi [\mnras] {10.1093/mnras/sts345}, \href {https://ui.adsabs.harvard.edu/abs/2013MNRAS.429..423M} {429, 423}

\bibitem[\protect\citeauthoryear{{Mihalas}}{{Mihalas}}{1970}]{Mihalas1970}
{Mihalas} D.,  1970, {Stellar atmospheres}

\bibitem[\protect\citeauthoryear{{Moe} \& {Di Stefano}}{{Moe} \& {Di Stefano}}{2017}]{Moe2017}
{Moe} M.,  {Di Stefano} R.,  2017, \mn@doi [\apjs] {10.3847/1538-4365/aa6fb6}, \href {https://ui.adsabs.harvard.edu/abs/2017ApJS..230...15M} {230, 15}

\bibitem[\protect\citeauthoryear{{Nishimura}, {Takiwaki}  \& {Thielemann}}{{Nishimura} et~al.}{2015}]{Nishimura2015}
{Nishimura} N.,  {Takiwaki} T.,   {Thielemann} F.-K.,  2015, \mn@doi [\apj] {10.1088/0004-637X/810/2/109}, \href {https://ui.adsabs.harvard.edu/abs/2015ApJ...810..109N} {810, 109}

\bibitem[\protect\citeauthoryear{{Nissen}, {Chen}, {Carigi}, {Schuster}  \& {Zhao}}{{Nissen} et~al.}{2014}]{Nissen2014}
{Nissen} P.~E.,  {Chen} Y.~Q.,  {Carigi} L.,  {Schuster} W.~J.,   {Zhao} G.,  2014, \mn@doi [\aap] {10.1051/0004-6361/201424184}, \href {https://ui.adsabs.harvard.edu/abs/2014A&A...568A..25N} {568, A25}

\bibitem[\protect\citeauthoryear{{Nomoto}}{{Nomoto}}{1987}]{Nomoto1987}
{Nomoto} K.,  1987, \mn@doi [\apj] {10.1086/165716}, \href {https://ui.adsabs.harvard.edu/abs/1987ApJ...322..206N} {322, 206}

\bibitem[\protect\citeauthoryear{{Nordlander}, {Amarsi}, {Lind}, {Asplund}, {Barklem}, {Casey}, {Collet}  \& {Leenaarts}}{{Nordlander} et~al.}{2017}]{Nordlander2017}
{Nordlander} T.,  {Amarsi} A.~M.,  {Lind} K.,  {Asplund} M.,  {Barklem} P.~S.,  {Casey} A.~R.,  {Collet} R.,   {Leenaarts} J.,  2017, \mn@doi [\aap] {10.1051/0004-6361/201629202}, \href {https://ui.adsabs.harvard.edu/abs/2017A&A...597A...6N} {597, A6}

\bibitem[\protect\citeauthoryear{{Pagel}}{{Pagel}}{2009}]{Pagel2009}
{Pagel} B. E.~J.,  2009, {Nucleosynthesis and Chemical Evolution of Galaxies}

\bibitem[\protect\citeauthoryear{{Pakmor}, {Kromer}, {Taubenberger}, {Sim}, {R{\"o}pke}  \& {Hillebrandt}}{{Pakmor} et~al.}{2012}]{Pakmor2012}
{Pakmor} R.,  {Kromer} M.,  {Taubenberger} S.,  {Sim} S.~A.,  {R{\"o}pke} F.~K.,   {Hillebrandt} W.,  2012, \mn@doi [\apjl] {10.1088/2041-8205/747/1/L10}, \href {https://ui.adsabs.harvard.edu/abs/2012ApJ...747L..10P} {747, L10}

\bibitem[\protect\citeauthoryear{{Pakmor} et~al.,}{{Pakmor} et~al.}{2022}]{Pakmor2022}
{Pakmor} R.,  et~al., 2022, \mn@doi [\mnras] {10.1093/mnras/stac3107}, \href {https://ui.adsabs.harvard.edu/abs/2022MNRAS.517.5260P} {517, 5260}

\bibitem[\protect\citeauthoryear{{Pakmor}, {Seitenzahl}, {Ruiter}, {Sim}, {R{\"o}pke}, {Taubenberger}, {Bieri}  \& {Blondin}}{{Pakmor} et~al.}{2024}]{Pakmor2024}
{Pakmor} R.,  {Seitenzahl} I.~R.,  {Ruiter} A.~J.,  {Sim} S.~A.,  {R{\"o}pke} F.~K.,  {Taubenberger} S.,  {Bieri} R.,   {Blondin} S.,  2024, \mn@doi [\aap] {10.1051/0004-6361/202449637}, \href {https://ui.adsabs.harvard.edu/abs/2024A&A...686A.227P} {686, A227}

\bibitem[\protect\citeauthoryear{{Palla}}{{Palla}}{2021}]{Palla2021}
{Palla} M.,  2021, \mn@doi [\mnras] {10.1093/mnras/stab293}, \href {https://ui.adsabs.harvard.edu/abs/2021MNRAS.503.3216P} {503, 3216}

\bibitem[\protect\citeauthoryear{{Pietrow}, {Hoppe}, {Bergemann}  \& {Calvo}}{{Pietrow} et~al.}{2023}]{Pietrow2023}
{Pietrow} A.~G.~M.,  {Hoppe} R.,  {Bergemann} M.,   {Calvo} F.,  2023, \mn@doi [\aap] {10.1051/0004-6361/202346387}, \href {https://ui.adsabs.harvard.edu/abs/2023A&A...672L...6P} {672, L6}

\bibitem[\protect\citeauthoryear{{Pignatari} et~al.,}{{Pignatari} et~al.}{2016}]{Pignatari2016}
{Pignatari} M.,  et~al., 2016, \mn@doi [\apjs] {10.3847/0067-0049/225/2/24}, \href {https://ui.adsabs.harvard.edu/abs/2016ApJS..225...24P} {225, 24}

\bibitem[\protect\citeauthoryear{{Popa}, {Hoppe}, {Bergemann}, {Hansen}, {Plez}  \& {Beers}}{{Popa} et~al.}{2023}]{Popa2023}
{Popa} S.~A.,  {Hoppe} R.,  {Bergemann} M.,  {Hansen} C.~J.,  {Plez} B.,   {Beers} T.~C.,  2023, \mn@doi [\aap] {10.1051/0004-6361/202245503}, \href {https://ui.adsabs.harvard.edu/abs/2023A&A...670A..25P} {670, A25}

\bibitem[\protect\citeauthoryear{{Prantzos}, {Abia}, {Limongi}, {Chieffi}  \& {Cristallo}}{{Prantzos} et~al.}{2018}]{Prantzos2018}
{Prantzos} N.,  {Abia} C.,  {Limongi} M.,  {Chieffi} A.,   {Cristallo} S.,  2018, \mn@doi [\mnras] {10.1093/mnras/sty316}, \href {https://ui.adsabs.harvard.edu/abs/2018MNRAS.476.3432P} {476, 3432}

\bibitem[\protect\citeauthoryear{{Reichert}, {Obergaulinger}, {Aloy}, {Gabler}, {Arcones}  \& {Thielemann}}{{Reichert} et~al.}{2023}]{Reichert2023}
{Reichert} M.,  {Obergaulinger} M.,  {Aloy} M.~{\'A}.,  {Gabler} M.,  {Arcones} A.,   {Thielemann} F.~K.,  2023, \mn@doi [\mnras] {10.1093/mnras/stac3185}, \href {https://ui.adsabs.harvard.edu/abs/2023MNRAS.518.1557R} {518, 1557}

\bibitem[\protect\citeauthoryear{{Rolfs} \& {Rodney}}{{Rolfs} \& {Rodney}}{1988}]{Rolfs1988}
{Rolfs} C.~E.,  {Rodney} W.~S.,  1988, {Cauldrons in the cosmos : nuclear astrophysics}

\bibitem[\protect\citeauthoryear{{Romano}, {Matteucci}, {Zhang}, {Papadopoulos}  \& {Ivison}}{{Romano} et~al.}{2017}]{Romano2017}
{Romano} D.,  {Matteucci} F.,  {Zhang} Z.~Y.,  {Papadopoulos} P.~P.,   {Ivison} R.~J.,  2017, \mn@doi [\mnras] {10.1093/mnras/stx1197}, \href {https://ui.adsabs.harvard.edu/abs/2017MNRAS.470..401R} {470, 401}

\bibitem[\protect\citeauthoryear{{R{\"o}pke} \& {De Marco}}{{R{\"o}pke} \& {De Marco}}{2023}]{Roepke2023}
{R{\"o}pke} F.~K.,  {De Marco} O.,  2023, \mn@doi [Living Reviews in Computational Astrophysics] {10.1007/s41115-023-00017-x}, \href {https://ui.adsabs.harvard.edu/abs/2023LRCA....9....2R} {9, 2}

\bibitem[\protect\citeauthoryear{{Rosswog}, {Liebend{\"o}rfer}, {Thielemann}, {Davies}, {Benz}  \& {Piran}}{{Rosswog} et~al.}{1999}]{Rosswog1999}
{Rosswog} S.,  {Liebend{\"o}rfer} M.,  {Thielemann} F.~K.,  {Davies} M.~B.,  {Benz} W.,   {Piran} T.,  1999, \mn@doi [\aap] {10.48550/arXiv.astro-ph/9811367}, \href {https://ui.adsabs.harvard.edu/abs/1999A&A...341..499R} {341, 499}

\bibitem[\protect\citeauthoryear{{Rosswog}, {Korobkin}, {Arcones}, {Thielemann}  \& {Piran}}{{Rosswog} et~al.}{2014}]{Rosswog2014}
{Rosswog} S.,  {Korobkin} O.,  {Arcones} A.,  {Thielemann} F.~K.,   {Piran} T.,  2014, \mn@doi [\mnras] {10.1093/mnras/stt2502}, \href {https://ui.adsabs.harvard.edu/abs/2014MNRAS.439..744R} {439, 744}

\bibitem[\protect\citeauthoryear{{Ruiter}}{{Ruiter}}{2020}]{Ruiter2020}
{Ruiter} A.~J.,  2020, in {Barstow} M.~A.,  {Kleinman} S.~J.,  {Provencal} J.~L.,   {Ferrario} L.,  eds,  IAU Symposium Vol. 357, White Dwarfs as Probes of Fundamental Physics: Tracers of Planetary, Stellar and Galactic Evolution. pp 1--15 (\mn@eprint {arXiv} {2001.02947}), \mn@doi{10.1017/S1743921320000587}

\bibitem[\protect\citeauthoryear{{Sana} et~al.,}{{Sana} et~al.}{2012}]{Sana2012}
{Sana} H.,  et~al., 2012, \mn@doi [Science] {10.1126/science.1223344}, \href {https://ui.adsabs.harvard.edu/abs/2012Sci...337..444S} {337, 444}

\bibitem[\protect\citeauthoryear{{Sanders}, {Belokurov}  \& {Man}}{{Sanders} et~al.}{2021}]{Sanders2021}
{Sanders} J.~L.,  {Belokurov} V.,   {Man} K. T.~F.,  2021, \mn@doi [\mnras] {10.1093/mnras/stab1951}, \href {https://ui.adsabs.harvard.edu/abs/2021MNRAS.506.4321S} {506, 4321}

\bibitem[\protect\citeauthoryear{{Seitenzahl}, {Cescutti}, {R{\"o}pke}, {Ruiter}  \& {Pakmor}}{{Seitenzahl} et~al.}{2013}]{Seitenzahl2013}
{Seitenzahl} I.~R.,  {Cescutti} G.,  {R{\"o}pke} F.~K.,  {Ruiter} A.~J.,   {Pakmor} R.,  2013, \mn@doi [\aap] {10.1051/0004-6361/201322599}, \href {https://ui.adsabs.harvard.edu/abs/2013A&A...559L...5S} {559, L5}

\bibitem[\protect\citeauthoryear{{Serenelli}, {Weiss}, {Cassisi}, {Salaris}  \& {Pietrinferni}}{{Serenelli} et~al.}{2017}]{Serenelli2017}
{Serenelli} A.,  {Weiss} A.,  {Cassisi} S.,  {Salaris} M.,   {Pietrinferni} A.,  2017, \mn@doi [\aap] {10.1051/0004-6361/201731004}, \href {https://ui.adsabs.harvard.edu/abs/2017A&A...606A..33S} {606, A33}

\bibitem[\protect\citeauthoryear{{Sharda} \& {Krumholz}}{{Sharda} \& {Krumholz}}{2022}]{Sharda2022}
{Sharda} P.,  {Krumholz} M.~R.,  2022, \mn@doi [\mnras] {10.1093/mnras/stab2921}, \href {https://ui.adsabs.harvard.edu/abs/2022MNRAS.509.1959S} {509, 1959}

\bibitem[\protect\citeauthoryear{{Shen}, {Kasen}, {Miles}  \& {Townsley}}{{Shen} et~al.}{2018}]{Shen2018}
{Shen} K.~J.,  {Kasen} D.,  {Miles} B.~J.,   {Townsley} D.~M.,  2018, \mn@doi [\apj] {10.3847/1538-4357/aaa8de}, \href {https://ui.adsabs.harvard.edu/abs/2018ApJ...854...52S} {854, 52}

\bibitem[\protect\citeauthoryear{{Siegel} \& {Metzger}}{{Siegel} \& {Metzger}}{2017}]{Siegel2017}
{Siegel} D.~M.,  {Metzger} B.~D.,  2017, \mn@doi [\prl] {10.1103/PhysRevLett.119.231102}, \href {https://ui.adsabs.harvard.edu/abs/2017PhRvL.119w1102S} {119, 231102}

\bibitem[\protect\citeauthoryear{{Siegel}, {Barnes}  \& {Metzger}}{{Siegel} et~al.}{2019}]{Siegel2019}
{Siegel} D.~M.,  {Barnes} J.,   {Metzger} B.~D.,  2019, \mn@doi [\nat] {10.1038/s41586-019-1136-0}, \href {https://ui.adsabs.harvard.edu/abs/2019Natur.569..241S} {569, 241}

\bibitem[\protect\citeauthoryear{{Sieverding}, {Kresse}  \& {Janka}}{{Sieverding} et~al.}{2023}]{Sieverding2023}
{Sieverding} A.,  {Kresse} D.,   {Janka} H.-T.,  2023, \mn@doi [\apjl] {10.3847/2041-8213/ad045b}, \href {https://ui.adsabs.harvard.edu/abs/2023ApJ...957L..25S} {957, L25}

\bibitem[\protect\citeauthoryear{{Smiljanic} et~al.,}{{Smiljanic} et~al.}{2014}]{Smiljanic2014}
{Smiljanic} R.,  et~al., 2014, \mn@doi [\aap] {10.1051/0004-6361/201423937}, \href {https://ui.adsabs.harvard.edu/abs/2014A&A...570A.122S} {570, A122}

\bibitem[\protect\citeauthoryear{{Sneden}, {Cowan}  \& {Gallino}}{{Sneden} et~al.}{2008}]{Sneden2008}
{Sneden} C.,  {Cowan} J.~J.,   {Gallino} R.,  2008, \mn@doi [\araa] {10.1146/annurev.astro.46.060407.145207}, \href {https://ui.adsabs.harvard.edu/abs/2008ARA&A..46..241S} {46, 241}

\bibitem[\protect\citeauthoryear{{Sneden}, {Cowan}, {Kobayashi}, {Pignatari}, {Lawler}, {Den Hartog}  \& {Wood}}{{Sneden} et~al.}{2016}]{Sneden2016}
{Sneden} C.,  {Cowan} J.~J.,  {Kobayashi} C.,  {Pignatari} M.,  {Lawler} J.~E.,  {Den Hartog} E.~A.,   {Wood} M.~P.,  2016, \mn@doi [\apj] {10.3847/0004-637X/817/1/53}, \href {https://ui.adsabs.harvard.edu/abs/2016ApJ...817...53S} {817, 53}

\bibitem[\protect\citeauthoryear{{Steffen}, {Prakapavi{\v{c}}ius}, {Caffau}, {Ludwig}, {Bonifacio}, {Cayrel}, {Ku{\v{c}}inskas}  \& {Livingston}}{{Steffen} et~al.}{2015}]{Steffen2015}
{Steffen} M.,  {Prakapavi{\v{c}}ius} D.,  {Caffau} E.,  {Ludwig} H.~G.,  {Bonifacio} P.,  {Cayrel} R.,  {Ku{\v{c}}inskas} A.,   {Livingston} W.~C.,  2015, \mn@doi [\aap] {10.1051/0004-6361/201526406}, \href {https://ui.adsabs.harvard.edu/abs/2015A&A...583A..57S} {583, A57}

\bibitem[\protect\citeauthoryear{{Stockinger} et~al.,}{{Stockinger} et~al.}{2020}]{Stockinger2020}
{Stockinger} G.,  et~al., 2020, \mn@doi [\mnras] {10.1093/mnras/staa1691}, \href {https://ui.adsabs.harvard.edu/abs/2020MNRAS.496.2039S} {496, 2039}

\bibitem[\protect\citeauthoryear{{Storm} \& {Bergemann}}{{Storm} \& {Bergemann}}{2023}]{Storm2023}
{Storm} N.,  {Bergemann} M.,  2023, \mn@doi [\mnras] {10.1093/mnras/stad2488}, \href {https://ui.adsabs.harvard.edu/abs/2023MNRAS.525.3718S} {525, 3718}

\bibitem[\protect\citeauthoryear{{Storm} et~al.,}{{Storm} et~al.}{2024}]{Storm2024}
{Storm} N.,  et~al., 2024, \mn@doi [\aap] {10.1051/0004-6361/202348971}, \href {https://ui.adsabs.harvard.edu/abs/2024A&A...683A.200S} {683, A200}

\bibitem[\protect\citeauthoryear{{Sukhbold}, {Ertl}, {Woosley}, {Brown}  \& {Janka}}{{Sukhbold} et~al.}{2016}]{Sukhbold2016}
{Sukhbold} T.,  {Ertl} T.,  {Woosley} S.~E.,  {Brown} J.~M.,   {Janka} H.~T.,  2016, \mn@doi [\apj] {10.3847/0004-637X/821/1/38}, \href {https://ui.adsabs.harvard.edu/abs/2016ApJ...821...38S} {821, 38}

\bibitem[\protect\citeauthoryear{{Takahashi}, {Witti}  \& {Janka}}{{Takahashi} et~al.}{1994}]{Takahashi1994}
{Takahashi} K.,  {Witti} J.,   {Janka} H.~T.,  1994, \aap, \href {https://ui.adsabs.harvard.edu/abs/1994A&A...286..857T} {286, 857}

\bibitem[\protect\citeauthoryear{{Takiwaki}, {Kotake}  \& {Sato}}{{Takiwaki} et~al.}{2009}]{Takiwaki2009}
{Takiwaki} T.,  {Kotake} K.,   {Sato} K.,  2009, \mn@doi [\apj] {10.1088/0004-637X/691/2/1360}, \href {https://ui.adsabs.harvard.edu/abs/2009ApJ...691.1360T} {691, 1360}

\bibitem[\protect\citeauthoryear{{Taubenberger}}{{Taubenberger}}{2017}]{Taubenberger2017}
{Taubenberger} S.,  2017, in {Alsabti} A.~W.,  {Murdin} P.,  eds, , Handbook of Supernovae.
p.~317, \mn@doi{10.1007/978-3-319-21846-5_37}

\bibitem[\protect\citeauthoryear{{Temim} et~al.,}{{Temim} et~al.}{2024}]{Temim2024}
{Temim} T.,  et~al., 2024, \mn@doi [\apjl] {10.3847/2041-8213/ad50d1}, \href {https://ui.adsabs.harvard.edu/abs/2024ApJ...968L..18T} {968, L18}

\bibitem[\protect\citeauthoryear{{Timmes}, {Woosley}  \& {Weaver}}{{Timmes} et~al.}{1995}]{Timmes1995}
{Timmes} F.~X.,  {Woosley} S.~E.,   {Weaver} T.~A.,  1995, \mn@doi [\apjs] {10.1086/192172}, \href {https://ui.adsabs.harvard.edu/abs/1995ApJS...98..617T} {98, 617}

\bibitem[\protect\citeauthoryear{{Toonen}, {Perets}  \& {Hamers}}{{Toonen} et~al.}{2018}]{Toonen2018}
{Toonen} S.,  {Perets} H.~B.,   {Hamers} A.~S.,  2018, \mn@doi [\aap] {10.1051/0004-6361/201731874}, \href {https://ui.adsabs.harvard.edu/abs/2018A&A...610A..22T} {610, A22}

\bibitem[\protect\citeauthoryear{{Voronov}, {Yakovleva}  \& {Belyaev}}{{Voronov} et~al.}{2022}]{Voronov2022}
{Voronov} Y.~V.,  {Yakovleva} S.~A.,   {Belyaev} A.~K.,  2022, \mn@doi [\apj] {10.3847/1538-4357/ac46fd}, \href {https://ui.adsabs.harvard.edu/abs/2022ApJ...926..173V} {926, 173}

\bibitem[\protect\citeauthoryear{{Wanajo}, {Janka}  \& {M{\"u}ller}}{{Wanajo} et~al.}{2011}]{Wanajo2011}
{Wanajo} S.,  {Janka} H.-T.,   {M{\"u}ller} B.,  2011, \mn@doi [\apjl] {10.1088/2041-8205/726/2/L15}, \href {https://ui.adsabs.harvard.edu/abs/2011ApJ...726L..15W} {726, L15}

\bibitem[\protect\citeauthoryear{{Wanajo}, {M{\"u}ller}, {Janka}  \& {Heger}}{{Wanajo} et~al.}{2018}]{Wanajo2018}
{Wanajo} S.,  {M{\"u}ller} B.,  {Janka} H.-T.,   {Heger} A.,  2018, \mn@doi [\apj] {10.3847/1538-4357/aa9d97}, \href {https://ui.adsabs.harvard.edu/abs/2018ApJ...852...40W} {852, 40}

\bibitem[\protect\citeauthoryear{{Wang} \& {Burrows}}{{Wang} \& {Burrows}}{2024}]{Wang2024a}
{Wang} T.,  {Burrows} A.,  2024, \mn@doi [\apj] {10.3847/1538-4357/ad5009}, \href {https://ui.adsabs.harvard.edu/abs/2024ApJ...969...74W} {969, 74}

\bibitem[\protect\citeauthoryear{{Wang} et~al.,}{{Wang} et~al.}{2024}]{Wang2024b}
{Wang} E.~X.,  et~al., 2024, \mn@doi [\mnras] {10.1093/mnras/stae385}, \href {https://ui.adsabs.harvard.edu/abs/2024MNRAS.528.5394W} {528, 5394}

\bibitem[\protect\citeauthoryear{{Watson} et~al.,}{{Watson} et~al.}{2019}]{Watson2019}
{Watson} D.,  et~al., 2019, \mn@doi [\nat] {10.1038/s41586-019-1676-3}, \href {https://ui.adsabs.harvard.edu/abs/2019Natur.574..497W} {574, 497}

\bibitem[\protect\citeauthoryear{{Whelan} \& {Iben}}{{Whelan} \& {Iben}}{1973}]{Whelan1973}
{Whelan} J.,  {Iben} Icko J.,  1973, \mn@doi [\apj] {10.1086/152565}, \href {https://ui.adsabs.harvard.edu/abs/1973ApJ...186.1007W} {186, 1007}

\bibitem[\protect\citeauthoryear{{Wongwathanarat}, {Janka}, {M{\"u}ller}, {Pllumbi}  \& {Wanajo}}{{Wongwathanarat} et~al.}{2017}]{Wongwathanarat2017}
{Wongwathanarat} A.,  {Janka} H.-T.,  {M{\"u}ller} E.,  {Pllumbi} E.,   {Wanajo} S.,  2017, \mn@doi [\apj] {10.3847/1538-4357/aa72de}, \href {https://ui.adsabs.harvard.edu/abs/2017ApJ...842...13W} {842, 13}

\bibitem[\protect\citeauthoryear{{Woosley}}{{Woosley}}{2018}]{Woosley2018}
{Woosley} S.~E.,  2018, \mn@doi [\apj] {10.3847/1538-4357/aad044}, \href {https://ui.adsabs.harvard.edu/abs/2018ApJ...863..105W} {863, 105}

\bibitem[\protect\citeauthoryear{{Woosley} \& {Weaver}}{{Woosley} \& {Weaver}}{1995}]{Woosley1995}
{Woosley} S.~E.,  {Weaver} T.~A.,  1995, \mn@doi [\apjs] {10.1086/192237}, \href {https://ui.adsabs.harvard.edu/abs/1995ApJS..101..181W} {101, 181}

\bibitem[\protect\citeauthoryear{{Woosley}, {Arnett}  \& {Clayton}}{{Woosley} et~al.}{1973}]{Woosley1973}
{Woosley} S.~E.,  {Arnett} W.~D.,   {Clayton} D.~D.,  1973, \mn@doi [\apjs] {10.1086/190282}, \href {https://ui.adsabs.harvard.edu/abs/1973ApJS...26..231W} {26, 231}

\bibitem[\protect\citeauthoryear{{Woosley}, {Wilson}, {Mathews}, {Hoffman}  \& {Meyer}}{{Woosley} et~al.}{1994}]{Woosley1994}
{Woosley} S.~E.,  {Wilson} J.~R.,  {Mathews} G.~J.,  {Hoffman} R.~D.,   {Meyer} B.~S.,  1994, \mn@doi [\apj] {10.1086/174638}, \href {https://ui.adsabs.harvard.edu/abs/1994ApJ...433..229W} {433, 229}

\bibitem[\protect\citeauthoryear{{Woosley}, {Heger}  \& {Weaver}}{{Woosley} et~al.}{2002}]{Woosley2002}
{Woosley} S.~E.,  {Heger} A.,   {Weaver} T.~A.,  2002, \mn@doi [Reviews of Modern Physics] {10.1103/RevModPhys.74.1015}, \href {https://ui.adsabs.harvard.edu/abs/2002RvMP...74.1015W} {74, 1015}

\bibitem[\protect\citeauthoryear{{Yakovleva}, {Voronov}  \& {Belyaev}}{{Yakovleva} et~al.}{2016}]{Yakovleva2016}
{Yakovleva} S.~A.,  {Voronov} Y.~V.,   {Belyaev} A.~K.,  2016, \mn@doi [\aap] {10.1051/0004-6361/201628659}, \href {https://ui.adsabs.harvard.edu/abs/2016A&A...593A..27Y} {593, A27}

\bibitem[\protect\citeauthoryear{{Yakovleva}, {Belyaev}  \& {Bergemann}}{{Yakovleva} et~al.}{2020}]{Yakovleva2020}
{Yakovleva} S.~A.,  {Belyaev} A.~K.,   {Bergemann} M.,  2020, \mn@doi [Atoms] {10.3390/atoms8030034}, \href {https://ui.adsabs.harvard.edu/abs/2020Atoms...8...34Y} {8, 34}

\bibitem[\protect\citeauthoryear{{Zhao}, {Zhao}, {Chu}, {Jing}  \& {Deng}}{{Zhao} et~al.}{2012}]{Zhao2012}
{Zhao} G.,  {Zhao} Y.-H.,  {Chu} Y.-Q.,  {Jing} Y.-P.,   {Deng} L.-C.,  2012, \mn@doi [Research in Astronomy and Astrophysics] {10.1088/1674-4527/12/7/002}, \href {https://ui.adsabs.harvard.edu/abs/2012RAA....12..723Z} {12, 723}

\bibitem[\protect\citeauthoryear{{Zhao} et~al.,}{{Zhao} et~al.}{2016}]{Zhao2016}
{Zhao} G.,  et~al., 2016, \mn@doi [\apj] {10.3847/1538-4357/833/2/225}, \href {https://ui.adsabs.harvard.edu/abs/2016ApJ...833..225Z} {833, 225}

\makeatother
\end{thebibliography}




\appendix

\section{Line formation in 1D hydrostatic in LTE and 3D radiation-hydrodynamic model atmospheres in NLTE}
\label{app:1d_vs_3d}

\begin{figure}
\includegraphics[width=1\columnwidth]{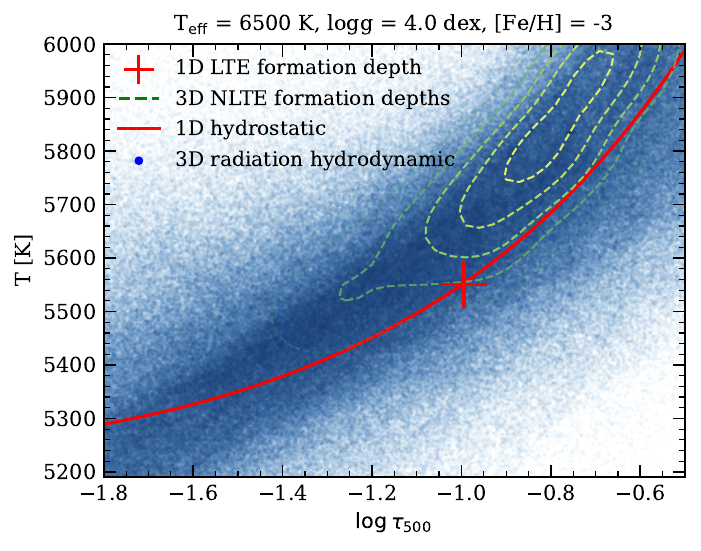}
\caption{{Zoomed-in temperature structure of the 1D MARCS (red line) and the 3D RHD \texttt{Stagger} model atmosphere (blue distribution) for $\teff = 6500$ K, $\logg = 4.0$ dex, and [Fe/H] $= -3$. The red cross indicates the formation height of the Ni I line at 5476 \AA~as a function of $\log \tau_{500}$ in 1D LTE; similarly, the formation heights in 3D NLTE are indicated by green contours. Here the formation height is defined as the location where the line core's contribution function contributes the most.}}
 \label{fig:1d_vs_3d_temperature}
\end{figure}


\begin{figure}
\includegraphics[width=1\columnwidth]{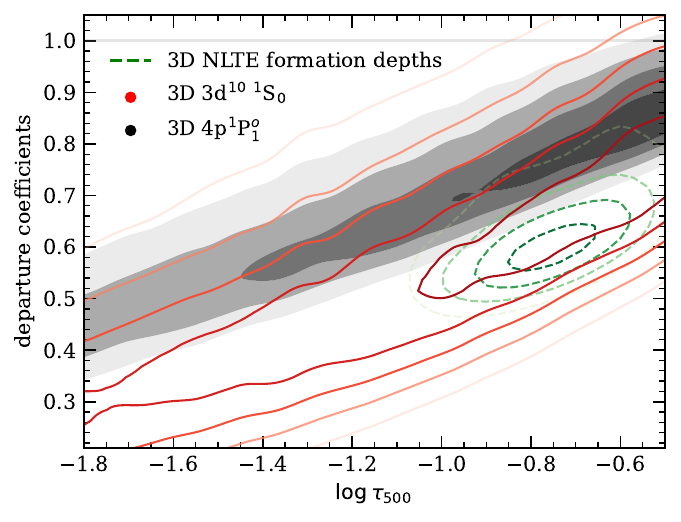}
\caption{{3D NLTE departure coefficients for the same Ni line at 5476.90~\AA~and model atmosphere as in Fig. \ref{fig:1d_vs_3d_temperature}. The departure coefficients of the lower level \Ni{3d^{10}}{1}{S}{}{0} are shown as red contours, and those of the upper level \Ni{4p}{1}{P}{o}{1} are shown as black contours. Green contours indicate the formation height of this line.}}
 \label{fig:1d_vs_3d_dep_coef}
\end{figure}


{To explore the origin of large 3D NLTE effects in the lines of iron-peak species, we performed spectrum synthesis  calculations of Ni~I line at 5476~\AA~(1.83~eV - 4.09~eV, Table \ref{tab:lines_data_ni}) using a 1D hydrostatic model atmosphere under LTE assumption and compared it to the 3D RHD model atmosphere with NLTE included. This line was chosen as a representative line, for which large 3D NLTE - 1D LTE effects are obtained, similarly to Fig. \ref{fig:spectra_3d_1d} and Fig. \ref{fig:cont_func_3d_both}.}

{Fig. \ref{fig:1d_vs_3d_temperature} shows the zoomed-in temperature structures of 1D (red curve) and 3D (blue dots) model atmospheres as a function of $\log \tau_{500}$. While the subsequent discussion is primarily limited to the effect of the temperature structure and effect of departure coefficients, we emphasise that there is also a contribution from the velocity and pressure differences between 3D RHD and 1D hydrostatic models, although these effects are much smaller compared to the effect of temperature variations. In this Fig. \ref{fig:1d_vs_3d_temperature}, the red cross indicates the line formation depth for the Ni I line at 5476 \AA~for 1D LTE at around $\log \tau_{500} \approx -1.0$. However, in 3D NLTE radiation transfer calculations the line forms over a range of optical depths and thus the line formation depths are plotted in green contours. The peak is located much deeper in the atmosphere at around $\log \tau_{500} \approx -0.9 \textrm{ to } -0.7$. Thus the temperature difference between the formation heights amounts to $\approx 300$ K. The main reason for NLTE effect is the radiation field \citep{Mihalas1970, Asplund2005, Bergemann2012a, Kubat2014} and its influence on the line opacity and source function. Hence in 3D NLTE, at the formation height of the Ni I line, the strong continuum UV radiation field \textit{defines} the rates and, hence, the atomic level populations. Fig. \ref{fig:1d_vs_3d_dep_coef} demonstrates exactly that. The lower and upper level populations are plotted in respectively red and black contours. Similarly to Fig. \ref{fig:1d_vs_3d_temperature}, the green contour is plotted in the rough location where the line formation occurs from the lower level at around $\log \tau_{500} \approx -0.9 \textrm{ to } -0.7$. At this depth, the energy states of the Ni I line are strongly under-populated ($b_i$ are much less than unity, Fig. \ref{fig:1d_vs_3d_dep_coef}) due to radiative over-ionization and line pumping, the classical NLTE effects \citep{Bruls1992}. Therefore, the line opacity ($\kappa^l \sim b_i < 1$), but also the line source function ($S^l / B^l_\nu \approx b_j / b_i > 1$) exceeds the Planck function (LTE). Both effects lead to the weakening of the Ni I line compared to the 1D LTE case \citep{Bergemann2014}, resulting in positive NLTE abundance corrections for the Ni I line. In contrast, the 1D LTE line is so much stronger, as its formation is purely set by the temperature (and other quantities), and especially at the line core the temperature is low enough ($\sim 5500$ K) to allow for a significantly higher population of the Ni I states involved in the atomic transition. One can almost mimic the very large 3D NLTE effect by e.g. using a significantly hotter MARCS model atmosphere. In that case, the EW difference between 1D LTE and 3D NLTE line profiles would reduce significantly. This is a toy experiment, and of course in realistic physical simulations as presented in this study for 7 chemical elements, much more complex distributions of line opacities and the source functions are obtained that is due to a highly non-linear correlations between the physical parameters in the coupled solution of hundreds or even thousands of statistical equilibrium equations (one for each energy state in our NLTE atoms and molecules) and radiation transfer equations in inhomogeneous 3D radiation-hydrodynamics models. In other words, there is obviously \textit{no single analytic solution or relation that can be used directly to explain the behaviour of the line shape in 1D NLTE or 3D NLTE. This is in contrast to 1D LTE line formation which is fully defined by the Saha-Boltzmann statistics applied to smooth 1D distributions of temperature and pressure with depth}. Hence, this experiment shall only be viewed as a qualitative demonstration of the underlying physical effect and shall not be extrapolated to other physical conditions, atomic or molecular lines, or to other chemical elements. Therefore for the detailed discussion of line formation we refer the readers to the papers where NLTE model atoms were introduced and validated (see Sec. \ref{subsec:spectrum_synthesis} in the main text).}

\section{3D NLTE and 1D NLTE abundance differences}
\label{app:nlte_corrections}

\subsection{Mn}

\begin{figure*}
\includegraphics[width=1\textwidth]{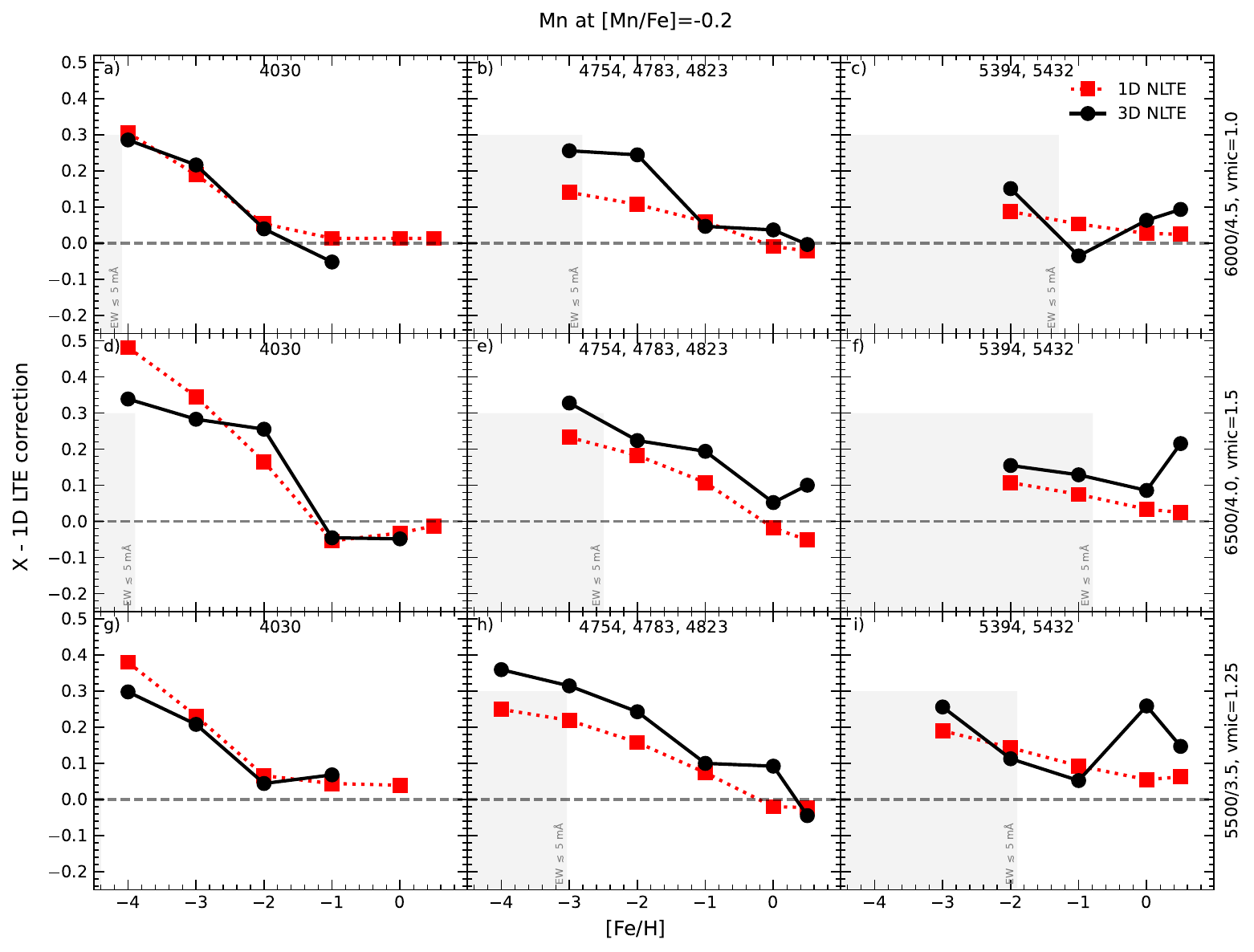}
\caption{Average Mn corrections for 1D LTE abundances at [Mn/Fe] = -0.2. Corrections are plotted for 1D NLTE and 3D NLTE respectively in red squares and black dots for three line groups in each column. Each row represents a different atmosphere, top to bottom: $\teff=6000$ K, $\logg=4.5$ dex, $\vmic = 1.00$ km s$^{-1}$ (main-sequence); $\teff=6500$ K, $\logg=4.0$ dex, $\vmic = 1.50$ km s$^{-1}$ (turn-off); $\teff=5500$ K, $\logg=3.5$ dex, $\vmic = 1.25$ km s$^{-1}$ (subgiant). $\vmic$ only refers to the value used in 1D models. A vertical shaded area represents a rough indication where a line becomes too weak to be measured reliably (EW $\leq$5m\AA).}
 \label{fig:mn_corrections}
\end{figure*}

Our derived 1D NLTE corrections for Mn in Fig. \ref{fig:mn_corrections} are similar to the previously derived ones in \citet{Bergemann2019}. 3D NLTE have similar trends, with a slight amplifications of effects in some regimes, mostly for turn-off stars. Out of the plotted lines, 4030 and 5394/5432 would be visible in the 4MOST HR spectra. Their corrections easily reach +0.30 dex in the most metal-poor regimes. 

\subsection{Co}

\begin{figure*}
\includegraphics[width=1\textwidth]{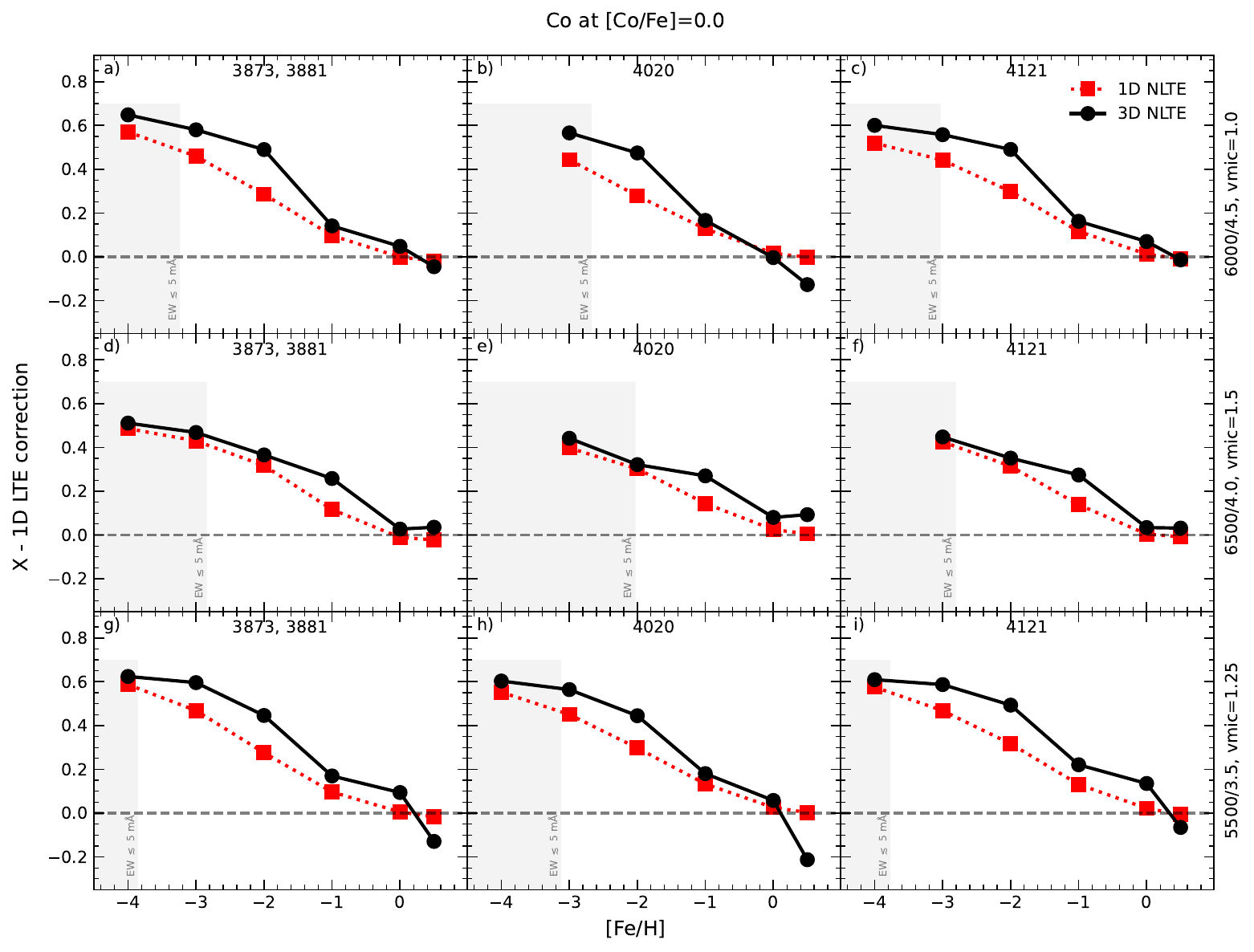}
\caption{Same as Fig. \ref{fig:mn_corrections}, but for Co at [Co/Fe] = 0.0.}
 \label{fig:co_corrections}
\end{figure*}

For Co the trend is very similar to Mn. Fig. \ref{fig:co_corrections} represents those corrections, with most lines reaching corrections of at least +0.3 dex at [Fe/H] = -2, whereas 3D amplifies the NLTE correction even further by up to +0.2 dex. Our corrections are smaller than the old atom \citet{Bergemann2010b}, but similar to the updated model atom in \citet{Yakovleva2020} due to the updated collisions which bring the NLTE effects down.

\subsection{Ni}

\begin{figure*}
\includegraphics[width=1\textwidth]{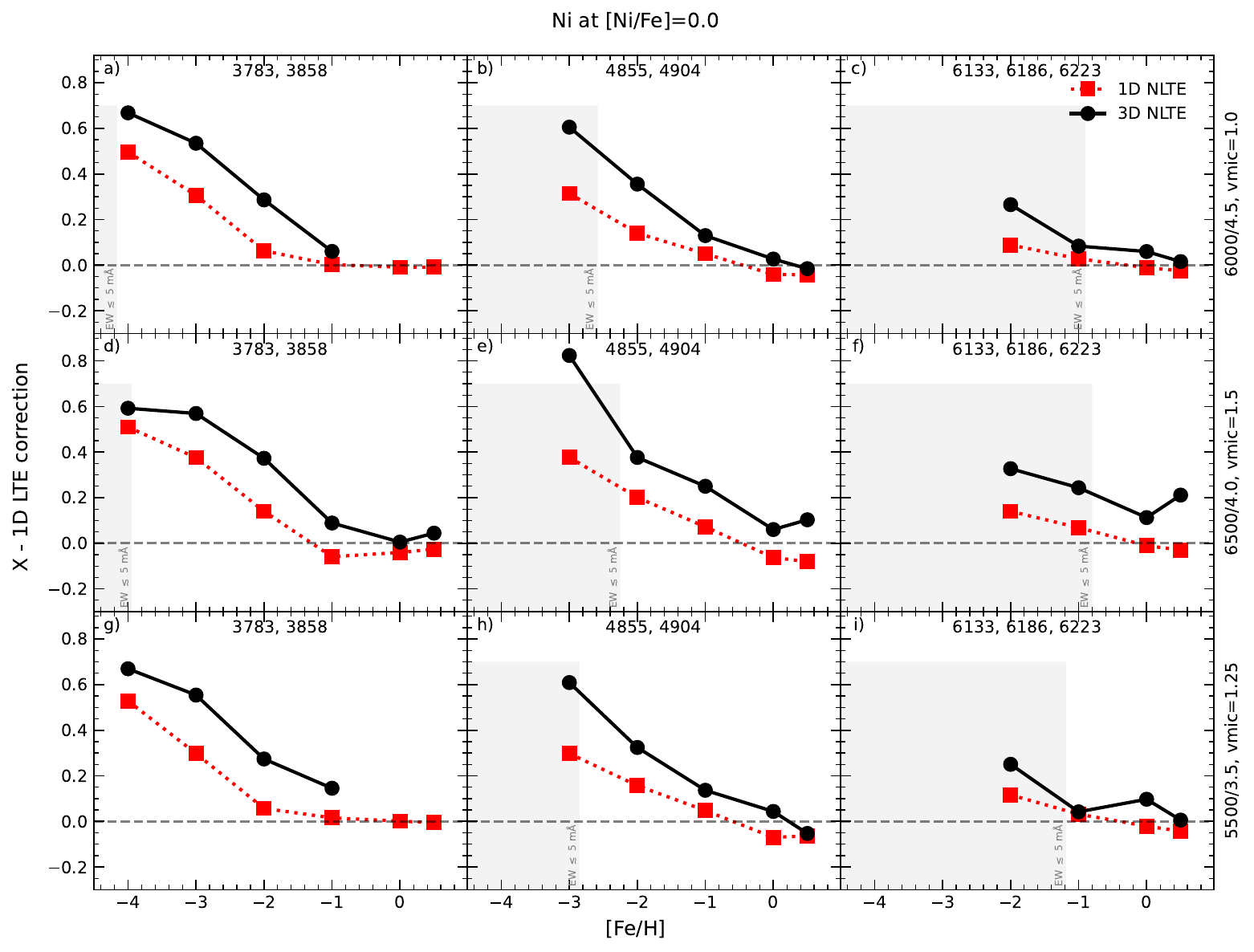}
\caption{Same as Fig. \ref{fig:mn_corrections}, but for Ni at [Ni/Fe] = 0.0.}
 \label{fig:ni_corrections}
\end{figure*}

Similarly to \citet{Eitner2023}, we get positive Ni corrections for our diagnostic lines as seen in Fig. \ref{fig:ni_corrections}. 3D amplifies this effect even further, especially in the metal-poor regime, in some cases by more than 0.2 dex compared to 1D NLTE.

\subsection{Sr}

\begin{figure*}
\includegraphics[width=1\textwidth]{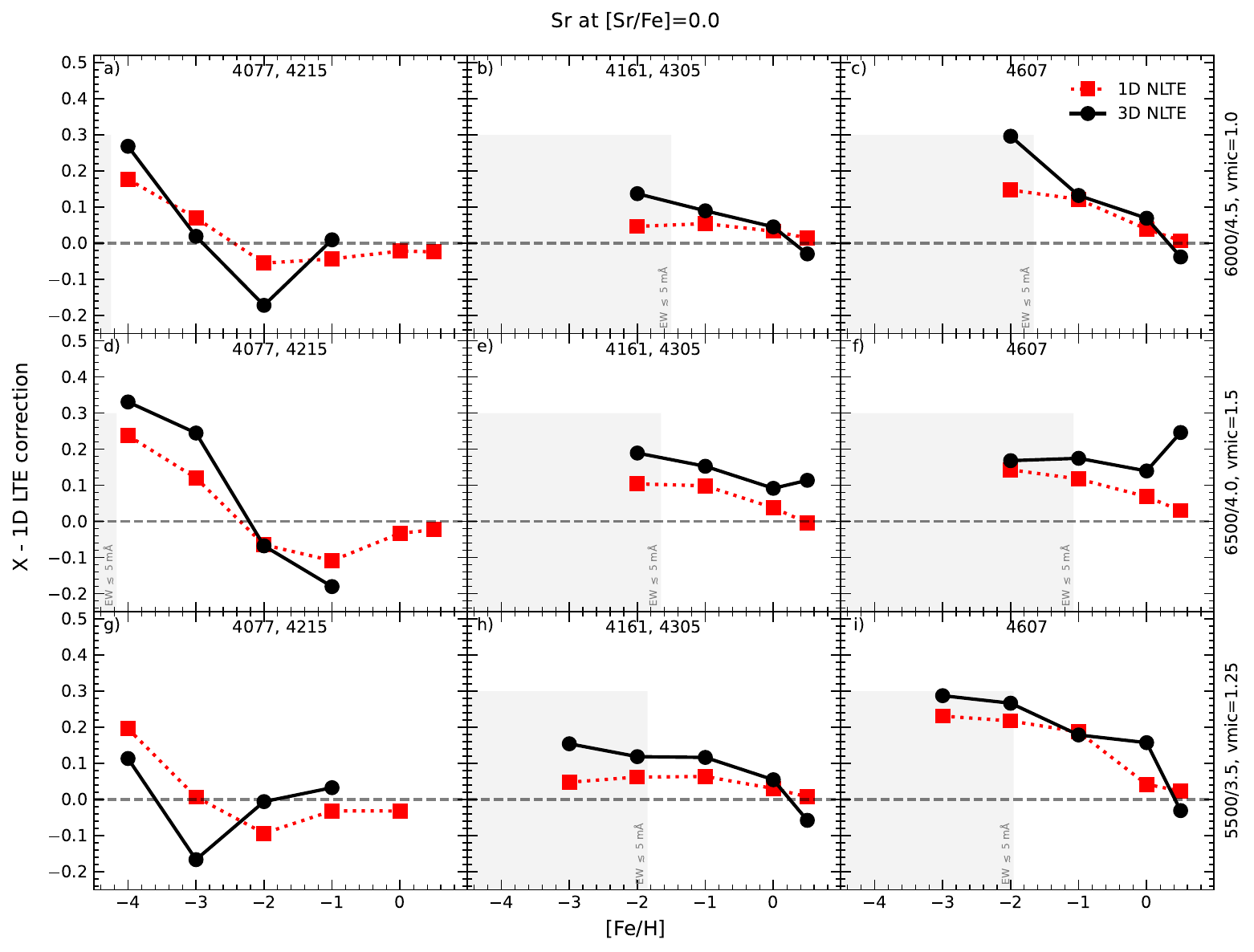}
\caption{Same as Fig. \ref{fig:mn_corrections}, but for Sr at [Sr/Fe] = 0.0.}
 \label{fig:sr_corrections}
\end{figure*}

Fig. \ref{fig:sr_corrections} {shows} corrections for the three diagnostic Sr line multiplets. 1D NLTE corrections are similar to the ones from \citep{Bergemann2012b}, where the line 4607 has positive corrections for all metallicities, while 4077 has negative and then positive corrections towards lower [Fe/H] values. 3D amplifies NLTE the effects even further, resulting in corrections that can reach 0.2-0.3 dex in some regimes.

\subsection{Y}

\begin{figure*}
\includegraphics[width=1\textwidth]{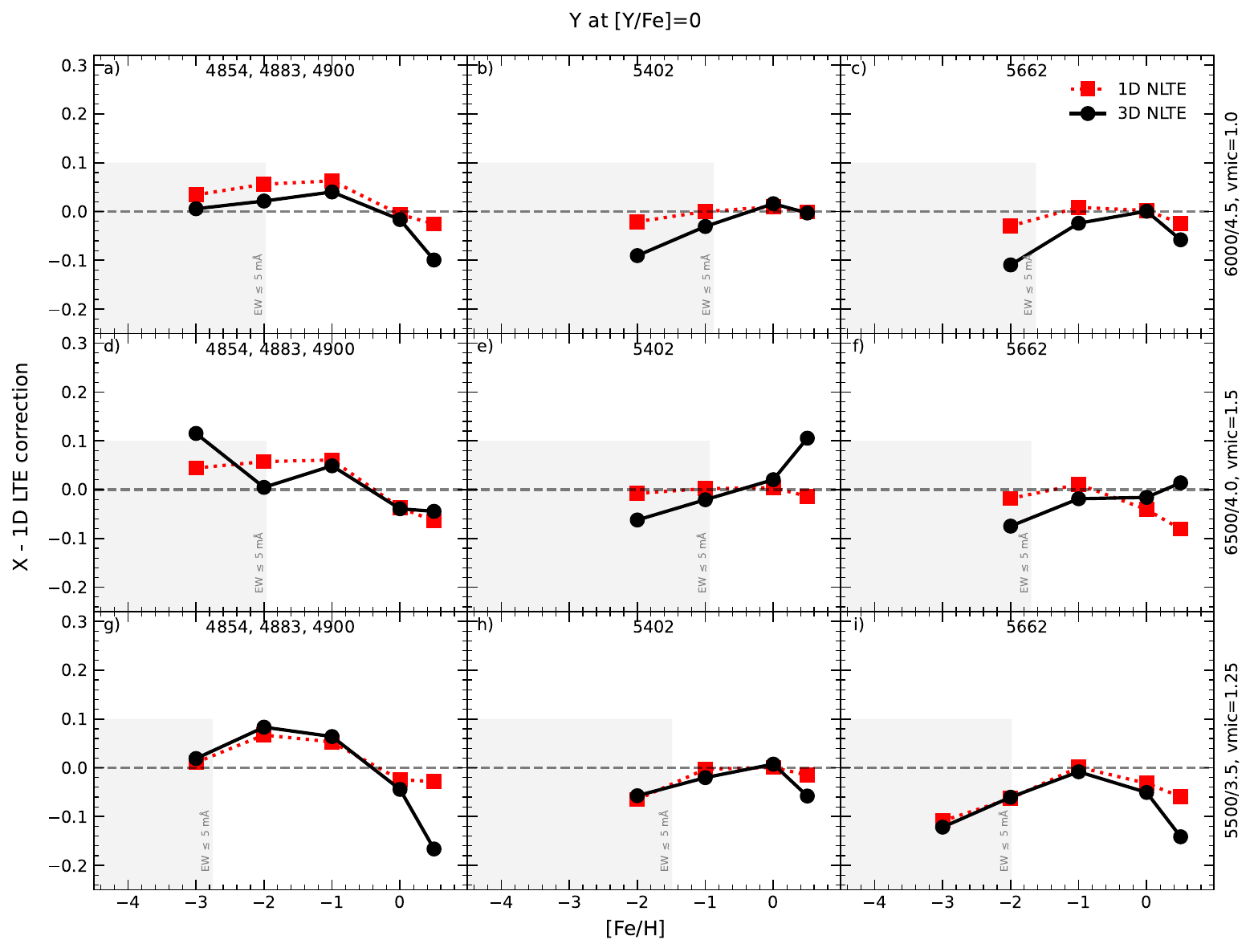}
\caption{Same as Fig. \ref{fig:mn_corrections}, but for Y at [Y/Fe] = 0.0.}
 \label{fig:y_corrections}
\end{figure*}

Y corrections for main optical lines can be seen in the Fig. \ref{fig:y_corrections}. As already discussed previously in \citet{Storm2023}, 1D NLTE yttrium corrections are small positive for theses lines of up to 0.10-0.15 dex. 3D NLTE doesn't change our analysis here. Lines 5200 and 5402 would be the typical diagnostic 4MOST HR lines. They are rather weak at the lower metallicity and have small negative corrections down to -0.15 dex.

\subsection{Ba}

\begin{figure*}
\includegraphics[width=0.75\textwidth]{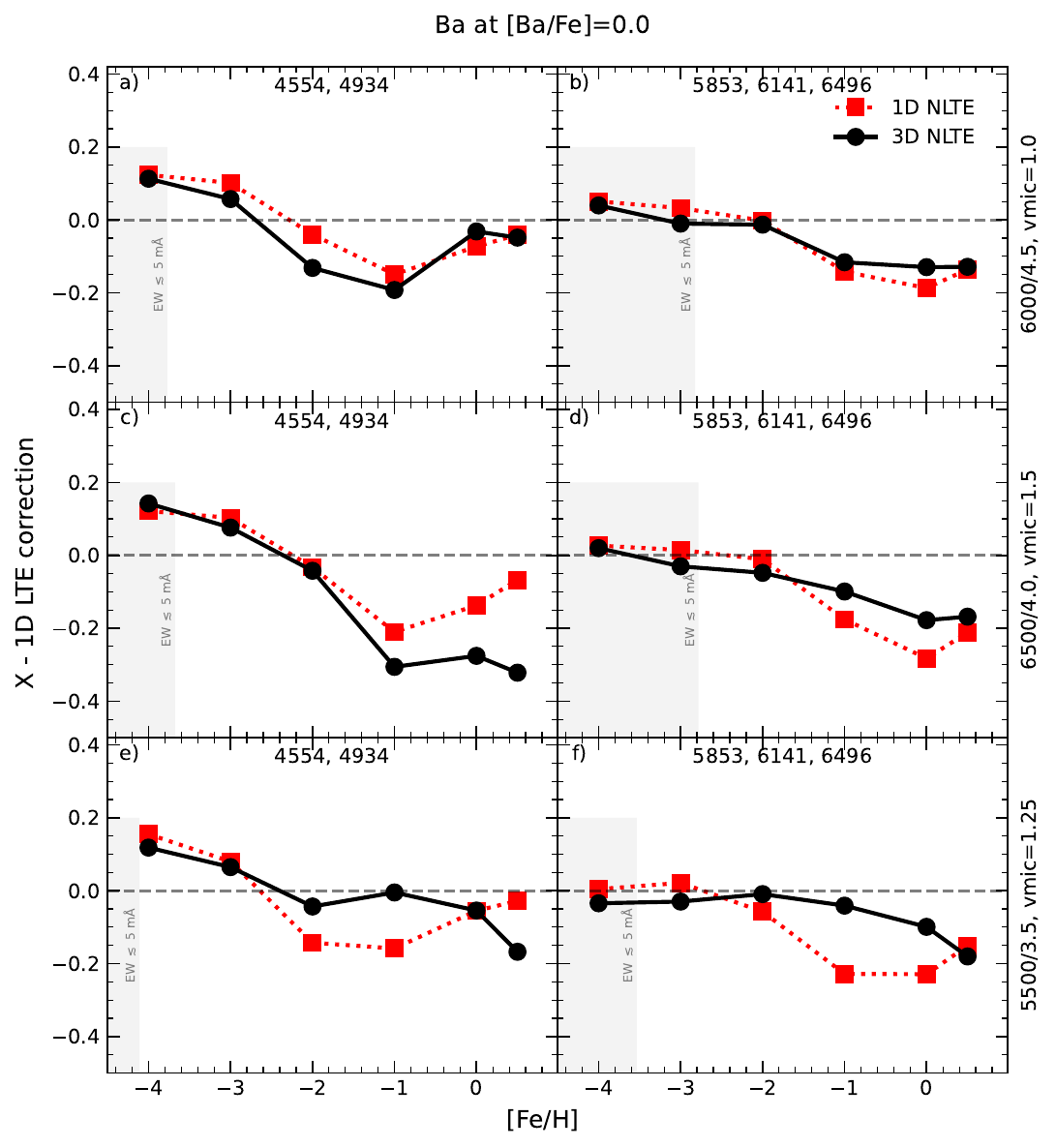}
\centering
\caption{Same as Fig. \ref{fig:mn_corrections}, but for Ba at [Ba/Fe] = 0.0}
 \label{fig:ba_corrections}
\end{figure*}

There are two typical diagnostic multiplets of the Ba lines, both exhibit quantitatively similar NLTE behaviour in Fig. \ref{fig:ba_corrections}: the correction is negative in the sub-solar regime that inverts to positive one at [Fe/H] $\leq$ -2 to -3. The similar qualitative behaviour can be seen in \citep{Korotin2015}, although their corrections were computed only for [Fe/H] = -2. \citep{Gallagher2020} also suggested that Ba is heavily affected by 3D effects. We can see a similar effect here, where 1D and 3D NLTE show different behaviour in some regimes. The 1D LTE EW of Ba is heavily affected by $\vmic$, so the 3D NLTE corrections can differ by 0.2 dex for each 0.25 km s$^{-1}$ difference in $\vmic$.

\subsection{Eu}

\begin{figure*}
\includegraphics[width=0.75\textwidth]{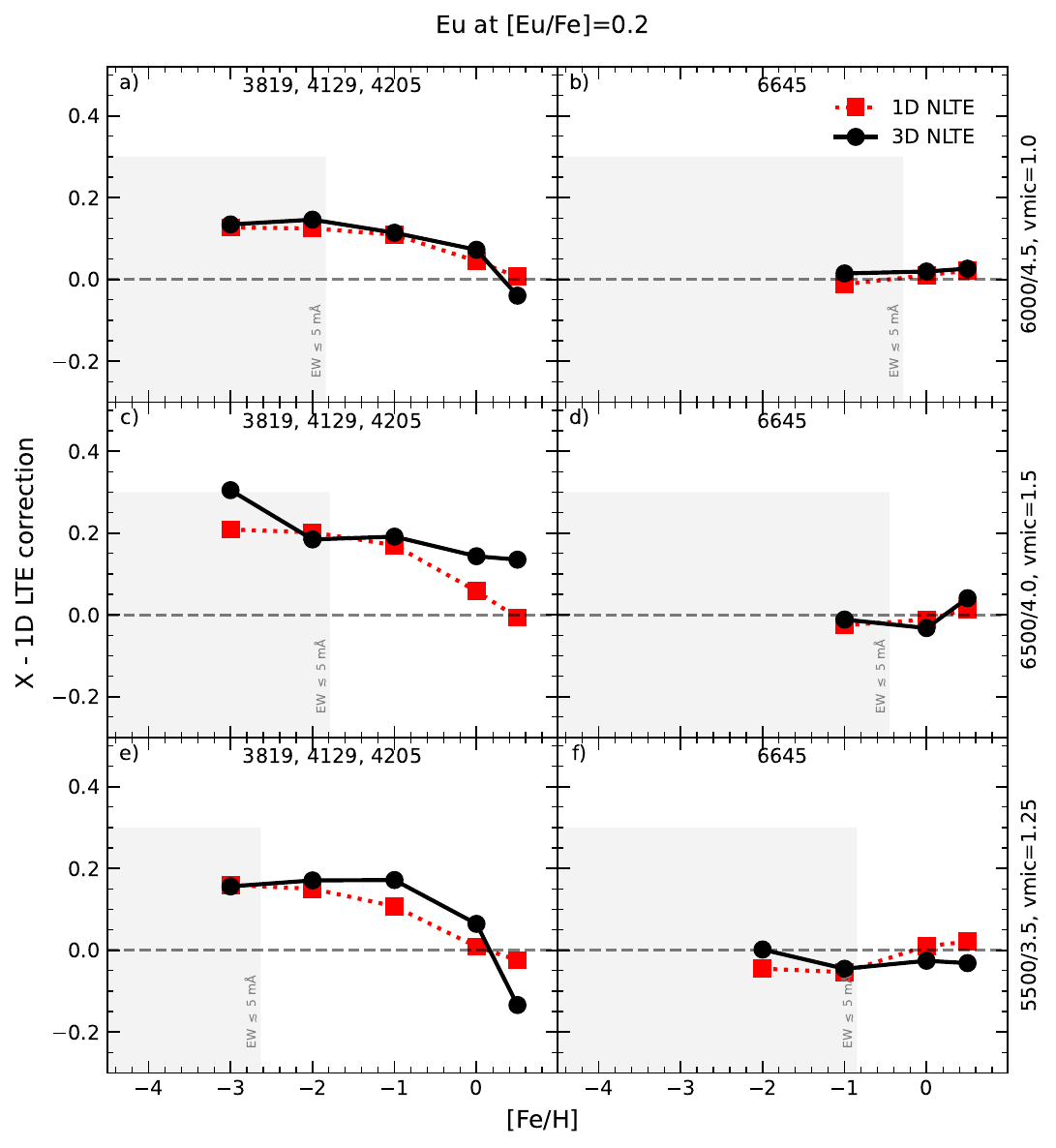}
\centering
\caption{Same as Fig. \ref{fig:mn_corrections}, but for Eu at [Eu/Fe] = 0.2.}
 \label{fig:eu_corrections}
\end{figure*}

Fig. \ref{fig:eu_corrections} {shows} the corrections for Eu for typical diagnostic lines. Most of these lines will be within the 4MOST HR windows as well. Here we used a slightly enhanced [Eu/Fe] = 0.2, which is more representative of typically measured super-solar Eu abundance in metal-poor stars, including our sample. The typical diagnostic line 6645 in main-sequence stars has weak NLTE negative effects, roughly down to -0.1 dex. This is consistent with the recent study of metal-poor stars in \citet{Guo2025}. 3D NLTE are similar down to [Fe/H] = -2. The low-excitation potential lines 3819, 4129 and 4205 are typically used in metal-poor stars, since they are more heavily blended and saturated at higher metallicities. NLTE corrections for this line are positive, reaching +0.2 dex in metal-poor stars. In general, 3D NLTE has similar effects to 1D NLTE, only slightly enhancing the correction. 

\section{Testing of Ni model atom}
\label{app:ni_comparison}

\begin{figure}
\includegraphics[width=1\columnwidth]{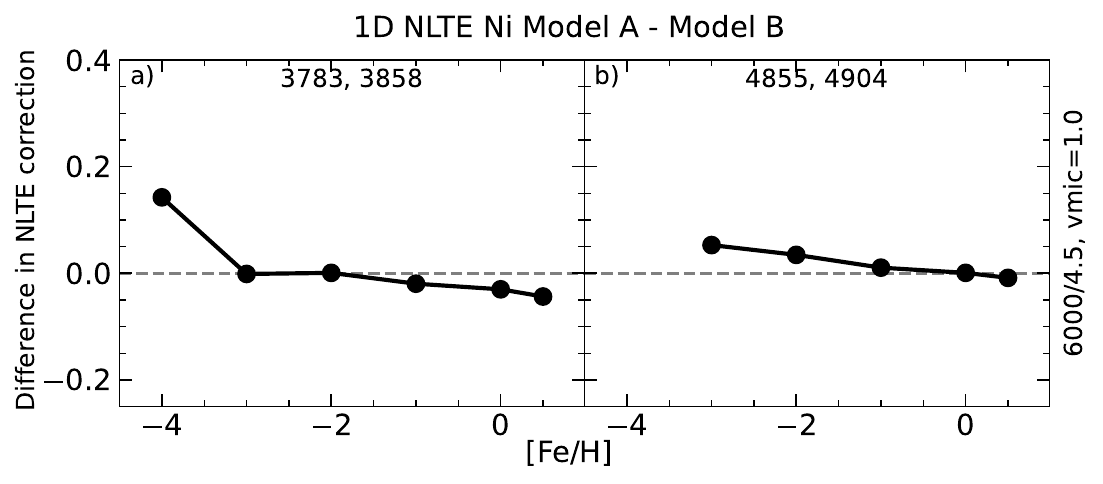}
\caption{Difference between model atom filled with (model A) and without (model B) Drawin's recipe of 1D NLTE [Ni/Fe] abundance {corrections} for two difference model atmospheres.}
 \label{fig:ni_sh0sh1_corrections}
\end{figure}

Given such a new and strong Ni correction, we carefully investigated the NLTE model atom in further detail. First, we opted to do our tests in 1D NLTE, to minimise any potential effects from 3D models. Just like in \citet{Eitner2023}, our baseline model atom ("Model A") includes Ni+H reactions using quantum-mechanical data from \citet{Voronov2022} and any missing collisional data is filled with Drawin's recipe \citep{Drawin1968}. In our tests, the extra Drawin's collisions would keep a lot of upper levels closer to LTE for the relevant transitions. This would mean that the source function deviates more from the Planck function \citep[since $S / B\approx b_j / b_i$, see details in][]{Bergemann2014}, resulting in stronger positive NLTE corrections. For comparison we plotted the difference of 1D NLTE corrections without Drawin's collisions ("Model B") in the Fig. \ref{fig:ni_sh0sh1_corrections} for two diagnostic lines. The difference between Model A - Model B is typically $< 0.05$ dex at [Fe/H] $\geq -3$. For the low excitation potential it reaches 0.15 dex at [Fe/H] $< -3$. Thus, we get strong positive NLTE correction with both of our model atoms, even without Drawin's recipe. 

\begin{figure}
\includegraphics[width=0.8\columnwidth]{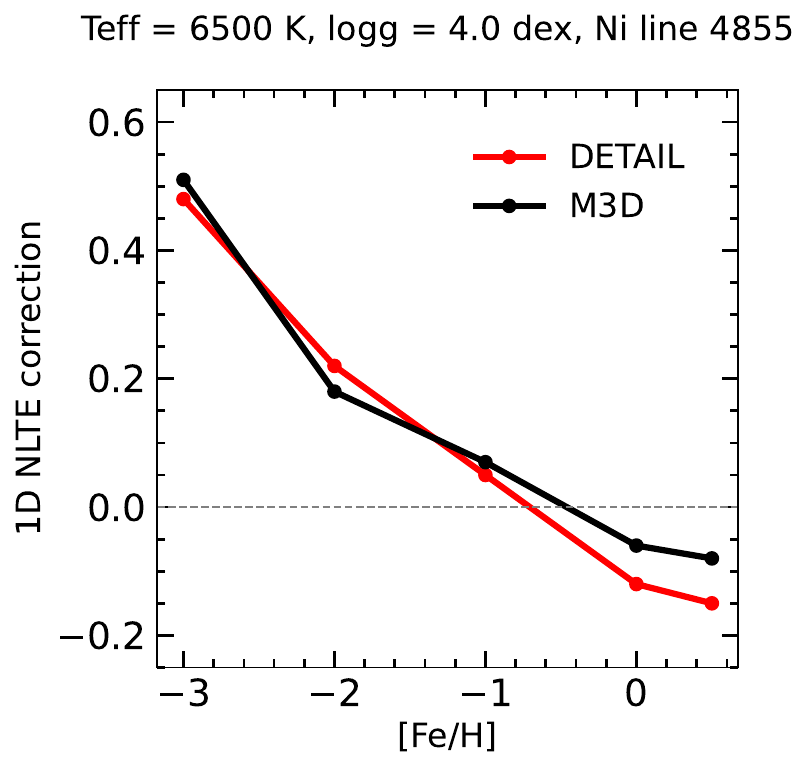}
\caption{Comparison of 1D NLTE corrections calculated in \textsc{detail} and \textsc{multi3d at dispatch} for Ni line 4855 \AA in the model atmosphere $\teff = 6500$ K, $\logg = 4.0$ dex.}
 \label{fig:ni_detail_vs_m3d}
\end{figure}

Finally, we also did the comparison with the alternative code \textsc{detail} in 1D, represented in Fig. \ref{fig:ni_detail_vs_m3d}. The codes have reassuringly an excellent agreement. The differences reach at most 0.07 dex and NLTE correction trends are identical. 
\section{Discussion on the stripped massive binary yields}
\label{app:binary_yields_discussion}
In nature, the effects of binary stellar evolution, which are thought to be ubiquitous among massive stars, are far more complex than the simple binary stripping considered in \cite{Farmer2023}. There have simply never been grids of stellar yields computed accounting for different metallicities or binary evolution channels other than binary stripping. Even for the case of binary stripping through stable mass transfer, \cite{Farmer2023} only considered the evolution of the primary. As our coverage of different binary stellar evolution channels expands, we will have a far more realistic and complete picture of the sort of variation -- and effect -- that accounting for it can have on our GCE models. A further caveat is that the supernova yields of \cite{Farmer2023} were computed by fixing the explosion energy, which results in unrealistically small remnant masses and all of their stars exploding. In nature, the explosion energy will be a function of the preceding stellar evolution, implying that a potentially crucial effect of the preceding binary stellar evolution is not propagated to the supernova yields. This simplification results in unreasonably small remnant masses in \cite{Farmer2023}, and therefore higher than normal supernova yields in general. A more realistic supernova model would also likely change the abundance ratios of the ejecta. 

As a final caveat, we note that in this work, we neglect the yield of the secondary star, which will affect the binary stellar yields. Preliminary work (Kemp et al., in prep) suggests that accounting for the yields of the secondary stars in the binary stellar yield calculations will drive up iron peak yields per unit star forming material relative to the stellar yields of the primary reported in \cite{Farmer2023} and used in this work, but not enough to account for our observations. 

We conclude that current binary stellar models cannot account for the significant discrepancies between contemporary GCE models and iron peak elements at low metallicities. However, we note that this could be due to several factors, including the extreme level of incompleteness in our coverage of binary stellar evolution channels, metallicity effects, or the highly simplified supernova physics that the underlying binary stellar yields rely upon.

\section{Comparing 1D NLTE corrections of \textsc{multi1d} and \textsc{multi3d at dispatch}}
\label{app:multi3d_vs_multi1d}

\begin{table*}
\makegapedcells
\begin{minipage}{\linewidth}
\renewcommand{\footnoterule}{} 
\setlength{\tabcolsep}{3pt}
\caption{Comparison of EW and corrections between \textsc{multi1d} (M1D) \citet[originally published in][]{Carlsson1986} with recent updates our group described in \citet{Bergemann2019, Gallagher2020} and \textsc{multi3d at dispatch} (M3D) \citep{Eitner2024} for Mn at [Mn/Fe] = -0.12 for the same model atmospheres. The test shows that 1D NLTE corrections are within 0.05 dex for most and 0.1 dex for all configurations.}
\label{tab:m1d_vs_m3d_comparison}     
\begin{center}\makebox[\textwidth][c]{
    \begin{tabular}{ccc|r|rr||r|rr|rr|rr}

\toprule
\multirow{4}{*}{\rotatebox{90}{teff/logg/$\vmic$ }} & \multirow{4}{*}{\rotatebox{90}{[Fe/H]}}
           &      &  \multicolumn{6}{c|}{1D}                 & \multicolumn{4}{c}{3D}  \\
     &     &      & \multicolumn{3}{c||}{M1D} &\multicolumn{3}{c|}{M3D}&\multicolumn{4}{c}{M3D}    \\
     &     &      & LTE & \multicolumn{2}{c||}{NLTE} &LTE & \multicolumn{2}{c|}{NLTE} & \multicolumn{2}{c|}{LTE} & \multicolumn{2}{c}{NLTE} \\
     &     & line &  EW &                 EW & corr & EW &                 EW & corr &                EW & corr &                EW & corr \\

\midrule

\multirow{15}{*}{\rotatebox{90}{6000 / 4.0 / 1.0}}
         &    \multirow{3}{*}{\rotatebox{90}{0.5}} 
                  & 4030 & 316.39 & 311.14 & 0.02 &  324.74 & 325.20 &  0.00 & 305.02 & 0.07 & 278.10  &  0.17  \\
         & &        5394 & 121.10 & 115.47 & 0.07 &  118.57 & 116.42 &  0.03 & 106.47 & 0.14 &  96.51  &  0.26  \\
         & &        6013 & 111.81 & 111.05 & 0.01 &  112.11 & 113.92 & -0.03 & 109.28 & 0.05 & 104.06  &  0.14  \\
         \cline{2-13}
         &    \multirow{3}{*}{\rotatebox{90}{0}} 
                  & 4030 & 254.17 & 249.37 & 0.03 &  260.03 & 260.31 & 0.00 & 250.86 &  0.05 & 238.91  &  0.11  \\
         &        & 5394 &  74.80 &  68.16 & 0.07 &   72.28 &  69.78 & 0.03 &  78.91 & -0.07 &  60.62  &  0.13  \\
         &        & 6013 &  80.14 &  76.87 & 0.05 &   79.91 &  79.79 & 0.00 &  80.47 & -0.01 &  71.39  &  0.13  \\
         \cline{2-13}
         &   \multirow{3}{*}{\rotatebox{90}{-1}} 
                  & 4030 & 159.53 & 156.29 & 0.04 &  153.56 & 154.30 & -0.01 & 186.20 & -0.36 & 151.98 &  0.02  \\
         & &        5394 &  13.79 &  11.18 & 0.10 &   12.80 &  11.20 &  0.06 &  26.95 & -0.47 &  10.33 &  0.10  \\
         & &        6013 &  21.67 &  17.89 & 0.10 &   20.68 &  18.24 &  0.07 &  20.95 & -0.01 &  14.68 &  0.17  \\
         \cline{2-13}
         &   \multirow{3}{*}{\rotatebox{90}{-2}} 
                  & 4030 & 102.89 & 101.47 & 0.04 &  101.64 & 100.35 &  0.03 & 149.88 & -1.15 & 101.50 &  0.01  \\
         & &        5394 &   1.74 &   1.24 & 0.14 &    1.58 &   1.19 &  0.12 &   8.95 & -1.84 &   0.97 &  0.19  \\
         & &        6013 &   2.88 &   1.92 & 0.17 &    2.68 &   1.88 &  0.15 &   2.85 & -0.02 &   1.45 &  0.23  \\
         \cline{2-13}
         &   \multirow{3}{*}{\rotatebox{90}{-3}} 
                  & 4030 &  61.83 &  44.16 & 0.33 &   59.16 &  43.25 &  0.28 & 109.51 & -0.92 &  32.63 &  0.47  \\
         & &        5394 &   0.19 &   0.09 & 0.26 &    0.17 &   0.09 &  0.27 &   0.88 & -1.42 &   0.05 &  0.40  \\
         & &        6013 &   0.31 &   0.17 & 0.22 &    0.29 &   0.17 &  0.20 &   0.24 &  0.08 &   0.12 &  0.28  \\
\hline\hline
\multirow{15}{*}{\rotatebox{90}{5000 / 3.0 / 1.5}}
         &    \multirow{3}{*}{\rotatebox{90}{0.5}} 
                  & 4030 & 650.18 & 624.16 & 0.05 &  741.84 & 720.98 &  0.03 & 762.90 & -0.02 & 732.89 &  0.02  \\
         & &        5394 & 223.93 & 215.27 & 0.18 &  222.39 & 216.29 &  0.12 & 223.68 & -0.02 & 204.50 &  0.37  \\
         & &        6013 & 173.77 & 170.48 & 0.06 &  175.26 & 175.59 & -0.01 & 168.13 &  0.12 & 158.93 &  0.26  \\
         \cline{2-13}
         &    \multirow{3}{*}{\rotatebox{90}{0}}   
                  & 4030 & 530.89 & 504.61 & 0.06 &  599.84 & 588.13 &  0.02 & 535.61 &  0.12 & 513.72 &  0.16  \\
         &        & 5394 & 194.54 & 181.82 & 0.22 &  192.83 & 186.23 &  0.11 & 183.46 &  0.16 & 163.86 &  0.50  \\
         &        & 6013 & 141.25 & 136.47 & 0.08 &  142.02 & 141.68 &  0.01 & 129.25 &  0.20 & 122.62 &  0.30  \\
         \cline{2-13}
         &   \multirow{3}{*}{\rotatebox{90}{-1}} 
                  & 4030 & 376.86 & 350.80 & 0.09 &  332.20 & 317.84 &  0.05 & 314.80 &  0.08 & 276.93 &  0.21  \\
         & &        5394 & 117.07 & 100.73 & 0.17 &  115.35 & 104.85 &  0.11 & 122.27 & -0.07 &  90.44 &  0.27  \\
         & &        6013 &  70.76 &  61.21 & 0.12 &   69.98 &  64.16 &  0.07 &  64.74 &  0.07 &  51.77 &  0.24  \\
         \cline{2-13}
         &   \multirow{3}{*}{\rotatebox{90}{-2}} 
                  & 4030 & 209.37 & 198.35 & 0.09 &  195.77 & 190.09 &  0.05 & 218.89 & -0.14 & 168.37 &  0.27  \\
         & &        5394 &  26.95 &  20.61 & 0.13 &   25.53 &  20.46 &  0.11 &  67.70 & -0.75 &  19.00 &  0.15  \\
         & &        6013 &  13.34 &  10.05 & 0.13 &   12.77 &  10.10 &  0.11 &  14.11 & -0.04 &   6.97 &  0.24  \\
         \cline{2-13}
         &   \multirow{3}{*}{\rotatebox{90}{-3}} 
                  & 4030 & 122.27 & 120.35 & 0.11 &  123.34 & 120.28 &  0.06 & 162.96 & -0.68 & 114.65 &  0.25  \\
         & &        5394 &   3.19 &   2.09 & 0.17 &    2.93 &   2.08 &  0.14 &  17.50 & -1.91 &   1.68 &  0.21  \\
         & &        6013 &   1.49 &   0.91 & 0.19 &    1.39 &   0.92 &  0.17 &   1.53 & -0.03 &   0.63 &  0.28  \\
\bottomrule
\end{tabular}}
\end{center}
\end{minipage}
\end{table*} 

\begin{figure*}
\includegraphics[width=1\textwidth]{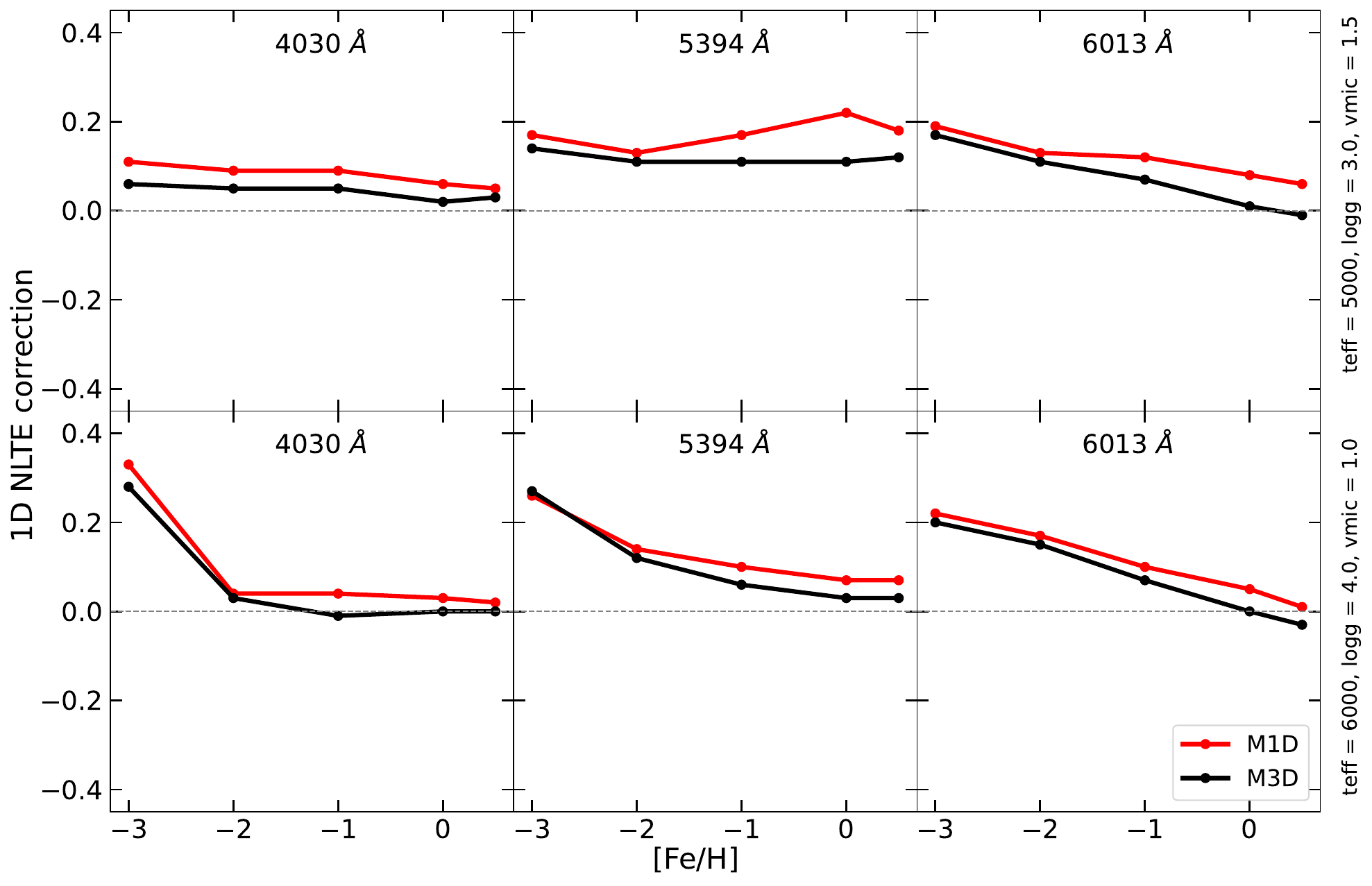}
\caption{Comparison of \textsc{multi1d} to \textsc{multi3d at dispatch} 1D NLTE corrections for 3 different lines and 2 MARCS model atmospheres from Table \ref{tab:m1d_vs_m3d_comparison}. The difference is always within 0.1 dex.}
 \label{fig:appendix_m1d_vs_m3d}
\end{figure*}

\section{Corrected lines for all elements in 3D NLTE}
\label{app:lines}

\begin{table}
\begin{minipage}{\linewidth}
\renewcommand{\footnoterule}{} 
\setlength{\tabcolsep}{3pt}
\caption{Line properties from literature sources that were used for NLTE calculations. The log$gf$ is taken directly from the model atom and does not reflect the most up-to-date value. Utilising sources: 
 $^1$: \citet{Bonifacio2009}     
 $^2$: \citet{Hansen2013}
 $^3$: \citet{Bensby2014} 
 $^4$: \citet{Battistini2015} 
 $^5$: \citet{Battistini2016} 
 $^6$: \citet{Zhao2016}
 $^7$: \citet{Li2022}
 $^8$: \citet{Mardini2024}
}
\label{tab:lines_data}     
\begin{center}
\begin{tabular}{lcccccrr}
\noalign{\smallskip}\hline\noalign{\smallskip}  
$\lambda$ [\AA] & States & $g_l$ & $g_u$ &  $\Elow$ & $\Eup$ & log$gf$ & Utilising  \\
 & && &  [eV] &  [eV] & & Sources \\
\noalign{\smallskip}\hline\noalign{\smallskip}
Mn I      &  & &  &  &  \\
3823.51   & \Mn{a}{6}{D}{}{7/2}  - \Mn{z}{6}{F}{o}{9/2} & 8 &10 & 2.14 & 5.38 & 0.058 & $^8$ \\
4030.76   & \Mn{a}{6}{S}{}{5/2}  - \Mn{z}{6}{P}{o}{7/2} & 6 & 8 & 0.00 & 3.08 &-0.497 & $^1$ $^8$\\
4055.54   & \Mn{a}{6}{D}{}{7/2}  - \Mn{z}{6}{D}{o}{7/2} & 8 & 8 & 2.14 & 5.20 &-0.077 & $^7$ \\
4754.03   & \Mn{z}{8}{P}{o}{5/2} - \Mn{e}{8}{S}{}{7/2}  & 6 & 8 & 2.28 & 4.89 &-0.080 & $^8$     \\
4783.42   & \Mn{z}{8}{P}{o}{7/2} - \Mn{e}{8}{S}{}{7/2}  & 8 & 8 & 2.30 & 4.89 & 0.044 & $^7$ $^8$  \\
4823.51   & \Mn{z}{8}{P}{o}{9/2} - \Mn{e}{8}{S}{}{7/2}  &10 & 8 & 2.32 & 4.89 & 0.136 & $^1$ $^7$ $^8$\\
5394.67   & \Mn{a}{6}{S}{}{5/2}  - \Mn{z}{8}{P}{o}{7/2} & 6 & 8 & 0.00 & 2.30 &-3.503 & $^4$ \\
5432.54   & \Mn{a}{6}{S}{}{5/2}  - \Mn{z}{8}{P}{o}{5/2} & 6 & 6 & 0.00 & 2.28 &-3.795 & $^4$ \\
6013.49   & \Mn{z}{6}{P}{o}{3/2} - \Mn{e}{6}{S}{}{5/2}  & 4 & 6 & 3.07 & 5.13 &-0.354 & $^4$ \\
Co I      &  & &  &  &  \\  
3845.47  & \Co{a}{2}{F}{}{7/2} - \Co{y}{2}{G}{o}{9/2} & 8 &10 & 0.92 & 4.15 & 0.097 & $^1$ $^8$\\
3873.11  & \Co{b}{4}{F}{}{9/2} - \Co{z}{4}{D}{o}{7/2} &10 & 8 & 0.43 & 3.63 &-0.830 & $^8$ \\
3881.87  & \Co{b}{4}{F}{}{5/2} - \Co{z}{4}{D}{o}{3/2} & 6 & 4 & 0.58 & 3.77 &-1.452 & $^8$ \\
3995.31  & \Co{a}{2}{F}{}{7/2} - \Co{y}{4}{G}{o}{9/2} & 8 &10 & 0.92 & 4.03 &-0.370 & $^1$ $^8$\\
4020.90  & \Co{b}{4}{F}{}{9/2} - \Co{z}{4}{F}{o}{9/2} &10 &10 & 0.43 & 3.51 &-1.737 & $^8$ \\
4092.85  & \Co{b}{4}{P}{}{1/2} - \Co{z}{4}{S}{o}{3/2} & 2 & 4 & 2.01 & 5.04 &-1.597 & $^7$ \\ 
4110.53  & \Co{a}{2}{F}{}{5/2} - \Co{z}{2}{F}{o}{5/2} & 6 & 6 & 1.05 & 4.06 &-2.618 & $^8$ \\
4121.32  & \Co{a}{2}{F}{}{7/2} - \Co{z}{2}{G}{o}{9/2} & 8 &10 & 0.92 & 3.93 &-0.729 & $^1$ $^7$ $^8$\\
5301.04  & \Co{a}{4}{P}{}{5/2} - \Co{y}{4}{D}{o}{5/2} & 6 & 6 & 1.71 & 4.05 &-2.081 & $^4$ \\ 
5352.04  &\Co{z}{4}{G}{o}{11/2}- \Co{f}{4}{F}{}{9/2}  &12 &10 & 3.58 & 5.89 &-0.097 & $^4$ \\ 
5647.23  & \Co{a}{2}{P}{}{3/2} - \Co{y}{2}{D}{o}{5/2} & 4 & 6 & 2.28 & 4.47 &-1.364 & $^4$ \\ 
Sr I      &  & &  &  &  \\
4607.33 & \Sr{5s^2}{1}{S}{}{0}  - \Sr{5p}{1}{P}{o}{1}   &  1 &  3 & 0.00 & 2.69 & 0.283 & $^2$ $^5$  \\
Sr II      & & & &  &   \\
4077.71 & \Sr{5s}{2}{S}{}{1/2}  - \Sr{5p}{2}{P}{o}{3/2}   &  2 &  4 & 0.00 & 3.04 & 0.150 & $^1$ $^2$ $^6$ $^7$ $^8$ \\
4161.79 & \Sr{5p}{2}{P}{o}{1/2} - \Sr{6s}{2}{S}{}{1/2}    &  2 &  2 & 2.94 & 5.92 &-0.470 & $^6$ \\
4215.52 & \Sr{5s}{2}{S}{}{1/2}  - \Sr{5p}{2}{P}{o}{1/2}   &  2 &  2 & 0.00 & 2.94 &-0.170 & $^6$ $^7$ $^8$ \\
Y II      & & & &  &   \\
4374.93 &    \Y{a}{1}{D}{}{2} - \Y{z}{1}{D}{o}{2}     &  5 &  5 & 0.41 & 3.24 & 0.150 & $^7$ \\
4854.86 &    \Y{a}{3}{F}{}{2} - \Y{z}{3}{D}{o}{1}     &  5 &  3 & 0.99 & 3.54 &-0.240 & $^3$ $^7$ \\
4883.68 &    \Y{a}{3}{F}{}{4} - \Y{z}{3}{D}{o}{3}     &  9 &  7 & 1.08 & 3.62 & 0.150 & $^3$ $^7$ \\
4900.12 &    \Y{a}{3}{F}{}{3} - \Y{z}{3}{D}{o}{2}     &  7 &  5 & 1.03 & 3.56 &-0.020 & $^3$ $^7$ \\
5087.42 &    \Y{a}{3}{F}{}{4} - \Y{z}{3}{F}{o}{4}     &  9 &  9 & 1.08 & 3.52 &-0.180 & $^3$ \\
5200.41 &    \Y{a}{3}{F}{}{2} - \Y{z}{3}{F}{o}{2}     &  5 &  5 & 0.99 & 3.38 &-0.620 & $^3$ \\
5402.77 &    \Y{b}{1}{D}{}{2} - \Y{z}{1}{F}{o}{3}     &  5 &  7 & 1.84 & 4.13 &-0.350 & $^3$ \\
5662.92 &    \Y{a}{1}{G}{}{4} - \Y{z}{1}{F}{o}{3}     &  9 &  7 & 1.94 & 4.13 & 0.320 & $^3$ \\
Ba II      & & & &  &    \\
4554.03 &    \Ba{6s}{2}{S}{}{1/2} - \Ba{6p}{2}{P}{o}{3/2}   &  2 &  4 & 0.00 & 2.72 & 0.140 & $^1$ $^3$ $^6$ $^7$ $^8$ \\
4934.08 &    \Ba{6s}{2}{S}{}{1/2} - \Ba{6p}{2}{P}{o}{1/2}   &  2 &  2 & 0.00 & 2.51 &-0.160 & $^6$ $^7$ $^8$ \\
5853.68 &    \Ba{5d}{2}{D}{}{3/2} - \Ba{6p}{2}{P}{o}{3/2}   &  4 &  4 & 0.60 & 2.72 &-0.908 & $^3$ $^6$ \\
6141.71 &    \Ba{5d}{2}{D}{}{5/2} - \Ba{6p}{2}{P}{o}{3/2}   &  6 &  4 & 0.70 & 2.72 &-0.032 & $^3$ $^6$ $^7$ \\
6496.90 &    \Ba{5d}{2}{D}{}{3/2} - \Ba{6p}{2}{P}{o}{1/2}   &  4 &  2 & 0.60 & 2.51 &-0.407 & $^3$ $^6$ $^7$ \\
Eu II      &  &  & & &   \\ 
3819.67 &    \Eu{a}{9}{S}{o}{4} - \Eu{z}{9}{P}{}{5}   &  9 & 11 & 0.00 & 3.25 & 0.510 & \\
4129.73 &    \Eu{a}{9}{S}{o}{4} - \Eu{z}{9}{P}{}{4}   &  9 &  9 & 0.00 & 3.00 & 0.220 & $^5$ $^6$ $^7$ \\
4205.04 &    \Eu{a}{9}{S}{o}{4} - \Eu{z}{9}{P}{}{3}   &  9 &  7 & 0.00 & 2.95 & 0.210 & $^6$ $^7$ \\
6645.10 &    \Eu{a}{9}{D}{o}{6} - \Eu{z}{9}{P}{}{5}   & 13 & 11 & 1.38 & 3.25 & 0.120 &  \\
\noalign{\smallskip}\hline\noalign{\smallskip}
\end{tabular}
\end{center}
\end{minipage}
\end{table}

\begin{table}
\begin{minipage}{\linewidth}
\renewcommand{\footnoterule}{} 
\setlength{\tabcolsep}{3pt}
\caption{Same as Table \ref{tab:lines_data}, but only for Ni lines. Utilising sources: 
 $^1$: \citet{Bonifacio2009}     
 $^2$: \citet{Hansen2013}
 $^3$: \citet{Bensby2014} 
 $^4$: \citet{Battistini2015} 
 $^5$: \citet{Battistini2016} 
 $^6$: \citet{Zhao2016}
 $^7$: \citet{Li2022}
 $^8$: \citet{Mardini2024}
}
\label{tab:lines_data_ni}    
\begin{center}
\begin{tabular}{lcccccrr}
\noalign{\smallskip}\hline\noalign{\smallskip}  
$\lambda$ [\AA] & States & $g_l$ & $g_u$ &  $\Elow$ & $\Eup$ & log$gf$ & Utilising  \\
 & && &  [eV] &  [eV] & & Sources \\
\noalign{\smallskip}\hline\noalign{\smallskip}
Ni I      &  & &  &  &  \\  
3423.71   & \Ni{4s}{3}{D}{}{1}    - \Ni{4p}{3}{D}{o}{1}  & 3 & 3 & 0.21 & 3.83 &-0.675 & $^8$ \\
3437.28   & \Ni{4s^2~}{3}{F}{}{4} - \Ni{4p}{5}{F}{o}{4}  & 9 & 9 & 0.00 & 3.61 &-1.729 & $^8$ \\
3452.89   & \Ni{4s}{3}{D}{}{2}    - \Ni{4p}{5}{F}{o}{3}  & 5 & 7 & 0.11 & 3.70 &-0.460 & $^8$ \\
3472.55   & \Ni{4s}{3}{D}{}{2}    - \Ni{4p}{3}{D}{o}{3}  & 5 & 7 & 0.11 & 3.68 &-1.855 & $^8$ \\
3483.78   & \Ni{4s^2~}{3}{F}{}{2} - \Ni{4p}{3}{D}{o}{1}  & 5 & 3 & 0.28 & 3.83 &-1.198 & $^8$ \\
3492.96   & \Ni{4s}{3}{D}{}{2}    - \Ni{4p}{3}{P}{o}{1}  & 5 & 3 & 0.11 & 3.66 &-0.216 & $^8$ \\
3500.85   & \Ni{4s^2~}{3}{F}{}{3} - \Ni{4p}{3}{D}{o}{2}  & 7 & 5 & 0.17 & 3.71 &-1.336 & $^8$ \\
3519.77   & \Ni{4s^2~}{3}{F}{}{2} - \Ni{4p}{3}{F}{o}{2}  & 5 & 5 & 0.28 & 3.80 &-1.550 & $^8$ \\
3524.54   & \Ni{4s}{3}{D}{}{3}    - \Ni{4p}{3}{P}{o}{2}  & 7 & 5 & 0.03 & 3.54 & 0.044 & $^8$ \\
3566.37   & \Ni{4s}{1}{D}{}{2}    - \Ni{4p}{1}{D}{o}{2}  & 5 & 5 & 0.42 & 3.90 &-0.266 & $^8$ \\
3597.70   & \Ni{4s}{3}{D}{}{1}    - \Ni{4p}{3}{P}{o}{1}  & 3 & 3 & 0.21 & 3.66 &-1.080 & $^8$ \\
3783.53   & \Ni{4s}{1}{D}{}{2}    - \Ni{4p}{5}{F}{o}{3}  & 5 & 7 & 0.42 & 3.70 &-0.898 & $^8$ \\
3807.14   & \Ni{4s}{1}{D}{}{2}    - \Ni{4p}{3}{D}{o}{3}  & 5 & 7 & 0.42 & 3.68 &-2.079 & $^1$ $^8$\\  
3858.30   & \Ni{4s}{1}{D}{}{2}    - \Ni{4p}{3}{F}{o}{3}  & 5 & 7 & 0.42 & 3.63 &-0.865 & $^1$ $^8$\\  
4831.18   & \Ni{4p}{5}{F}{o}{4}   - \Ni{5s}{5}{F}{}{3}   & 9 & 7 & 3.61 & 6.17 &-0.214 & $^3$ \\  
4855.41   & \Ni{4p}{3}{P}{o}{2}   - \Ni{4d}{2}{3/2}{}{2} & 5 & 5 & 3.54 & 6.09 & 0.078 & $^8$ \\ 
4904.41   & \Ni{4p}{3}{P}{o}{2}   - \Ni{4d}{2}{1/2}{}{1} & 5 & 3 & 3.54 & 6.07 &-0.016 & $^3$ $^8$\\  
4980.17   & \Ni{4p}{5}{F}{o}{4}   - \Ni{4d}{2}{9/2}{}{5} & 9 &11 & 3.61 & 6.09 &-0.363 & $^8$ \\
5035.36   & \Ni{4p}{3}{F}{o}{3}   - \Ni{4d}{2}{9/2}{}{4} & 7 & 9 & 3.63 & 6.10 & 0.362 & $^8$ \\
5102.97  & \Ni{4s^2}{1}{D}{}{2}   - \Ni{4p}{3}{F}{o}{3}  & 5 & 7 & 1.68 & 4.11 &-2.837 & $^3$ \\  
5137.07  & \Ni{4s^2}{1}{D}{}{2}   - \Ni{4p}{1}{P}{o}{1}  & 5 & 3 & 1.68 & 4.09 &-1.843 & $^8$ \\
5468.10  & \Ni{4p}{1}{F}{o}{3}    - \Ni{4d}{2}{7/2}{}{3} & 7 & 7 & 3.85 & 6.11 &-1.775 & $^3$ \\  
5476.90  & \Ni{3d^{10}}{1}{S}{}{0}- \Ni{4p}{1}{P}{o}{1}  & 1 & 3 & 1.83 & 4.09 &-0.369 & $^1$ $^8$\\  
5578.72  & \Ni{4s^2}{1}{D}{}{2}   - \Ni{4p}{1}{D}{o}{2}  & 5 & 5 & 1.68 & 3.90 &-2.704 & $^3$ $^8$\\  
5587.85  & \Ni{4s^2}{3}{P}{}{2}   - \Ni{4p}{3}{D}{o}{3}  & 5 & 7 & 1.94 & 4.15 &-2.203 & $^3$ \\  
5593.74  & \Ni{4p}{1}{D}{o}{2}    - \Ni{4d}{2}{7/2}{}{3} & 5 & 7 & 3.90 & 6.11 &-0.682 & $^3$ \\  
5748.35  & \Ni{4s^2}{1}{D}{}{2}   - \Ni{4p}{3}{D}{o}{1}  & 5 & 3 & 1.68 & 3.83 &-3.242 & $^3$ \\  
5754.66  & \Ni{4s^2}{3}{P}{}{2}   - \Ni{4p}{1}{P}{o}{1}  & 5 & 3 & 1.94 & 4.09 &-2.288 & $^3$ $^8$ \\  
5846.99  & \Ni{4s^2}{1}{D}{}{2}   - \Ni{4p}{3}{F}{o}{2}  & 5 & 5 & 1.68 & 3.80 &-3.314 & $^3$ \\  
6007.31  & \Ni{4s^2}{1}{D}{}{2}   - \Ni{4p}{5}{F}{o}{2}  & 5 & 5 & 1.68 & 3.74 &-3.740 & $^3$ \\  
6133.96  & \Ni{4p}{3}{F}{o}{4}    - \Ni{4d}{2}{5/2}{}{3} & 9 & 7 & 4.09 & 6.11 &-1.917 & $^3$ \\  
6186.71  & \Ni{4p}{3}{F}{o}{3}    - \Ni{4d}{2}{5/2}{}{3}   & 7 & 7 & 4.11 & 6.11 &-0.880 & $^3$ \\  
6223.98  & \Ni{4p}{3}{F}{o}{3}    - \Ni{4d}{2}{9/2}{}{4}   & 7 & 9 & 4.11 & 6.10 &-0.835 & $^3$ \\  
6314.66  & \Ni{4s^2}{3}{P}{}{2}   - \Ni{4p}{1}{D}{o}{2}    & 5 & 5 & 1.94 & 3.90 &-2.418 & $^3$ \\  
6322.17  & \Ni{4p}{3}{D}{o}{3}    - \Ni{4d}{2}{7/2}{}{3}   & 7 & 7 & 4.15 & 6.11 &-1.115 & $^3$ \\  
6327.60  & \Ni{4s^2}{1}{D}{}{2}   - \Ni{4p}{3}{F}{o}{3}    & 5 & 7 & 1.68 & 3.63 &-3.211 & $^3$ \\  
6378.25  & \Ni{4p}{3}{D}{o}{3}    - \Ni{4d}{2}{9/2}{}{4}   & 7 & 9 & 4.15 & 6.10 &-0.926 & $^3$ \\  
6482.80  & \Ni{4s^2}{3}{P}{}{2}   - \Ni{4p}{1}{F}{o}{3}    & 5 & 7 & 1.94 & 3.85 &-2.948 & $^3$ \\  
6598.60  & \Ni{4p}{3}{F}{o}{2}    - \Ni{4d}{2}{7/2}{}{3}   & 5 & 7 & 4.24 & 6.11 &-0.821 & $^3$ \\ 
6643.63  & \Ni{4s^2}{1}{D}{}{2}   - \Ni{4p}{3}{P}{o}{2}    & 5 & 5 & 1.68 & 3.54 &-2.415 & $^3$ $^8$ \\  
6767.77  & \Ni{3d^{10}}{1}{S}{}{0}- \Ni{4p}{3}{P}{o}{1}    & 1 & 3 & 1.83 & 3.66 &-1.922 & $^3$ $^8$\\  
6772.32  & \Ni{4p}{3}{P}{o}{1}    - \Ni{5s}{2}{3/2}{}{2}   & 3 & 5 & 3.66 & 5.49 &-0.797 & $^3$ \\  
6842.04  & \Ni{4p}{3}{P}{o}{1}    - \Ni{5s}{2}{3/2}{}{1}   & 3 & 3 & 3.66 & 5.47 &-1.374 & $^3$ \\  
7110.90  & \Ni{4s^2}{3}{P}{}{2}   - \Ni{4p}{3}{D}{o}{3}    & 5 & 7 & 1.94 & 3.68 &-2.895 & $^3$ \\  
7715.58  & \Ni{4p}{5}{F}{o}{3}    - \Ni{5s}{2}{5/2}{}{2}   & 7 & 5 & 3.70 & 5.30 &-0.357 & $^3$ \\  
7748.89  & \Ni{4p}{3}{D}{o}{2}    - \Ni{5s}{2}{5/2}{}{2}   & 5 & 5 & 3.71 & 5.30 &-0.185 & $^3$ \\  
7788.94  & \Ni{4s^2}{3}{P}{}{1}   - \Ni{4p}{3}{P}{o}{2}    & 3 & 5 & 1.95 & 3.54 &-2.208 & $^3$ \\  
\noalign{\smallskip}\hline\noalign{\smallskip}
\end{tabular}
\end{center}
\end{minipage}
\end{table}

\section{Overview of Relevant Nucleosynthesis Processes}

In this section we will remind the reader about the relevant nucleosynthesis processes and based on \citet{Clayton1968, Krane1988, Rolfs1988, Woosley1995, Woosley2002, Sneden2008, Pagel2009, Boeltzig2016}.

\subsection{Hydrogen burning}

\begin{figure*}
\includegraphics[width=1.0\textwidth]{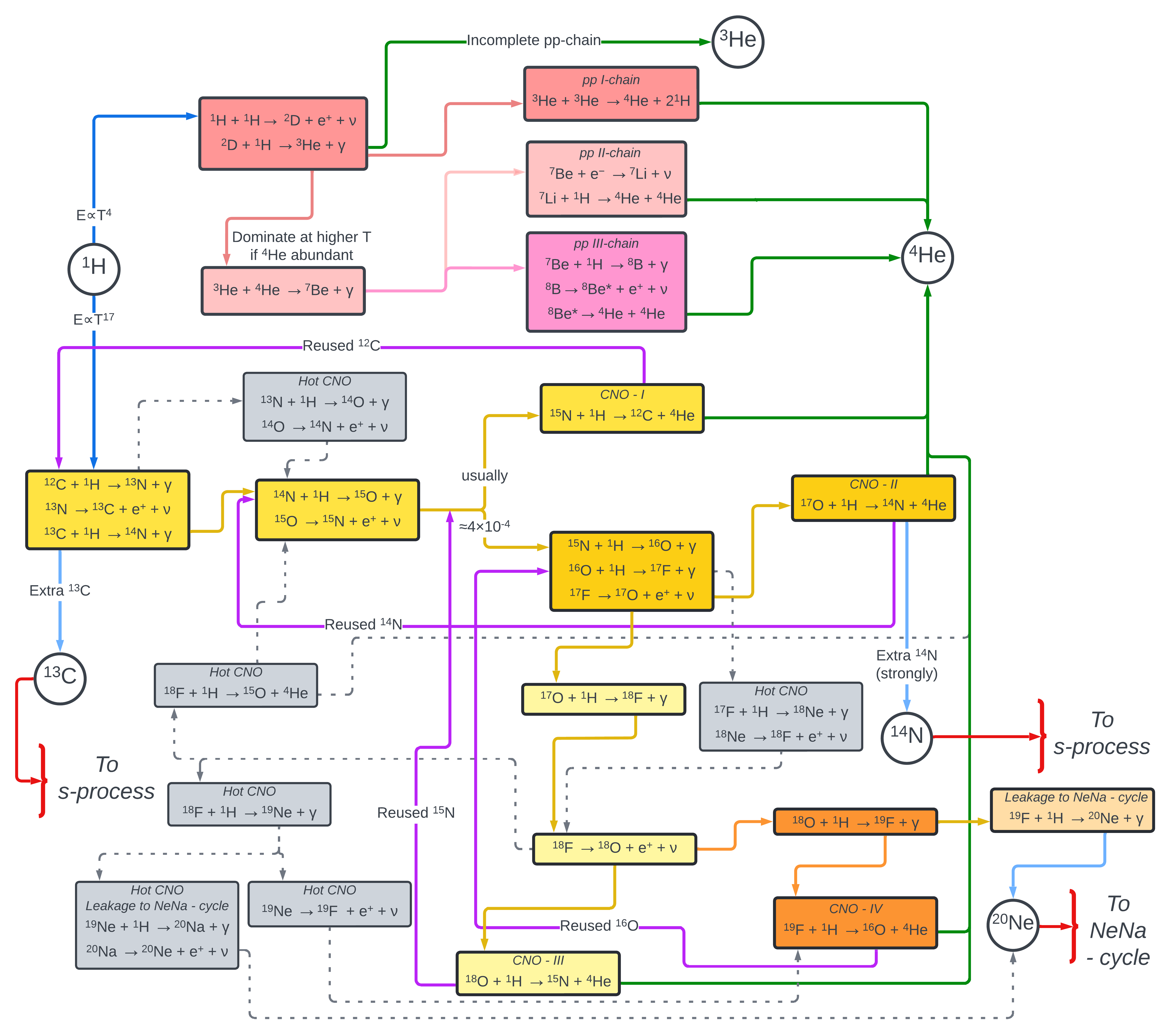}
\caption{Hydrogen burning stage diagram. Upper part of the diagram explains pp-chains, while the bottom one focuses on the CNO. "Hot CNO" in grey happens only at really high temperatures, typically not achieved in main-sequence stars. Further usage $^{13}$C, $^{14}$N and $^{20}$Ne to other diagrams are mentioned using red brackets.}
 \label{fig:nucleo_alt}
\end{figure*}

Fig. \ref{fig:nucleo_alt} shows main nuclear reactions during the hydrogen burning phase. At cooler temperatures of roughly $T < 1.8 \times 10^7$ K, the pp-chain contributes the most to energy production, by burning H into He. The CNO-cycle itself is a catalytic reaction, in which carbon, nitrogen and oxygen act as catalysts for the net result of fusing four $^1\text{H}$ into one $^4\text{He}$. This reaction is considered to be a more dominant energy production source at hotter temperatures compared to the pp-chain. The most dominant CNO-I cycle has the following reactions:
\begin{align}
^{12}\text{C} + ^{1}\text{H} &\rightarrow ^{13}\text{N} + \gamma \\
^{13}\text{N} &\rightarrow ^{13}\text{C} + e^+ + \nu \\
^{13}\text{C} + ^{1}\text{H} &\rightarrow ^{14}\text{N} + \gamma \\
\label{eq:n14}
^{14}\text{N} + ^{1}\text{H} &\rightarrow ^{15}\text{O} + \gamma \\
^{15}\text{O} &\rightarrow ^{15}\text{N} + e^+ + \nu \\
\label{eq:n15}
^{15}\text{N} + ^{1}\text{H} &\rightarrow ^{12}\text{C} + ^4\text{He} 
\end{align}

Thus the initially consumed carbon-12 is returned at the end of the reaction, hence the isotope is a catalyst. However, in rare (about 0.04 per cent cases), reaction \ref{eq:n15} instead continues further:
\begin{align}
^{15}\text{N} + ^{1}\text{H} &\rightarrow ^{16}\text{O} + \gamma \\
^{16}\text{O} + ^{1}\text{H} &\rightarrow ^{17}\text{F} + \gamma \\
^{17}\text{F} &\rightarrow ^{17}\text{O} + e^+ + \nu \\
^{17}\text{O} + ^{1}\text{H} &\rightarrow ^{14}\text{N} + ^4\text{He}
\end{align}

Which produces $^{14}$\text{N} that can be reused back into reaction \ref{eq:n14}. This also results in an accumulation of this isotope in the star. For the similar reason, this also increases the abundance of $^{13}$\text{C}, although in smaller fashion. Both of these elements are important to the s-process, since they can continue nuclear reactions that release free neutrons.

\subsection{Carbon production}
\label{subsec:carbon_prod}

\begin{figure*}
\includegraphics[width=1.0\textwidth]{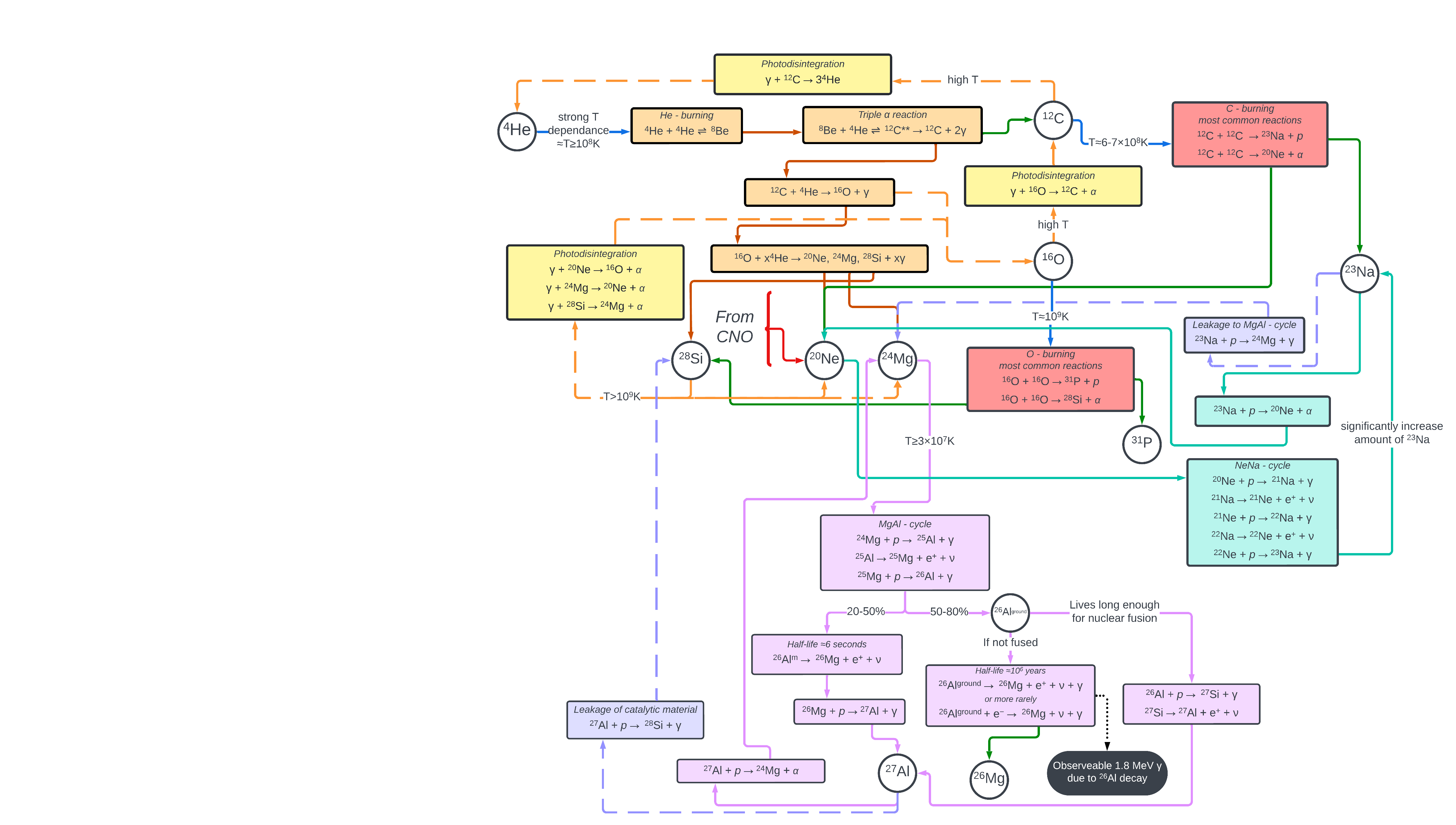}
\caption{Helium burning to carbon and oxygen, together with C- and O-burning. Note that NeNa- and MgAl-cycles may happen during hydrogen burning stage for second-generation stars.}
 \label{fig:nucleo2_alt}
\end{figure*}

Fig. \ref{fig:nucleo2_alt} demonstrates reactions beyond hydrogen-burning. Production of carbon is naturally assumed to arise from triple $\alpha$-reaction. However, a three-body reaction is rather improbable. Although it has a short lifetime, enough of $^8\text{Be}$ can accumulate in an equilibrium during an interaction of two $\alpha$-particles: 
\begin{equation}
\label{eq:1}
^4\text{He} + ^4\text{He} \rightleftharpoons ^8\text{Be} 
\end{equation}

Despite the low concentration of $^8\text{Be}$, the highly dense helium gas interacts with it to produce carbon:
\begin{equation}
^8\text{Be} + ^4\text{He} \rightleftharpoons ^{12}\text{C}^{*} \rightarrow ^{12}\text{C} + 2\gamma 
\end{equation}

This helium burning can continue in a similar fashion to produce heavier elements such as $^{16}\text{O}$, $^{20}\text{Ne}$, etc. The energy released $\epsilon_{3\alpha}$ has a very strong temperature dependence of $\epsilon_{3\alpha} \propto T^{40}$, at $T \approx 10^8$~K. In contrast, the concentration of carbon can decrease due to this reaction or C-burning. Thus, it is important to also take into account whether carbon is released from the star before further nuclear reactions.

\subsection{Iron-peak element production}
\label{sec:iron_peak_prod}

At the end of the carbon and oxygen burning, the most dominant heavy elements are $^{28}\text{Si}$, $^{32}\text{S}$ and $^{24}\text{Mg}$. The required temperature to fuse these particles together is too high to realistically achieve; instead intense $\gamma$ flux results in photodisintegration of nuclei with the smallest binding energies in ($\gamma$, $p$)\footnote{notation A(B,C)D means A + B $\rightarrow$ D + C; here A and D are omitted and can be replaced by appropriate elements}, ($\gamma$, $n$) and ($\gamma$, $\alpha$) reactions for their protons, neutrons and alpha particles. Inverse recaptures attempt to establish the so-called nuclear statistical equilibrium (NSE). However, many photoejected particles are captured by nuclei with higher binding energy, resulting in a so-called "photodisintegration rearrangement". Over time, $\alpha$-capture reactions such as 

\begin{align}
^{28}\text{Si} + \alpha &\rightleftharpoons ^{32}\text{S} + \gamma \\
^{32}\text{S} + \alpha &\rightleftharpoons ^{36}\text{Ar} + \gamma \\
^{36}\text{Ar} + \alpha &\rightleftharpoons ^{40}\text{Ca} + \gamma , \text{ etc}...
\end{align}

result in iron-peak element (Cr, Mn, Fe, Co and Ni) production. Some nuclei undergo $\beta$-decay, which tend to decrease the total proton-to-neutron ratio Z$/$N of the gas. However, the time for rearrangement depends on the temperature, hence higher temperatures preserve the initial value of Z$/$N. At lower temperatures $^{56}\text{Ni}$ is the dominant final isotope, while at higher ones it can be instead $^{54}\text{Fe} + 2p$. If temperature increases during this rearrangement, instead of staying roughly constant, then the final distribution of iron-peak elements is broader as it is reached at higher temperature. 

However, other potential processes play a big role during the explosive nucleosynthesis. For example, high $\alpha$ particles concentration during the alpha-rich freeze-out cause creation of heavier elements that are typically not expected during the NSE \citep{Woosley1973}. Thus one can get heavier elements or other specific isotopes.

\subsection{s-process}

\begin{figure*}
\includegraphics[width=1.0\textwidth]{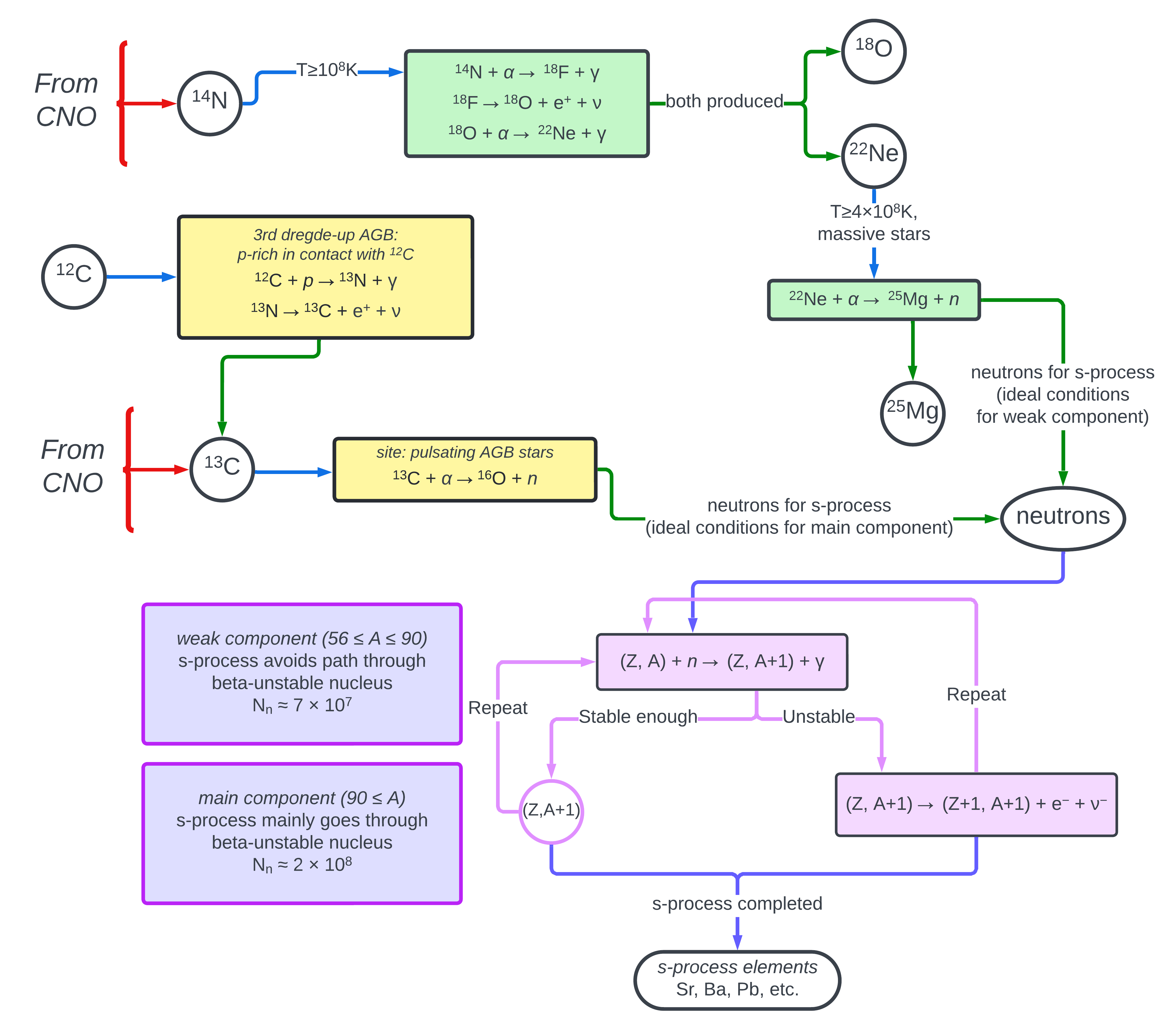}
\caption{s-process with some neutron-producing nucleosynthesis reactions. The bottom s-process is cyclical. "Stable enough" means that the isotope's half-life is much longer than the neutron-capture timescale.}
 \label{fig:nucleo4_alt}
\end{figure*}

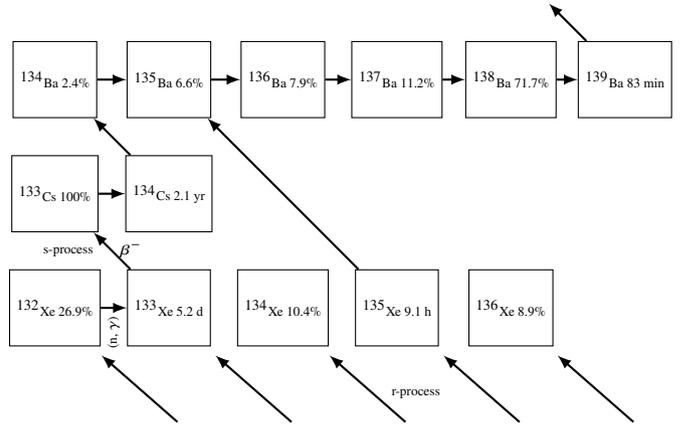
\begin{figure}
\centering
\begin{tikzpicture}[scale=0.5]
  \tikzstyle{sn}=[rectangle, draw, minimum size=1.0cm, text centered]
  \tikzstyle{process}=[-latex, thick]
  
  \node[sn] (Xe132) at (0,0) {\tiny{$^{132}$Xe 26.9\%}};
  \node[sn] (Xe133) at (3,0) {\tiny{$^{133}$Xe 5.2 d}};
  \node[sn] (Xe134) at (6,0) {\tiny{$^{134}$Xe 10.4\%}};
  \node[sn] (Xe135) at (9,0) {\tiny{$^{135}$Xe 9.1 h}};
  \node[sn] (Xe136) at (12,0) {\tiny{$^{136}$Xe 8.9\%}};

  \node[sn] (Cs133) at (0,3) {\tiny{$^{133}$Cs 100\%}};
  \node[sn] (Cs134) at (3,3) {\tiny{$^{134}$Cs 2.1 yr}};

  \node[sn] (Ba134) at (0,6) {\tiny{$^{134}$Ba 2.4\%}};
  \node[sn] (Ba135) at (3,6) {\tiny{$^{135}$Ba 6.6\%}};
  \node[sn] (Ba136) at (6,6) {\tiny{$^{136}$Ba 7.9\%}};
  \node[sn] (Ba137) at (9,6) {\tiny{$^{137}$Ba 11.2\%}};
  \node[sn] (Ba138) at (12,6) {\tiny{$^{138}$Ba 71.7\%}};
  \node[sn] (Ba139) at (15,6) {\tiny{$^{139}$Ba 83 min}};

  \draw[process] (Xe132) -- (Xe133) node[midway, below] {\rotatebox{90}{\tiny{(n, $\gamma$)}}};

  \draw[process] (Xe133) -- (Cs133) node[midway, right] {\tiny{$\beta^{-}$}};
  \draw[process] (Xe135) -- (Ba135);
  \draw[process] (Cs133) -- (Cs134);
  \draw[process] (Cs134) -- (Ba134);
  \draw[process] (Ba134) -- (Ba135);
  \draw[process] (Ba135) -- (Ba136);
  \draw[process] (Ba136) -- (Ba137);
  \draw[process] (Ba137) -- (Ba138);
  \draw[process] (Ba138) -- (Ba139);
  \draw[process] (Ba139) -- (13,8);

  \node at (0.35,1.48) {\tiny{s-process}};
  \node (rp) at (9.5,-2.25) {\tiny{r-process}};

  \draw[process] (3.225, -3) -- (1.225, -1.25);
  \draw[process] (6.225, -3) -- (4.225, -1.25);
  \draw[process] (9.225, -3) -- (7.225, -1.25);
  \draw[process] (12.225, -3) -- (10.225, -1.25);
  \draw[process] (15.225, -3) -- (13.225, -1.25);

\end{tikzpicture}

\caption{Example of s- and r-process paths for Xenon (Xe), Cesium (Cs) and Barium (Ba). Adapted from \citet{Sneden2008}.}
\label{fig:sr_process}
\end{figure}

The s-process happens when an isotope captures a neutron. Thus it is first important to establish where the neutrons come from. As illustrated in Fig. \ref{fig:nucleo4_alt}, typically two main reactions are associated with the s-process: $^{22}$Ne($\alpha, n$)$^{25}$Mg and $^{13}$C($\alpha, n$)$^{16}$O. $^{22}$Ne can be produced from $^{14}$N reaction at high temperatures of $T\geq10^8$~K:
\begin{align}
^{14}\text{N} + \alpha &\rightarrow ^{18}\text{F} + \gamma \\
^{18}\text{F} &\rightarrow ^{18}\text{O} + e^+ + \nu \\
^{18}\text{O} + \alpha &\rightarrow ^{22}\text{Ne} + \gamma 
\end{align}

Where large enough concentrations of $^{14}$N have been produced during the CNO-cycle. On the other hand, the amount of $^{13}$C produced in CNO-cycle might be too small. Thus it is thought that they are produced in AGB stars during the third dredge-up event when $^{12}$C mixes with proton-rich area:
\begin{align}
^{12}\text{C} + p &\rightarrow ^{13}\text{N} + \gamma \\
^{13}\text{N} &\rightarrow ^{13}\text{C} + e^+ + \nu
\end{align}

These two reactions ma produce neutrons needed for the s-process. The actual neutron-capture process can be summarised using the reaction:
\begin{equation}
\label{eq:neutron_capture}
\text{(Z,A)} + n \rightarrow \text{(Z,A+1)} + \gamma ,
\end{equation}

\noindent where A is the mass number and Z is the atomic number of the element. This reaction continues until it becomes an unstable enough element (i.e. half-life is shorter than the neutron-capture timescale), resulting in a decay:
\begin{equation}
\text{(Z,A+1)} \rightarrow \text{(Z+1,A+1)} + e^{-} + \bar{\nu}
\end{equation}

An example of an s-process path can be seen in the Fig. \ref{fig:sr_process}. First an isotope $^{132}$Xe undergoes a neutron capture, resulting in a release of $\gamma$ (hence denoted as (n, $\gamma$)), moving to the right on the diagram. $^{133}$Xe is an unstable isotope with half-life of 5.2 days, hence it undergoes $\beta^{-}$ decay into an isotope $^{133}$Cs with a higher atomic number (an arrow moving to the left and up). This isotope continues to undergo neutron captures until radioactive $^{134}$Cs, which decays to a higher mass number isotope once again. 

That's why this is called a slow-process, because the isotope has a possibility to undergo $\beta^{-}$ after every capture stage, if the isotope has a sufficiently short half-life time. The exact threshold will depend on the site and neutrons densities, hence different amount of isotopes can be produced in the different s-process sites. Hence it is also often called a path along the valley of beta stability, since the isotopes will decay unless they are stable enough.

\subsection{r-process}

\begin{figure*}
\includegraphics[width=1.0\textwidth]{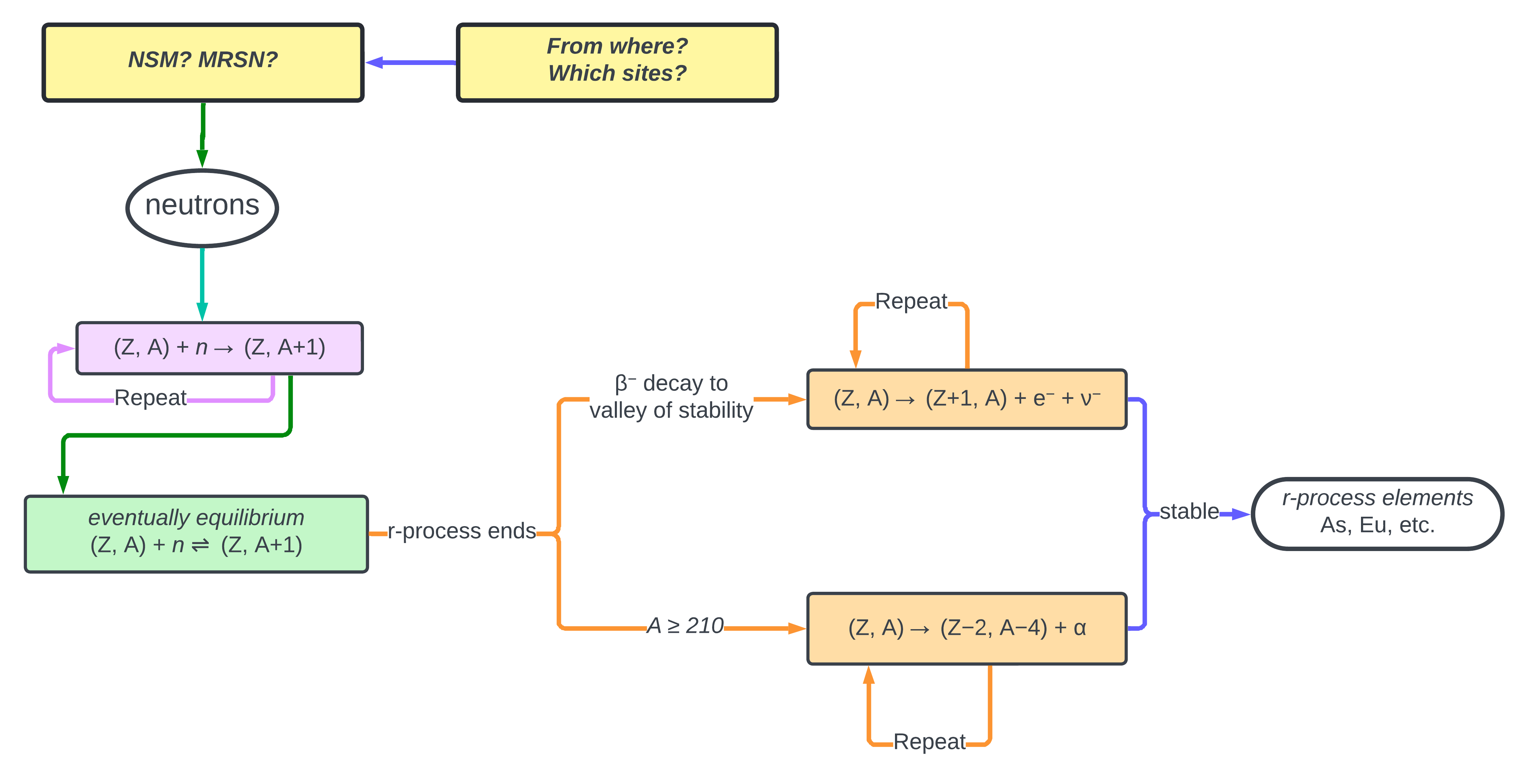}
\caption{A very simplified version of the r-process. Potentially more sites contribute to the neutrons, so only some of the potentially major ones are mentioned here.}
 \label{fig:nucleo5_alt}
\end{figure*}

The neutron sources for the r-process are much more heavily debated. Yet the actual r-process can be described exactly the same as in the equation \ref{eq:neutron_capture} (see also Fig. \ref{fig:nucleo5_alt}). However, the neutron densities in r-process sites are many magnitudes higher, so the isotopes keep undergoing neutron captures. Once the neutron densities drop, the isotopes typically undergo succession of $\beta^{-}$ decays. Fig. \ref{fig:sr_process} shows the final moments of the r-process, as the isotopes reach the stable isotopes after the beta decays. Something interesting that can be noticed is that the isotope $^{134}$Xe is stable. Hence r-process can produce it and cannot decay further to $^{134}$Ba. However, since $^{133}$Xe is radioactive, s-process would be very unlikely to produce the heavier isotope $^{134}$Xe. Hence, we have some isotopes that are produced only by r-process (here $^{134}$Xe), because they lie just beyond the s-process's reach. On the other hand, stable isotopes (in this case $^{134}$Ba) are "shielded" from the r-process by another stable isotope. 

Thus, we can have s, r, and sr isotopes, allowing to distinguish between different processes by looking at abundances of different elements (and in ideal case isotopes).


\bsp	
\label{lastpage}
\end{document}